\documentstyle{article}
\begin{document}

\title{The rotation curve of spiral galaxies and its cosmological
implications}

\author{E. Battaner \& E. Florido \\ Dpto. F\'{\i}sica Te\'orica y del
Cosmos. Universidad de Granada. Spain}
\maketitle

\begin{abstract}

We review the topic of rotation curves of spiral galaxies emphasizing
the standard interpretation as evidence for the existence of dark
matter halos. Galaxies other than spirals and late-type dwarfs may
also possess great amounts of dark matter, and therefore ellipticals,
dwarf spirals, lenticulars and polar ring galaxies are also
considered. Furthermore, other methods for determining galactic dark matter, such
as those provided by binaries, satellites or globular clusters, have to
be included. Cold dark matter hierarchical models constitute the
standard way to explain rotation curves, and thus the problem becomes
just one aspect of a more general theory explaining structure and
galaxy formation. Alternative theories also are included. In the
magnetic model, rotation curves could also be a particular aspect of
the whole history of cosmic magnetism during different epochs of
the Universe. Modifications of Newtonian Dynamics provide another
interesting possibility which is discussed here. 

\end{abstract}


\vskip 1.5cm
\tableofcontents
\vskip 1.5cm

\section{Introduction}

There is a wide consensus that the rotation curves of 
spiral galaxies constitute an observational proof -perhaps the best 
proof- for the existence of dark matter in the Universe. Dark matter 
is of evident interest in Cosmology, hence the interest of a review
on the topic, in this case from a post-graduate didactic point of view.

Dark matter is not an exotic or sophisticated hypothesis. Neutrinos, 
brown dwarfs and
black holes are all candidates to be identified with dark matter
and are, nonetheless, classical concepts in Physics, introduced by 
fully established theories. High Energy Physics actually predicts a number
of particles that do not interact with photons. We cannot 
claim that all the existing matter
emits or absorbs photons.

However, although the necessity of dark matter was proposed more than 
60 years ago (Zwicky, 1937), this hypothesis is still not whole-heartedly accepted by some 
workers. This scepticism is never explicitly expressed, but is 
subjacent, implicitly revealed in sentences such as
"a galaxy {\it{has}} a halo". We should rather say  "a galaxy {\it{is}} a halo" 
as a galaxy's mass may be at least 10 times its visible mass,
the visible element 
then being a mere minor component, only important for us because it is what
we see.

(We must, however, accept that when we state that if a galaxy {\it{is}} a
halo, the discovery of just one exception, i.e. of a visible galaxy
with no halo, would represent a serious problem of interpretation, a problem as
fundamental as  finding the light of a firefly not associated with a firefly.)

Once the  existence of dark matter is recognized as a 
conservative possibility, let us establish a difference between the
problem of dark matter in the Universe and the problem of dark matter
in galaxies. We implicitly assume throughout
the paper that $\Omega \sim 1$, and that visible matter only 
contributes with $\Omega_{V} \sim 0.003$. The total matter contributes
either with $\Omega_M \sim 1$, as classically assumed, or with $\Omega_M \sim
0.3$, coherent with the more recently assumed scenario deduced from the
observations of early supernovae, leading to the re-accelerating Universe, the non-vanishing
cosmological constant or other identifications of dark energy (e.g. Turner, 1999).
The assumption of a plane Universe, $\Omega \sim 1$, is based on theoretical
ideas about inflation, observations of large-scale dynamics, the interpretation
of the CMB and even on some philosophical arguments. Our position on dark matter
is based on the difference between $\Omega_M \sim 0.3$ and $\Omega_{V}
\sim 0.003$, as observed. We only see, in the best of cases, 1\% of the matter.
This is the basic fact, rather than the existence and possible observation
of dark matter in galaxies, that demonstrates that dark matter exists.

Rotation curves and other dynamical effects in galaxies suggest
that total galactic mass is about 10 times larger than that
observed. Even in this case, we should still find about 90\% of dark
matter elsewhere in the Universe, which means, given the relative uncertainty in
all these figures, that practically all the required dark
matter lies outside galaxies. The possible galactic contribution to dark
matter is negligible to close the Universe. Therefore, the
existence of galactic dark matter is clearly very important for our
knowledge of galaxies and their dynamics, but not so decisive for the
cosmological problem of identifying its nature and amount.

This notion allows us to make a more objective analysis of the topic of dark matter
in galaxies, once it has been partially disconnected from the cosmological one.
Rotation curves of spirals, and many other observations, are currently interpreted
as evidence of the existence of massive large dark halos, but a critical
analysis of the observations, and the theoretical interpretations involved, is
permanently necessary. Even if dark matter is a major ingredient in
the most widely accepted and orthodox picture of a galaxy,  
most authors only differing about its mass and size, we will see that the
hypothesis of a total
absence of galactic dark matter cannot be completely ruled out.

The problems involved in  determining galactic dark matter may be summarized as follows:
the internal regions of galaxies require little or no dark matter and
we must examine the external ones, where there are few stars and we must
observe gas to determine the gravitational potential. However gas, in these
regions where the gravitational force is low, may be influenced by magnetic
fields. Thus, it is convenient to look for stellar systems lying far 
from the galaxy, and particularly satellite or companion galaxies.
Then, however, it is very difficult to distinguish between galaxies
with a halo
and halo with galaxies, i.e. the hypothesis of a large common halo is difficult
to reject.

Though the rotation curves of spirals is the main topic to be discussed, we 
must pay attention to other methods of determining DM, in particular
those based on globular clusters, satellite galaxies, binary systems,
polar-ring galaxies and so on. Lensing by spiral galaxies could become
an important tool in the near future (Maller et al. 1999). We should 
also review the methods of DM
estimation in other types of galaxies. With respect to the DM problem, spirals
do not seem to be exceptional. 

Other models, which do not require DM to explain rotation curves, have
been reported in the literature and also require our attention in this
study. One such model is MOND (Modified Newtonian Dynamics) and another is the 
magnetic model. The authors
 have contributed to the development
of this latter model, which is commented in some detail at the end, together
with its cosmological implications. Despite the inclusion of MOND and
the magnetic approach as interesting possibilities, through out this
paper we adhere to the most conservative point of view based on DM. Nevertheless,
we emphasize the difficulties inherent to most methods and models.

The purpose of this review is mainly didactic, as a bibliographic
source for postgraduate courses, but also critical. This critical approach might
be considered unnecessary, in view of the wide acceptance of the dark
matter hypothesis by the scientific community; but apart from its didactic
interest, it is always pertinent to reconsider apparently solid
beliefs. In this respect, it is convenient to remind the reader of four historical
aspects related to the early adoption of the DM halo hypothesis, which were
based on arguments that eventually became open to discussion or
were made obsolete:

a) Kahn and Woltjer (1959) considered the dynamics of the double system
formed by M31 and the Milky Way and concluded that their motion would
require a binary system mass of $\ge 1.8 \times 10^{12}
M_\odot$, much greater than the visible matter in the two galaxies. Kahn
and Woltjer formed the opinion that there existed an as yet unobserved mass in some
invisible form. Although they identified this invisible mass with hot
gas rather than dark matter in its present sense (unobserved rather
than unobservable) this work gave a first proof of the missing mass. It
should be remembered, however, that this mass was considered to lie
either in M31 and the Milky Way or in the intergalactic space (either 
two halos or a common halo). It is still not clear whether this second
possibility can be completely ruled out.  Therefore, this paper
established the existence of dark
matter, but not necessarily within galaxies. The doubt remained: either in or in between.

b) Oort (1960, 1965) found evidence for the presence of dark matter in
the disk of our galaxy, although van dem Bergh (1999) wrote: ``However, late
in his life, Jan Oort told me that the existence of missing mass in
the galactic plane was never one of his most firmly held scientific
beliefs''. After a long debate since then (see Binney and Tremaine,
1987; Ashman 1992) the discussion finally seems to be closed. Using
HIPPARCOS data, Crez\'e et al. (1998) have found that there is no
evidence for dark matter in the disk: gas and stars perfectly account
for the gravitational potential.

c) Ostriker and Peebles (1973) suggested that spherical dark matter halos
around the visible component of the spiral galaxies were necessary to
suppress bar instabilities. However, their arguments did not
convince Kalnajs (1983) and Sellwood (1985), who showed that a
central bulge was equally efficient to stabilize disks.

d) Babcock (1939) observed that the stars in M31 were rotating at an
unexpectedly high velocity, indicating a high outer mass-to-light
ratio, although he also considered  other possibilities: either strong
dust absorption or, as he stated, ``new dynamical considerations are required, which
will permit of a smaller relative mass in the outer parts''. This
sentence, quoted by van dem Bergh (1999), is interesting, bearing in
mind that gravitation is the only force considered at present to
explain rotation curves, while other ``dynamical
considerations'' are ignored. Optical rotation curves of other galaxies were
obtained, until those published by Rubin et al. (1980), that were
considered to be clear evidence of dark matter in galaxies. Today, however,
many authors consider that optical rotation curves can be
explained without dark matter (even without rejecting its
contribution). For instance, Broeils and Courteau (1997) by means of
r-band photometry and $H_\alpha$ rotation curves for a sample of 290
spirals concluded that ``no dark halo is needed''.

Therefore, the basic initial arguments leading to the belief of dark matter
halos around spiral galaxies failed or were not conclusive. Only
the interpretation of the HI rotation curves by Bosma (1978) subsists
unmodified. 
For early thoughts on dark matter, the paper by
van dem Bergh (1999) and the Ph.D.  Thesis of Broeils (1992) are
essential reading.

In brief, there is a standard interpretation of the rotation curve
of spiral galaxies that is implicitly adopted throughout this paper,
but we should not ignore other possibilities.

\section{Rotation curves of spiral galaxies and dark matter}

At large distances from the galactic centre the gravitational
potential should be that produced by a central point mass and, in
the absence of forces other than gravitation, it should be expected
that $GM/R^2 = \theta^2/R$ ($G$, universal gravitation constant; $M$,
galactic mass; $R$, galactocentric radius; $\theta$, rotation
velocity), therefore $\theta \propto R^{-1/2}$, which is called, for
obvious reasons, the Keplerian rotation curve. This Keplerian decline was
not observed, but rather, flat rotation curves with $\theta$=cte were
obtained. Apparently, this has the direct implication that $M \propto R$,
thus depending on the quality of the telescope used. The ``Dark
Matter'' (DM) hypothesis interprets this result in the sense that the
Keplerian regime holds at much greater distances than those
at which we obtain observations. There should be great quantities of dark matter
extending far beyond the visible matter in a more or less spherically
symmetric DM halo. If its distribution is spherically symmetric, the
mass interior to a sphere of radius $R$ would be $M(R) \propto R$, so
that we obtain a first rough model of DM density distribution:
$\rho = (1/4\pi R^2)dM/dR = \theta/4\pi GR^2$, i.e. $\rho \propto
R^{-2}$, for distances far beyond the visible radius. This model is
obviously over simplified, as we will see, but it coincides with
the so called ``nonsingular isothermal'' profile
\begin{equation}
  \rho ={\rho_0 \over {1+ \left({R \over R_0}\right)^2}} 
\end{equation}
(with $\rho_0$ and $R_0$ being constants), one of the most frequently types
of halos.

The interpretation of rotation curves of spiral galaxies as evidence
of DM halos was probably first proposed by Freeman
(1970) who noticed that the expected Keplerian decline was not present
in NGC 300 and M33, and considered an undetected mass, with a
different distribution for the visible mass. The observation of flat
rotation curves was later confirmed and the DM hypothesis reinforced
by successive studies. Rubin, Ford and Thonnard (1980) and Bosma (1978,
1981a,b) carried out an extensive study, after which the existence of DM
in spiral galaxies was widely accepted. Van Albada et al. (1985)
analyzed the rotation of NGC 3198 and the distribution of its
hypothetical DM, concluding that this galaxy has a dark halo, in
agreement with the paper by Ostriker and Peebles (1987) about the
stability of disks. The rotation of spirals was soon considered the most
solid proof for the existence of DM in the Universe, particularly
important when it was later believed that $\Omega =1$. Other decisive
papers were produced by Begeman (1987) and Broeils (1992).

The initial conclusion could be schematized by considering the
rotation curve to be high and flat. If it is high, the dark halo should
be very massive; if it is flat, the dark halo should be very
large. Indeed, the flatness of the rotation curves  was explained
``too'' well, because if the disk
and halo had such different distributions, very careful matching was
required between the falling disk rotation curve and the rising halo
one. The curve was ``too'' flat; there was a ``disk-halo conspiracy''
(Bahcall and Casertano, 1985, van Albada and Sancisi,  1986).

The only explanation offered for this ``conspiracy'' is the adiabatic
compression of the halo material when the disk was formed (Barnes
1987, Blumenthal et al. 1986) (which is commented later) although Bosma
(1998) gave a list of galaxies for which this mechanism is not fully
operational. The disk-halo conspiracy is a problem that remains to be 
satisfactorily solved. The problem is not why curves are flat (not all are
flat) but why the transition from disk to halo domination is so
smooth.

Different procedures have been used to obtain dark matter
distribution: stellar distribution is determined from photometric observations
and must then be complemented with CO and HI observations
(with a correction factor to include the He mass) mainly for late spirals,
in order to assess the gas profiles. These data determine the densities of
bulge, disk and gas in the disk, or rather their contribution to the
rotation velocity through the so called ``circular velocity'',
$V_c(R)$, which would coincide with the true rotation velocity
$\theta$, if the component were dominant in the galaxy. The rotation
curve, $\theta(R)$, is determined mainly with 21 cm maps. The addition
of the different visible components does not, in general, coincide
with $\theta(R)$, from which the existence of a DM halo is deduced.

Then, to obtain its distribution, there are several different
techniques. One of the most widely used is the ``maximum disk
hypothesis'' (see for instance, Begeman, Broeils and Sanders,
1991). Here, the mass to light, M/L, ratio is fixed for the different visible
components, with values higher than about 10 being difficult to assign
to a stellar population. Then, the innermost regions are adjusted so
that the disk is able to produce the observed rotation curve without a
halo. The disk M/L obtained is then kept constant at all radii and the
circular velocity of the halo is then obtained for higher
radii. Another possibility is the so called ``best  fit'' technique. In this case,
it is necessary to adopt a halo profile defined with several
adjustable parameters. Most decompositions have adopted the isothermal
spherical profile. At present, it might be profitable to
investigate the alternative NFW profile, as this has a higher
theoretical justification (we will come back to this point in the
section devoted to theoretical models). The problem with the best fit
procedure is that the halo distribution function must be known 
although, in part, this is precisely what we want to obtain.

The maximum disk technique was introduced by van Albada and Sancisi
(1986). There are some psychological aspects to their introduction: 
``dark matter is a daring assumption; the intention is therefore to make
the halo as small as possible, at least in the traditional optical
best known innermost regions and reserve the exotic physics
for the outer radio observable regions. It can therefore be found as
noticeable that ``maximum disk'' fits are reasonable and do not very
much differ from other fits. This gives us a first information: the
inner parts do not require large amounts of dark matter''. This
conclusion was ``a
priori'' not obvious. At present, it is considered that the amounts of
dark and visible matter in the optical disks are similar, with not so
much DM being needed as in the outermost disks.

The basic initial description consisting of an innermost region in
which $\theta(R)$ increased linearly followed by a constant
$\theta$ in the outer region was soon considered too simple. Casertano
and van Gorkom (1991) found galaxies with declining rotation curves
and analyzed current observations to show that bright compact galaxies
have slightly declining rotation curves and that rising curves are
predominant in low-luminosity galaxies (see also Broeils, 1992). This
latter fact indicated that low-luminosity galaxies are more DM rich
and that, in general, there is an increase in the dark matter fraction
with decreasing luminosity (Persic and Salucci, 1988, 1990). 
Nearly all rotation curves belonging to
the different types of spirals  can be described by means of a 
single function, the so
called ``Universal Rotation Curve'' (Persic, Salucci and Stel, 1996;
Salucci and Persic, 1997) which is a successful fit of  galactic
astronomy that will be commented later.

\subsection{Some examples of typical rotation curves}

Figures 1, from the PhD Thesis of Begeman (1987), 
illustrates the photometric profiles used to obtain
the bulge and disk contributions to the  total rotation velocity, the
observational rotation curve and the circular velocity of the
halo. This is calculated from
\begin{equation}
  \theta =[V_{gas}^2 + V_{disk}^2 + V_{bulge}^2 + V_{halo}^2]^{1/2}
\end{equation}
where the V's are the circular velocities of the different components.
With the ``maximum disk'' hypothesis we previously estimate
$V_{disk}^2$ (in reality, also $V_{bulge}^2$), which determines M/L. It
is usual to assume that the disk is exponential; then the disk
circular velocity is calculated with the formula deduced by Casertano
(1983) and that of the bulge with the formula given by Kent (1986). If
the halo is spherically symmetric its circular velocity is simply
$V_{halo}^2 =GM(R)/R$.

Note that the bulge, stellar disk and gas disk are insufficient to
give the observed $\theta(R)$, and therefore either dark matter is needed
or forces other than gravity are involved. Note also that, even if the
maximum disk method gives reasonable results and real disks are
probably near maximum disks, we assume $V_{halo} =0$ in the central
region, which implies $\rho_{halo}=0$ in the central region, which is, 
from the point of
view of galaxy formation, highly improbable. This fact, however, does
not in practice introduce a serious problem.

On the other hand, we reproduce the rotation curve of the Sd galaxy
NGC 1560 from the Broeils thesis. Figure 2 shows the obtention of
the rotation curve from a velocity-position map. It is apparent that
the curve is always an increasing function with no sign of becoming flat,
which is typical of small galaxies. Fig 3 gives the contribution of the
different components, following three methods, the ``maximum disk'',
the ``best fit'' and the ``minimum disk''. This latter method is also
called the ``maximum halo'', with a fixed M/L of only 0.1 (measured in
solar units, $M_\odot/L_\odot=1$ for the
stellar disk, so that the disk gas makes the main contribution to
the observable mass. This could be a reasonable assumption, as dwarf galaxies may
be dominated by non-luminous material even in the innermost
region. The ``best fit'' method again assumes a spherical isothermal halo.

\subsection{Some problems}

Direct inspection of the input data reveals some
inherent problems in data handling. One is that this type of analysis
is made under the assumption that disks are exponential. This
``hypothesis'' is reasonable as a zeroth order description, but there
is an immoderate use of it in the literature. In Fig. 4 we see
several photometric profiles of two galaxies (Begeman 1987). Rather
than being an exception, the
case of NGC 5033 is fairly typical. An
extrapolation is needed in the bulge region (about 1 arcmin), there is
a ring
(about 1 arcmin) and a truncation at large radii (about 1 arcmin) and
the stellar disk is less than 6 arcmin in radius.

The truncation of the disk (van der Kruit, 1979; van der Kruit and
Searle, 1981a,b,1982a,b; Barteldrees and Dettmar, 1994) is very
noticeable, for instance, in the galaxy NGC 5033. In Fig. 4 we plot
the photometric surface brightness (in mag arcsec$^{-2}$) of this
galaxy, showing a spectacular truncation, and the decomposition of the
rotation curve following Begeman (1987). Is this disk really
exponential? 

Another problem may arise from the fact that the rotation curve is
usually measured at 21 cm but the stellar disk in the optical. The
stellar disk and the gas in the disk usually corotate, but due to
frequent mergers and the accretion of clouds, captures, etc, this is not
always the case. Unfortunately, non-corotation is more frequent 
than is generally assumed and
very often the rotation curve of stars and of the gas differ
greatly. Figures 5 and 6 are two examples taken from the Ph. D. thesis of
Vega-Beltran (1997). In NGC 3898 the ionized gas and the stars not only
counterrotate, but the shapes of the two rotation curves are quite
different. The spiral galaxy IC 4889 is another surprising example:
gas and stars do not counterrotate but the gas exhibits a typical flat
curve while the stars decline in a Keplerian-like fall-off. These two examples
are not an exception: important deviations from corotation are found
in about 14 out of 22 galaxies in the  
Vega-Beltran sample, where gas and star rotation curves were measured
independently.

There is another problem raised by Sofue and collaborators (Sofue,
1999; Sofue et al. 1999). These authors made a detailed measurement of the
innermost rotation curve based on millimeter CO observations, because
the frequent HI hole in many galaxies impedes a proper observation using
21 cm. Their rotation curves have a steeper central increase, followed
by a broad maximum in the disk and the characteristic flat rotation
due to the massive halo. As the interest of rotation curves lies
traditionally on the periphery, the central region has been
neglected. From the observational point of view, this
region is particularly difficult to study, specially in edge-on
galaxies; a novel technique, the so called ``envelope-tracing''
method, has been used, in contrast with other current methods. Its
logarithmic rotation curve of the Milky Way is reproduced in figure
7 and compared with the linear one of Clemens (1985) and Honma and
Sofue (1997). The inner curve is of Keplerian type due to a central
black hole. Observe that the curve should begin at the origin,
$\theta(0) =0$, but this very central steep increase has not yet been
observed. Other galaxies have been observed to have a central fast
rotation as in the Milky Way. These indicate the existence of dark
matter in the centre, probably that of a black hole, which is also
important in Cosmology, but the processes involved are different from those
affecting the problems considered  in this
paper. Sofue et al. present rotation curves of 50 spirals, with a steep
central rise, warning that previous curves could be incorrect
in the centres, with the outermost regions remaining unaltered. 

It is necessary
to study how these results modify the standard methods of interpreting
rotation data. For example, if the inner part of the rotation curve is
steeper (Swaters, Madore and Trewhella, 2000), then M/L is
understimated and, when using the maximum technique the luminous
matter contribution is also understimated.

\subsection{The Bosma relation}

Bosma (1978, 1981b) and Carignan et al. (1990) found a trend for the gas
distribution to have the same shape as DM distribution. This
correlation between gas and DM is puzzling and if real, has no easy
explanation in the light of present CDM models. Not only is there a
general trend, but several individual features found in the rotation
curve seems to correspond to features in gas circular
velocity. This can be observed in Fig. 2 for NGC 1560 and Fig.
8 for NGC 2460.

This fact has inspired a theory (commented in 
section 3.1), identifying the dark matter with an as
yet undetected dark gas. The magnetic hypothesis would provide
another explanation, as the rotation curve is due in part to magnetic
fields, which are generated by gas. A direct relation is
not obvious when very extended curves are obtained (Corbelli and
Salucci, 1999). Bosma (1998) himself states that this relation may not be
correct.

\subsection{The ``Universal rotation curves''}

Persic, Salucci and Stel (1996) and Salucci and Persic (1997) have
analyzed a large number of rotation curves mainly catalogued by Persic
and Salucci (1995) taking into account the $H\alpha$ data published by
Mathewson, Ford and Buchhorn (1992) and also adopting some radio rotation
curves. They claim that rotation curves can be fitted to what they
call the ``universal rotation curve'' not only for any luminosity,
but also for any type of galaxy, including spirals, low-surface-brightness,
ellipticals and dwarf irregular galaxies.  The existence of a
``universal'' rotation curve had previously been
claimed by Rubin et al. (1980). Let us then reproduce the
formulae of these ``universal rotation curves'', or PSS curves,
restricting ourselves to spirals.

Following Persic and Salucci (1997), rotation curves of spirals can be 
fitted by a combination of two
components. The first is an exponential thin disk, whose circular
velocity can be approximated in the range $0.04 R_{opt} < R \le
2R_{opt}$ as
\begin{equation}
  V_{disk}^2 = V^2(R_{opt})\beta {{1.97 x^{1.22}} \over {(x^2 +
  0.78^2)^{1.43}}} 
\end{equation}
where $x$ is a radial variable taking $R_{opt}$ as unit
\begin{equation}
  x = {R \over R_{opt}}
\end{equation}
$R_{opt}$ is the radius encircling 83\% of the light; for an
exponential disk $R_{opt} = 3.2 R_D$ where $R_D$ is the radial scale
length. $\beta$ is a constant that depends on the luminosity. This
function $V_{disk}$ does not give a Keplerian fall-off for $x
\rightarrow \infty$, nor is it the general expression of $V_{disk}$ for
exponential disks, but its application is restricted to a radial
range.

The other component is the halo, with a circular velocity expressed as 
\begin{equation}
  V_{halo}^2 = V^2(R_{opt}) (1-\beta)(1+a^2){x^2 \over {x^2+a^2}}
\end{equation}
where $a$  is another constant, also depending on the
luminosity. Then, the PSS curve is given by
\begin{equation}
  V = \left(V_{halo}^2 + V^2_{disk} \right)^{1/2}
\end{equation}
the contribution of a bulge therefore being considered negligible. The
constants $a$ and $\beta$ are functions of the galaxy's luminosity, the
best results being obtained for
\begin{equation}
   \beta =0.72 +0.44 \lg{L \over L_*}
\end{equation}
\begin{equation}
  a= 1.5 \left( {L \over L_*} \right)^{1/5} 
\end{equation}
where $L_* = 10^{10.4}L_\odot$. Then for a galaxy with luminosity
$L_*$, $a$ corresponds to a value of $x$ of the order of
$R_{opt}$, exactly $1.5 R_{opt}$. Note that these values
provide good fits, but for $L > 4.33 L_*$ give negative values of $V_{halo}^2$.

Hence
\begin{equation}
  M_{halo}(x) = M_{halo}(1) (1+a^2){x^3 \over {x^2 + a^2}} 
\end{equation}
(for a spherically symmetric halo)
and
\begin{equation}
  \rho ={1 \over {4\pi R^2}} {{dM} \over {dR}} \propto {{x^2+3a^2}
  \over {(x^2+a^2)^2}} 
\end{equation}

with very little in common with the NFW theoretical halos (see later). They are
reasonable, because 

\begin{itemize}
 \item The x-derivative of $\rho(x)$ vanishes at $x=0$, which is physically satisfactory.

 \item Then, for $x \ll a$ the density slowly decreases; 
 for $x=a$, the density is still 1/3 of the central value, i.e. there
 is a ``core'' of radius $a$, which is therefore called the halo core
 radius.

 \item The density does not vanish for any value of $x$, i.e. there is
 no sharp boundary. The density always decreases.

 \item For $x \gg a$, the density decreases as $x^{-2}$ (compared
 to $x^{-3}$ in NFW halos).
\end{itemize}

This is reminiscent of the non-singular isothermal sphere, with a faster
decrease from the centre out to the core radius, both of which for large
$x$ obey $\rho \propto x^{-2}$.

As $a \propto L^{1/5}$, i.e. low luminosity galaxies are
much more concentrated. For a galaxy with $L_*$ the core is of the
order of the optical radius.

A non-physical property of the PSS halo density profile 
is that $M$ does not converge for very large values of $x$,
but rather linearly increases with $x$, with the mass of any
halo being infinite. To surmount this difficulty the halo mass was
defined as that at
$R_{200}$, where $R_{200}$ is the radius of a sphere within which the
mean density is 200 times the mean density of the Universe, as also
defined in theoretical models. Then
\begin{equation}
  M_{halo} \approx M_{200} = M(R=R_{200}) = {4 \over 3} \pi
  R^3_{200} 200 \rho_c 
\end{equation}
where $\rho_c$ is the critical density of the Universe. We see
therefore that $M_{200} \propto R_{200}^3$. Following equation (5),
$V_{200}$ (the circular velocity at $R_{200}$) is a complicated
function of $R_{200}$, but according to these authors, it can be
approximated to
\begin{equation}
  R_{200} = 250 \left( {L \over L_*} \right)^{0.2}
\end{equation}
As the exponent, 0.2, is very small, the radii of the halos are
relatively independent of the luminosity. A galaxy with $10 L_*$ would
have a halo only 1.5 times larger than a galaxy with luminosity $L_*$
(this cannot be checked directly as also equation (5) cannot be applied
to galaxies with $L > 4.33 L_*$). The authors propose
\begin{equation}
  M_{200} = 2 \times 10^{12} \left( {L \over L_*} \right)^{0.5}
  M_\odot 
\end{equation}
(Note, however, that from $R_{200} \propto L^{0.2}$, together with the exact relation
$M_{200} \propto R_{200}^3$, we should obtain $M_{200} \propto
L^{0.6}$. The small difference in these exponents --0.5 and 0.6--
arises from the complexity of the problem). Therefore, the brighter
galaxies have a halo that is more massive, but only slightly larger. The
mass-luminosity ratio is then 
\begin{equation}
  {{M_{200}} \over L} \approx 75 \left( {L \over L_*} \right)^{-0.5}
\end{equation}
Brighter galaxies have smaller mass-to-light ratios, hence the dark
matter has more dominant effects in small or low-surface brightness
galaxies. We can also calculate the luminous to dark matter ratio
\begin{equation}
  {{M_{lum}} \over {M_{200}}} =0.05 \left( {L \over L_*} \right)^{0.8}
\end{equation}
Bright galaxies have relatively smaller dark matter halos, while
the very bright galaxies nearly reach the maximum $M_{lum}/M_{dark}$ ratio
($\sim 0.1$) established from primordial nucleosynthesis models for
the baryonic $\Omega_B$.

For $x \gg a$, a constant value of $V_{halo}$ is obtained $V_{halo}^2 =
V^2(R_{opt}) (1-\beta) (1+a^2)$. For galaxies with $L \sim L_*$, it is
obtained that $V_{200} \sim V(R_{opt})$ which is readily interpreted: if a
constant $V(R_{opt})$ is observed in a region already dominated by
dark matter, it should be related to the halo circular velocity at
large distances.

As $V^2_{200} = G M_{200}/R_{200} \propto L^{0.5}/L^{0.2} = L^{0.3}$
we should have $V_{200} \propto L^{0.15}$. In binary galaxies, which
are considered later, $V_{200}$ could be identified with the orbital
velocity of the secondary galaxy, statistically related to the
difference of the two projected velocities along the
line-of-sight. $L$ would be the luminosity of the primary. A
correlation between $L$ and $V_{200}$ has not been found
(e.g. Zaritsky, 1997). This is, in part, justified as the
exponent, 0.15, is so small that the orbital velocities are nearly
independent of the luminosities. Theoretical models also agree in
this respect.

Note that, if instead of $M_{200} \propto L^{0.5}$ as determined by these authors, we
had taken $M_{200} \propto L^{0.6}$ from the definition of
$M_{200}$ as mentioned above, we would have obtained $V_{200} \propto
L^{0.2}$, and $L \propto V^5_{200}$, closer to the observational
infrared Tully-Fisher relation $L \propto V^4(R_{opt})$, if
$V(R_{opt})$ were close to $V_{200}$.

White et al. (1983) and Ashman (1992) proposed $M_{200}/L \propto
L^{-3/4}$, in which case $R_{200} \propto L^{0.08}$, indicating a
lower dependence of $R_{200}$ on $L$, and $V_{200} \propto L^{0.08}$.
We will see later that these undetected correlations have a
natural explanation in the magnetic model of the rotation curves.

The universal rotation curves also give two relations
\begin{equation}
  {{R_{core}} \over {R_{200}}} = 0.075 \left({{M_{200}} \over
  {10^{12} M_\odot}} \right)^{0.6} 
\end{equation}
\begin{equation}
  \rho_{halo}(0) = 6.3 \times 10^4 \rho_c \left({{M_{200}} \over
  {10^{12} M_\odot}}\right)^{-1.3} 
\end{equation}

where $\rho_c$ is the critical density of the Universe. Hence,
  brighter galaxies have relatively large core radii and small values
  for the central halo density. Therefore, even the central region of 
  low-brightness galaxies is dominated by dark matter, while bright
  galaxies have their internal regions dominated by visible
  matter. These relations are important and confirmed by the NFW
  theoretical halos, even if the universal rotation curves do not have 
  much in common with those deduced by the former.

The same formulae are valid for low-surface-brightness galaxies. In
this case, the dark matter would be completely dominant, with a core
radius of about 5-6 kpc.

Summarizing, the most interesting fact in the fitting effort made by
these authors is that such a large variety of galactic types have
rotation curves which can be adjusted to a single universal rotation
curve (even for ellipticals, not considered here). This fitting
assumes the existence of a dark halo that does  not coincide with the
universal halo profiles obtained by most theoretical models but which
is very reasonable (except, perhaps, in that they have an infinite
mass, which in practice is not a real problem). The explanation of the
puzzling behaviour of binary systems, however, is still not completely
satisfactory.

There is another general comment to be made. The universal rotation
curve is a fitting problem. But this fitting should be interpreted in other
models in a different way. Therefore, even if terms like ``dark
matter'' and ``dark halos'' are used, this fitting does not prove the
existence of dark matter in galaxies.

Bosma (1998) considered that the notion of universal rotation curves
breaks down. He observed several galaxies with a high rotation velocity, 
but non-declining rotation curves. This could be due to the
inclusion in Persic and Salucci's sample of very inclined galaxies, where
opacity problems are difficult to handle when using $H_\alpha$
rotation curves. Verheijen (1997) also found 10 out of 30 galaxies in
the Ursa Major clusters for which the rotation curves do not fit 
the universal rotation curves. Despite all these exceptions the
scheme introduced by Persic, Salucci and collaborators, provides a
first fit that theoretical models should take into account.

\subsection{Dwarf irregular galaxies}

Dwarf irregular galaxies can be considered  extreme late-type
spirals, at least as concerns the rotation curve and the
associated dark matter. Many of them present well defined rotation
curves that can be obtained from HI measurements. With respect to normal
spirals, dwarf irregulars have the advantage of presenting a large
gaseous component, and so rotation curves can be traced better and
to much larger radii, up to 17 radial scale lengths. Indeed, the rim
of the halo has probably been detected (Ashman, 1992).

Kerr, Hindman and Robinson (1954) and Kerr and de Vaucouleurs (1955)
showed that the LMC and the SMC were rotating. This result was
extended to other irregulars showing that most late-type dwarf
galaxies rotate, although their rotation velocities are lower, of the
order of $\sim 60 km s^{-1}$. It was also established that the rotation
curve rose slowly to the last measured point. These galaxies
were soon considered ideal to study galactic dark matter, not
only because the absence of a bulge made analysis simpler, but
mainly because the rising curves greatly differed from the expected
Keplerian decline. First results (Carignan, 1985; Carignan, Sancisi
and van Albada, 1988) seemed to indicate that these galaxies have dark
matter properties similar to those of normal spirals and that the
inner parts do not require great amounts of dark matter. This trend
was not confirmed later, and the commonly accepted picture was that
the contribution of the disk is insignificant and that they are
dominated by dark matter at all radii. For more details of this
history see the thesis by Swaters (1999).

But this conclusion was based on a very small sample of galaxies. The
studies by Broeils (1992) and Cot\'e (1995) were based on only eight
late-type dwarf galaxies, a small number taking into account the large
spread of dark matter properties that this type of galaxies
presents. Recently, Swaters (1999) has carried out the greatest effort
made to date to systematically observe and analyze this problem. It is also
important to have a large sample observed and reduced with the same
techniques. To determine the dark matter amounts, it is necessary to
obtain photometric maps. In this study, this was done for 171 galaxies
at the 2.5m INT at La Palma. Of these, 73 were observed in HI with
the Westerbrok Synthesis Radio Telescope. Rotation curves were
obtained for 60 of them, and detailed dark matter models were
carried out for 35. Clearly, the results obtained in this work are
based on the largest and most homogeneous sample. In general,
these results did not confirm the previous widely accepted picture;
late-type dwarfs are not essentially different from normal
brighter spirals, which is more in agreement with the pioneering
interpretations.

Despite their apparent loss of symmetry, the exponential decline of
typical disks is usually observed in these HI rich galaxies. Swaters
found that the rotation curves flatten after about two disk scale
lengths. There are several galaxies with fairly flat rotation curves
with amplitudes as low as 60 kms$^{-1}$. The main difference with
rotation curves of spiral galaxies is that no cases of declining curves
were found, which was explained by the fact that these galaxies have
no bulge at all or only a small one while bright spirals with 
declining curves do have a large bulge.

The outer slope as a function of R-magnitude is plotted in figure 9
and includes both bright galaxies (from Broeils, 1992) and
galaxies belonging to the Ursa Major cluster (Verheijen, 1997). It is
seen that the variation in slopes is larger in late-type spirals. 

In Fig. 10 we reproduce the Tully-Fisher relation from Swaters
(1999) that extends the relation to fainter types. We observe $L
\propto V_{max}^\alpha$, where $\alpha \sim 4.4$. It is also observed
that late-type dwarfs rotate noticeably faster than predicted by the
Tully-Fisher relation.

In general, late-type galaxies are not dominated by dark matter within the
optical disk for radii less than about four scale lengths. The
required stellar mass-to-light ratio is however greater than in bright
spirals, of the order of 10, reaching values as high as 15. Maximum
disk models fit the obtained rotation curves reasonably well, but other
models cannot be excluded.

Many irregulars are satellites of bright
galaxies or at least of a small group of galaxies as in the Local
Group. Here, a gradation in the properties from dEs to
dIrrs would support the hypothesis that Irrs could eventually evolve
into dEs (e.g. Aparicio et al. 1997; Martinez-Delgado 1999). Phoenix
could be a clear example of an intermediate type. Moreover, dEs are
preferentially distributed close to the largest galaxies, while dIrrs
are found in the outskirts (the Magellanic Clouds are
exceptions) (van dem Bergh 1999).

Salucci and Persic (1997) proposed that the ``universal rotation
curve'' was also valid for dwarf irregulars, though then the large amount
of data in the thesis of Swaters was not available. In this case, the
calculation of $a$ and $\beta$ is given by different formulae:
\begin{equation}
  a= 0.93 \left({{V_{opt}} \over {63 kms^{-1}}} \right)^{-0.5} 
\end{equation}
\begin{equation}
  \beta = 0.08 \left({{V_{opt}} \over {63 kms^{-1}}} \right)^{1.2}
\end{equation}
if
\begin{equation}
  V_{opt} = 63 \left( {L \over {0.04 L_*}} \right)^{0.16} km s^{-1}
\end{equation}
Thus, for a bright dwarf irregular, $L \sim 0.04 L_*$, $V_{opt} \sim
63 km s^{-1}$, characteristic values are $a$ = 0.93 and
$\beta$=0.08. Therefore the core radius is nearly as large as the
optical radius and the contribution of the visible matter at the
optical radius is nearly negligible. If $L < 0.04L_*$, $\beta$ is even
lower and $a$ higher. Under this interpretation, therefore, dwarf
irregulars are very dark galaxies, have very dense halos and large
masses, obtainable with $8 \times 10^{10} (L/0.04L_*)^{1/3}$.

Salucci and Persic (1997) give a formula to estimate the total mass of
a galaxy as a function of its visible mass
\begin{equation}
  M_{200} = 3 \times 10^{12} \left( {{M_{visible}} \over {2\times
  10^{11} M_\odot}} \right)^{0.4} M_\odot 
\end{equation}
Thus, when $M_{visible}$ is large the
$M_{200}/M_{visible}$ ratio decreases. Merely to state that $M_{total}$ is
proportional to $L$, as is often done, would be to make a bad
assumption, worse than supposing that all galaxies have equal mass,
irrespective of their luminosity.

Though most studies of these galaxies conclude that moderate or
large amounts of dark matter are required, we cannot  exclude the
magnetic interpretation of the data for spiral galaxies, 
which should  also be taken into account for dwarf
irregulars. Under this interpretation, the higher magnetic fields
required imply larger escaping fluxes that are actually observed, for
instance in M82 (also associated with the ejection of magnetic fields
(Reuter et al., 1982; Kronberg and Lesch, 1997) or in NGC 1705 (Meurer,
Staveley-Smith and Killeen, 1998), a galaxy requiring specially
high DM central density (0.1 $M_\odot pc^{-3}$) and a large mass loss
rate of the order of 0.2-2 $M_\odot yr^{-1}$.

\subsection{The rotation curve of the Milky Way}

Paradoxically, the rotation curve of the nearest galaxy remains poorly
known. Extinction is too large to observe the stars and too small to
observe the gas. It is preferable to observe the gas, either at 
21 cm or at 2.7 mm, 
because it extends at much greater radii. Thus we must rely on the
corotation of both the stellar and the gaseous systems, an assumption
that is not always justified, as mentioned previously. The tangent-point method to
obtain the rotation curve for $R < R_0$, with $R_0$ being the solar
galactocentric distance, is well known and need not be repeated here in
detail. 

The points of the circle with a Galactic Centre-Sun diameter are
characterized by a radial velocity from the Sun equal to their
rotation velocity, and this velocity is determined by the fact that it
corresponds to the largest redshift (in the first quadrant) or the
largest blueshift (in the fourth quadrant). The different values at
the points of this circle give us the rotation curve.

However, this method only provides the rotation curve out to 8
kpc, but to analyze our dark halo and the mass of the Milky Way itself, this
is too small. To extend the rotation curve to larger galactocentric
distances, a variety of objects have been observed. These objects have
to be bright, to be observed from afar, their distance must be accurately
determined (which in practice is the largest source of error) and their
radial velocities must be easily obtainable. Carbon stars, OB stars, planetary
nebulae, cepheids and HII regions have been used to study the outer
Galaxy, but the errors are large and the maximum distance is less than
$2R_0$. A complete account of these attempts was given in the 
review by Fich and Tremaine (1991) and will not be  repeated here. There
is a crucial date (1965) prior to which, as reviewed by Schmidt (1965),
it was thought that the outer rotation curve was Keplerian and the
estimated mass of the Milky Way was about $2 \times 10^{11}
M_\odot$. After this year, various authors began to realize
that the outer curve was more or less flat, and the conclusion that
our Milky Way may contain large amounts of dark matter became more
and more widely accepted.

The best method to investigate the outer rotation curve was proposed
by Merrifield (1992), who considered that a ring in the
Milky Way with constant vertical scale length, $h_z$, has a variable
angular size when seen from the Sun; or in his own words ``much as
a person standing in a volcano might estimate his or her location
within the crater from the variations in the apparent height of the
walls in different directions''. It is easily obtained that
\begin{equation}
  {v_r \over {\sin{l} \cos{b}}} = {R_0 \over R} \theta(R)- \theta_0
  \equiv W(R) 
\end{equation}
where $v_r$ is the  radial velocity from the Sun, $\theta(R)$ is the
rotation at a point with galactocentric distance $R$, $\theta_0$ is
the rotation velocity at $R=R_0$ and $l$ and $b$ are the galactic
longitude and latitude. Therefore, if we have a data cube
$T_b(v_r, l, b)$, where $T_b$ is the HI brightness temperature, it is
possible to divide the cube into slices with constant $W(R)$ as defined in
the above equation. $W(R)$ only depends on $R$ and we may use $h_z$ to
know exactly what value of $R$ we are speaking about. From our
location in the Galaxy, the HI layer of thickness $h_z$ at some point
of radius $R$ will present an angular size in galactic latitude of
\begin{equation}
  h_b = 2 \tan^{-1} \left({{h_z/2R_0} \over {\cos{l} + [(R/R_0)^2 -
  \sin^2{l}]^{1/2}}} \right) 
\end{equation}
If we then take a constant-$W$ slice, obtain the variation in angular
width as a function of longitude $l$, and compare it by this formula,
we can calculate it by fitting the value of $R/R_0$ (and even $h_z/R_0$)
of the slice, and hence obtain $v_r(R/R_0)$ and $h_z(R/R_0)$.

There are some inherent problems. The orbits must be circular and
vertical shear must be absent, i.e. $v_r$ should not depend on $z$. The
galactic warp introduces further complications, although these can be
overcome. In such a way, Merrifield was able to reach points in the
Milky Way rotation curve out to about 20 kpc, or 2.5 $R_0$, with an
unprecedented degree of precision.

The results greatly depend on the values of $R_0$ and
$\theta_0$. Merrifield proposed $R_0 = 7.9 \pm 0.8 kpc$ and $\theta_0
= 200 \pm 10 kms^{-1}$, rather lower than usually recommended in other works,
to match other kinematic constraints and in line with the rotation
curves of similar galaxies. More recently, Honma and
Sofue (1996, 1997) have used Merrifield's method to estimate the
rotation curve, the geometry of the halo and the total mass of the
Milky Way, investigating their uncertainties. Errors in $R_0$ are
relatively unimportant because they just change the scaling in the
radial direction, but changes in $\theta_0$ produce highly different
interpretations of our hypothetical halo. In Fig. 11 we plot
their results for three different values of $\theta_0$: 220, 200 and
180 km s$^{-1}$.

The rotation velocity decreases beyond $2R_0$ for all
cases. Reasonable Keplerian fits are obtained for $R \ge 2R_0$ if
$\theta_0$ is in the range 200-207 kms$^{-1}$. If $\theta_0 < 200 km
s^{-1}$ the curve declines faster than Keplerian.

Using  $\theta_0 = 220 kms^{-1}$, as recommended by the IAU, the outer
rotation curve rises between $R= 1.1 R_0$ and $2R_0$, which is uncommon
in other galaxies of the same type, having a flat curve
within the optical disk. To obtain a flat rotation curve, $\theta_0$
should be as small as 192 km/s.

$R_0$ and $\theta_0$ are related to Oort's constants A and B (note
$A-B = \theta_0/R_0$), which are fairly well determined. 
($\theta_0/R_0$ could also be directly determined by means of the VLBI
determination of the proper motion of Sgr B2, taking two quasars
behind as reference. Accurate data, in this respect, will be available
in the near future.
Honma and Sofue (1996)  propose $\theta_0 =200 km$ and
therefore $R_0 = 7.6 kpc$, based on this result and those obtained by other
authors also claiming lower $\theta_0$ and $R_0$.

Assuming a spherical mass distribution they obtain for the mass of the
Milky Way a low value of $2.0 \pm 0.3 \times 10^{11} M_\odot$, which
is close to the early estimates.

The Keplerian rotation curve does not require dark matter beyond
$2R_0$, but it would still be necessary within $2R_0$, because an
exponential disk has a rotation curve declining beyond 2.2r, when $r$
is the radial scale length, in conflict with the flat rotation out to
$2R_0$. However, the dark matter needed could be much less than
previously calculated. On the other hand, the shape of the dark halo
would differ greatly from that theoretically deduced. 

\section{Dark matter in other galaxies}

Although this paper considers spiral galaxies and rotation curves, this
type of galaxies may have a lot in common with other types, and therefore
these must be commented as well, in particular from the point of view
of their dark matter content. Moreover, observations other than rotation
curves have inspired a long list of possible methods to test for dark matter.

\subsection{Dark matter in elliptical galaxies}
There are basically three methods for specifically estimating the mass
of an elliptical galaxy: 

a) From the stellar velocity dispersion.

b) From the neutral gas velocities found in the outermost region, 
in certain galaxies.

c) From the X-ray corona surrounding all ellipticals.

There also exist complementary methods, using observations of ionized gas 
in the central parts, globular clusters, gravitational lensing, theoretical
considerations about the bar instability and the chemical evolution.

The general conclusion, taking into account all these studies, could
be, in summary,
that dark matter amounts comparable to visible matter could be present
in the visible part of the galaxy, and that larger dark matter amounts, probably
as large as in spirals, are present in a halo surrounding the galaxy, but that,
in any case, the evidence of dark matter in ellipticals is  less than 
in the case of spirals. Even the complete absence of dark matter cannot
be easily ruled out.

Several reviews have been written on the topic (e.g. Ashman, 1992;
Trimble, 1987; de Zeeuw, 1992; Kent, 1990; Bertin and Stavielli, 1993).
Let us remember that the surface brightness of an elliptical galaxy can be 
fitted by de Vaucouleurs' law
\begin{equation}
  I(R) = I_e e^{-7.67\left( \left( R/R_e \right)^{1/4}-1 \right)}
\end{equation}
(de Vaucouleurs, 1948), where $R_e$ is the radius enclosing half of
the light and $I_e = I(R=R_e)$ is another constant. The value of $R_e$
is often used as a parameter that normalizes all lengths as does the radial
scale length in spirals. This law seems to be rather well matched, but
it is just one fitting which might be less appropriate for some subtypes
(Andreakis, Peletier and Balcells, 1995).

Let us comment on the three basic methods, and more briefly about other methods:

a) The observations of stellar velocity dispersion, interpreted in terms of
Jeans' equation or of the Virial theorem, can provide the total mass 
for $R < R_e$, or even at larger distances.

The Virial theorem for a spherical, steady-state, static isothermal elliptical 
galaxy reduces to the simple expression
\begin{equation}
  2R \approx {{GM} \over \sigma^2}
\end{equation}
where $R$ is an equivalent radius. For a given $R$, $\sigma^2 \propto
M$, because the stellar chaotic thermal velocities, quantified by the
velocity dispersion, $\sigma$, must prevent gravitational collapse.
The larger the mass, the larger the stellar velocities must be. This
formula gives a first approximate mass. In practice
however, much more sophisticated models than this one are used to interpret the
velocity dispersions. There is a ``degeneracy'' between the unknown
anisotropy and the unknown gravitational potential. If the anisotropy
of the orbits is known the potential can be determined, but not both
simultaneously. We should know if orbits are mainly circular, or
mainly radial, or something in between. The anisotropy is
characterized by the parameter $\beta$, which is defined later, in
Section 3.5.2.

Pioneering works by Binney, Davies and Illingworth (1990), van der Marel,
Binney and Davies (1990) and others have concluded that no gradients in M/L
were clearly appreciated and that no dark matter was needed to explain
the central surface brightness and the velocity dispersions. The M/L values
are of the order of 12h (Binney and Tremaine, 1987) (about 8 for h=0.65)
which is comparable to the solar neighbourhood values. It is slightly
higher, but this fact can be explained mainly by the absence of 
young stars in ellipticals. One-component models, without any halo, provide
a good zeroth-order description (Bertin, Saglia and Stiavelli, 1992).

Bertin, Saglia and Stiavelli (1992) also considered two-component spherically
symmetric collisionless self-consistent models, which were later used to 
interpret real data from 10 bright selected galaxies (Saglia, Bertin and
Stiavelli, 1992) and found some evidence for dark matter to be of the 
order of the visible mass. The presence of rotation and of tangential
anisotropy makes it difficult to draw firm conclusions. 

As in the case of spirals
with their rotation curve, a flat or slowly increasing velocity dispersion,
$\sigma(r)$, may indicate dark halos dominating the dynamics (Saglia
et al. 1993) but there is a surprisingly large variety of $\sigma$-profiles, 
some of which decrease outwards relatively fast. Therefore, Saglia
et al. did not find any compelling evidence of dark matter out to 1-2 $R_e$.
Carollo et al. (1995) observed flat or gently declining velocity dispersion
profiles in four elliptical galaxies, concluding that massive dark halos must be
present in three of the four galaxies, although no clear conclusion was obtained
for the fourth. Bertin et al. (1994) found that in a sample of 6 galaxies,
three of them were not suitable for reliable modelling, two of them
presented no evidence for dark matter and one (NGC 7796) appeared to
have a distinct
dark halo. But the conclusion that some galaxies have a dark halo
while others do not is problematic for understanding what an elliptical
galaxy is. De Paolis, Ingrosso and Strafella (1995) found that dark matter
inside $R_e$ is negligible with respect to the visible mass.

b) A small fraction of elliptical galaxies are surrounded by a ring of
neutral hydrogen, for instance, NGC 1052, NGC 4278 and NGC 5128. In
these cases, the determination of a dark matter halo is very
similar to its determination in spiral galaxies, from the rotation
curve. One of the best studied gaseous rings is that of IC 2006
 (Schweizer, van Gorkom and Seitzer, 1989). The neutral gas counter-rotates at a 
radius of 18.9 kpc (6.5 $R_e$) and has a total mass of $4.8 \times
10^8 M_\odot$. This galaxy also has a counter-rotating central mass of
ionized gas out to $\sim$5 kpc. These gaseous components of some ellipticals
have either been accreted or are the remnant of a merger from which the elliptical
was created.

Schweizer, van Gorkom and Seitzer (1989) found evidence for a DM halo in 
IC 2006 with twice the mass of the luminous matter within 6.5 $R_e$,
under the assumption that the HI ring is flat and circular. Bertola et al.
(1993) analyzed five elliptical galaxies, combining the M/L ratios obtained
with the inner ionized hydrogen component and the outer neutral hydrogen
ring. M/L is constant out to about $R_e$ with a moderate value of $3.5 \pm 0.9$
but becomes very large in the ring region. These authors found a similarity
in the distribution of dark matter in ellipticals and in spirals, suggesting
a similar picture for the origin of both.

As we will discuss later, magnetic fields may explain rotation curves
without requiring dark matter in spirals. Similar arguments can be considered
to interpret gaseous rings around ellipticals. In particular, a narrow ring
is pushed towards the centre more easily than a disk, because the outward
magnetic pressure force need not be compensated by a magnetic tension.
It is to be emphasized that the IC 2006 gaseous ring is very narrow,
and is not even resolved by VLA.

c) The most promising method to study dark matter in ellipticals is based
on the existence of X-ray halos. A hot X-ray emitting gas typically extends
out to 50 kpc (Forman, Jones and Tucker 1985). The probable origin of the gas is mass
loss from stars; supernovae heat it up to $\sim 10^7K$,
bremsstrahlung being the main cooling process (Binney and Tremaine, 1987).
Typical masses of this hot gas are $10^{10} M_\odot$.

Hydrostatic equilibrium is usually assumed for the gas. Then, for a
spherical DM halo
\begin{equation}
  M(R) = -{{kTR}\over{Gm}} \left[ {{d \ln{\rho}} \over {d \ln{R}}} +
         {{d \ln{T}} \over {d \ln{R}}} \right]
\end{equation}
where $\rho$ is the density of the gas. Once $M(R)$ is determined in
this way, we obtain the DM halo profile.

The gas is not in perfect hydrostatic equilibrium. The innermost 
gas in the X-ray halo is more efficiently cooled, because cooling is
proportional to the electron density, which is still high. An inwards
flow in the inner region is therefore to be expected (Binney and
Tremaine, 1987). Cooling flows have been observed (Mushotzky et
al. 1994) and models including radial flows have been developed (e.g. Ciotti et
al. 1991). The equilibrium probably breaks down in galaxies with low
X-ray-to-optical luminosity ratios. Nevertheless,
hydrostatic equilibrium is generally assumed.

In the above formula, the temperature profile $T(R)$ is not provided
by the observations with enough precision. The strengths of some X-ray
lines or the shape of the X-ray continuum should provide this
T-profile but, in practice, this is still rather problematic. For giant
cD galaxies, like M87, the temperature is exceptionally well
determined and the method provides more reliable results. For M87 the
data are spectacular: $M(R < 300 kpc) \sim 3 \times 10^{13} M_\odot$;
the mass-to-light ratio reaches a value of 750; about 95\% of M87 mass
is dark matter (Fabricant and Gorenstein, 1983; Stewart et al. 1984;
Binney and Cowie, 1981). However, cD galaxies may be exceptional; as
they lie at the centre of a rich cluster, the DM encountered could
belong to the cluster as a whole. Below, we address this problem in
Section 5.

Difficulties arise in the analysis of normal ellipticals. If $T(r)$ is
unknown, it is tempting to assume an isothermal distribution
(e.g. Forman, Jones and Tucker, 1985), which might be justifiable. 
Mushotzky et al. (1994) were able to obtain 6 points of $T(R)$ in NGC
4636, finding that $T$ was approximately constant. Moreover Matsushita
(1997) and Jones and Forman (1994) confirmed the constancy of
$T(R)$. High M/L ratios are in general obtained, in the range 10-80,
especially at large distances, but Trinchieri, Fabbiano and Canizares
(1986) concluded that DM halos were not absolutely required by the
data. Fabbiano (1989) also found much lower masses.

Furthermore, the contribution of unresolved discrete X-ray sources, such as
accreting binaries, complicates the analysis (de Paolis, Ingrosso and
Strafella, 1995), which could be related to the fact that the 
relative amount of DM is higher for X-ray bright ellipticals.

Models often take as a boundary condition that X-ray emitting gas is
confined by the cluster intergalactic gaseous pressure (Fabian et
al. 1986). Other authors assume a vanishing pressure at infinity
(Loewenstein and White, 1999).

The gas responsible for the X-ray emission cannot rotate very fast and
hence no dynamo can generate magnetic fields capable of affecting the
hydrostatic equilibrium. However, intergalactic magnetic fields could
have an influence as a boundary condition. For the intracluster
intergalactic space, with $n \sim 10^{-5} cm^{-3}$ and $T \sim 10^7 K$
the thermal pressure is of the order of $10^{-14} dyn cm^{-2}$. As
discussed below, cluster intergalactic fields are of the order of
$10^{-6}$ G, and therefore the magnetic energy density is of the order of
the thermal pressure. External magnetic fields could contribute to
confining the X-ray emanating hot gas, thus reducing the large amounts
of dark matter required. This external field would not act
isotropically and would produce eccentric X-ray isophotes, such as for
instance in NGC 720. Eilek (1999) makes suggestions about the
importance of magnetic fields in the dynamics of clusters which are
relevant to the dynamics of X-ray halos around giant ellipticals at
the centre of clusters, where the field can provide an important part
of the pressure support.

Buote and Canizares (1994) observed a different isophote geometry for
X-rays and for the optical in NGC 720. The X-ray isophotes are more
elongated and their major axes are misaligned by about $30^{o}$. If
the total matter were distributed as is the optical light, it could
not produce the observed ellipticities of the X-ray isophotes. They
interpreted this ellipticity as being produced by a dark matter halo and
developed a model that did not need the $T(R)$ profile, and which also favoured
the existence of a large dark matter halo. Davis and White (1996)
and Loewenstein and White (1999), too, developed methods not requiring the
temperature profile that imply DM halos.

d) In addition to these basic methods there are others that should be mentioned.
The image splitting of an individual gravitational lens system consisting
of an elliptical is only slightly sensitive to the existence of a DM
halo, and so, one cannot definitely discriminate between galaxies with and 
without halos, with some exceptions (Breimer and Sanders, 1993;
Kochanek, 1995). Indeed,
in three cases where the lens is clearly a single galaxy, there is no need
to consider any dark matter halo. Maoz and Rix (1993), however,
 deduce from gravitational lensing methods
that $M(R)$ increases linearly with $R$, as is typical in isothermal
halos. 

Globular clusters have been considered to deduce the existence of dark
matter halos in ellipticals, mainly in M87 (Huchra and Brodie, 1987;
Mould et al., 1990). They support the conclusions obtained by other
methods: models without dark halos do not fit the data in M87, but they cannot
be excluded in NBC 4472 (Mould et al. 1990). This problem is
considered in Section 2.6. Planetary nebulae have also been
considered in NBC 5128 by Ford et al. (1989) and by others, who found a radial increase
in ($M/L_B$) reaching values of about 10, although de Zeeuw (1992) suggested a lower 
gradient. Ciardullo and Jacoby (1993) deduced that the non-interacting elliptical
galaxy NGC 3379 has no dark matter halo, and that a constant M/L of 
about 7 explained the observations perfectly. Theoretical studies of 
bar instability (Stiavelli and Sparke 1991) and chemical evolution
(Matteuci 1992) were unable to unambiguously determine the presence of
a dark halo. 

In conclusion, elliptical galaxies could have dark matter halos similar
in mass and extent to those in spiral galaxies (Danziger, 1997), but the evidence is not 
so clear and it cannot even be completely rejected that they possess no
dark halo at all. As exceptions, in giant cD galaxies like M87, the
existence of large amounts of dark matter seems to be fully demonstrated.

\subsection{Lenticular Galaxies}

Like ellipticals, SO galaxies present problems in the detection of a dark
matter halo.
Like ellipticals, SO galaxies were considered to possess low amounts
of gas, but new techniques are able to observe sufficient quantities of
gas to determine the rotation curve and hence dark matter,
under the standard interpretation for spirals. The general conclusion
is that these galaxies also have dark matter halos, but that they may 
be relatively smaller, as these galaxies are bright.

Estimations of DM in SO galaxies are more promising in these
exceptional lenticulars that have a large amount of gas. The gas is distributed in
a large outer ring, at about twice the optical radius, often warped
with respect to the optical plane, and is possibly of external
origin. Van Driel and van Woerden (1991, 1997) have studied gas-rich
lenticulars. NGC 2787, 4262 and 5084 have large M/L ratios at twice
the optical radius (about 25-30). A problem encountered in this study was that no
surface photometry was available to carry out standard analysis,
except for NGC 4203. The large DM halos needed were surprising
for these gas-rich lenticulars, compared with normal
lenticulars. Magnetic fields could again introduce a different
interpretation as gaseous rings are subject to magnetic centripetal
forces.

NGC 404 has been measured in
HI by del Rio et al. (1999), who found a large $M_{HI}/L_B$ ratio of the
order of 0.2, mainly contained in two rings. The
most important fact under our present perspective is that this galaxy
declines with a near perfect Keplerian profile, the Keplerian
fit being characterized by a correlation coefficient of 0.9 (see Fig. 12
which was interpreted by these authors as due to a central mass
concentration with no dark matter).

Pignatelli and Galletta (1997) obtained dynamical models of SO and Sa
galaxies that do not need a dark matter halo.

\subsection{Dwarf spheroidal galaxies}

This type of galaxy is probably the most common in the
Universe. Despite their low luminosity, they may contain large amounts
of dark matter, and thus contribute greatly to the mass of the
Universe. However, dwarf spheroidals do not possess gas in the
periphery, as do bright
ellipticals. Therefore, the determination of DM is more problematic. There
are basically two methods for detecting dark matter in this type of
galaxies:

a) Tidal radii.- The dwarf spheroidal satellites of the Milky Way, for
instance, could become tidally disrupted if they did not have enough dark
matter, thus increasing autogravitation and preventing it.
Let us determine the radius of the satellite necessary for
autogravitation to match tidal disruption, by means of a rough
model. 

Suppose a satellite dwarf with mass $m$ and radius $r$ orbiting
around the primary galaxy with mass $M$, with $R$ being the distance
between the two galaxies. Suppose the dwarf divided into two halves. They
would attract one another with a force of the order of $Gm^2/r^2$. The
tidal disrupting force would be the difference in gravitational force
produced by the primary  ${{GMm}\over 2} \left( {1 \over {(R-r)^2}} -
{1 \over {(R+r)^2}} \right) \approx GMmr/r^3$. The two forces become
equal when
\begin{equation}
  r = R \left({m \over M}\right)^{1/3}
\end{equation}

More precise calculations (e.g. Binney and Tremaine, 1987) give the same
orders of magnitude. As the dwarf is not a rigid
body, at a galactocentric radius equal to this critical value, $r$,
stars would escape and would be trapped in the gravitational field of
the primary. Therefore, at a ``tidal radius'', $r$, the density should
drop to zero. From the observational point of view, the tidal radius
is difficult to determine, and so it must be obtained by extrapolation.
It seems to be too large, when $m$ is obtained from the surface
brightness with a constant $M/L$ relation.

Hodge and Michie (1969) first detected a greater than expected tidal
radius for Ursa Minor, and Faber and Lin (1983) first used this
procedure for estimating dark matter. More (1996) used
the observation of stars being removed in the tidal tails of dwarf
galaxies to conclude that their dark matter halos must be truncated at
400 pc, limiting their $M/L$ ratio to less than about 100. Burkert
(1997) investigated four dwarf spheroidal galaxies orbiting around
the Milky Way: Sextans, Carina, Ursa Minor and Draco, considering
their tidal radii, and concluded that Sextans is dark matter dominated
but not the other three.

It should be emphasized that the mass obtained depends on the third power
of the estimated tidal radius, which is an important source of
errors. On the other hand, what is actually estimated is the
$m/M$ ratio. If the Milky Way mass were overestimated, so would be the
mass of the satellite. If the Milky Way contained no dark matter 
the satellite in turn would not require this component.

b) The velocity dispersion of the stellar system.- This method has
much in common with that used for the central parts of other
galaxies. But in others, the analysis is complemented with
peripherical effects which  are now completely absent. The study of
the dark matter in dwarf spheroidals basically rests on the assumption
that its distribution and that of the stars are similar, a condition
which is probably unrealistic, unless this type of galaxy is the
only exception. Nevertheless, most workers on this topic generally
agree that  these galaxies have
a large DM content, maybe 10 times
higher than luminous matter; indeed, ratios of 100 have been reported as
well as
very high central condensations in the range ($0.1-1 M_\odot pc^{-3}$)
(Mateo et al. 1992).

The possibility that all dwarf galaxies may have the same mass, despite their
large luminosity range, has been proposed (see Ashman, 1983). Mateo
(1997) obtains a mass of $2 \times 10^7 M_\odot$ for all dwarf
spheroidals irrespective of their luminosity. Salucci and Persic
(1997) obtain $M\propto L^{1/4}$.
Among the
observational difficulties we should also mention the fact that
velocity dispersions are low, of the order of 10 $km s^{-1}$,
and therefore, a high spectral resolution is required. 

If luminous and dark matter have the same spatial distribution, it is
easy to deduce the central density and the central mass-to-luminosity
ratio. From the Virial theorem, we straightforwardly deduce
\begin{equation}
  \rho_0 = {{ 9 \sigma_0^2} \over {4\pi G R_c^2}}
\end{equation}
where $\rho_0$ is the central density, $\sigma_0$ the central velocity
dispersion and $R_c$ an equivalent radius, or core radius, identifiable
with the radius at which the surface brightness is one half the
central value. Within this radius, the mass would be $\rho_0 {4 \over
3} \pi R_c^3$ and the luminosity, $\Sigma_0 \pi R_c^2$, where
$\Sigma_0$ would be the observable central surface brightness. Hence,
the $M/L$ fraction would be
\begin{equation}
  {M \over L} = {{3 \sigma_0^2} \over {\pi G \Sigma_0 r_c}}
\end{equation}
The constant should be $9/2\pi$, instead $3/\pi$, as deduced by more
detailed calculations (Richstone and Tremaine, 1986; Ashman, 1993).

An important study was made by Kuhn and Miller (1989), in which dwarf
spheroidal galaxies were not considered as virialized systems. These
galaxies could be unbound and losing stars. Numerical simulations
have been carried out by Kroupa (1997) and Klesen and Kroupa (1998) in
which the dwarf satellites are partially disrupted in perigalactic
passages; orbiting condensations can be identified after this event
near the satellite. Then, they might contain no dark matter at all and
yet present a high stellar velocity dispersion. The tidal
disruption of a satellite produces a remnant that contains about 1\%
of the initial mass.

Summarizing, dwarf spheroidal galaxies could contain
large amounts and concentrations of dark matter, but severe
observational difficulties prevent their precise determination. Even
models with no dark matter at all cannot be excluded.

\subsection{Polar ring galaxies}

Some galaxies possess a polar ring, most of which are of type
SO. The most widely accepted explanation for the formation of polar ring
galaxies is that accreted gas settles onto orbits that are more
frequently contained either within the equatorial plane or in polar
planes. As Polar Ring Galaxies are typically SO galaxies, and the ring
is gaseous, blue and undergoing star formation, their nature does not
greatly differ from gas-rich SO galaxies, commented above, but the importance 
of Polar Ring Galaxies with respect to the problem of dark matter
halos is that information about the overall gravitational potential
can be obtained in two planes: the disk and the ring planes, thus
potentially constraining the shape of the halo.

Sackett (1999) has reviewed the topic of the shape of halos, with
Polar Ring Galaxies being one of the most interesting techniques to determine it. Assuming
it to be a triaxial ellipsoid, the ovalness (b/a) and the flattening
(c/a), where $a \ge b \ge c$, are to be determined. Intrinsic ovalness
of the density distribution in the disk can be used to trace the
non-axisymmetry of the halo in any face-on spiral galaxy (Rix and
Zaritsky, 1995) finding a value for $b/a \sim 0.85$. This figure 
together with those obtained by other methods
summarized by Sackett (1999) indicates that $b/a > 0.7$, so that the
unobserved halo could have a higher axisymmetry.

However, the flatness is more difficult to assess, and polar ring galaxies are
specially suitable for this purpose. This point is particularly important because
it can provide information about the baryonic and dissipative nature of the halo
dark matter (Pfenniger, Combes and Martinet, 1994).

Polar rings have very large radii, of about 20  stellar disk radial
scale lengths, and therefore the perturbing influences of central luminous
components are less important, and observations would provide the
flatness $c/a$ of the dark halo. This task has been carried out by
several authors since the pioneering work by Schweizer, Whitmore and Rubin
(1983) including more recent analyses by Combes
and Arnaboldi (1996) and others (see the review by Sackett, 1999, and
references therein).

From the analysis of polar ring galaxies, Sackett concludes that halos 
are highly flattened, $0.3 \le c/a \le
0.6$, which coincides with  a similar
conclusion from the flattening of the X-ray halos of elliptical
galaxies (Buote and Canizares, 1998). Observations of gravitational
lensing (Kochanek, 1995) also suggest greatly flattened halos.

Other methods to determine c/a not based on PRG have been reported. 
The conclusions are
model dependent and, in some cases, are even based on hypotheses
that are not completely demonstrated. For instance, the 
analysis of warps (New at al, 1998)
and of flaring (Olling and Merrifield, 1997; Becquaert and Combes,
1998) are based on
interesting models, but which are not free of
alternative explanations. 

Ashman (1992) points out that polar ring galaxies are
unusual objects and therefore their hypothetical halos may be
atypical. For instance, the merging process from which they have 
originated could have given rise to a flattened halo. Alternatively,
the settling of a polar ring in the accretion process may require a
flattened halo, in which case, the scarcity of polar ring galaxies
would suggest that most halos are spherical. The influence of magnetic 
fields on the dynamics of these
rings may be non ignorable.

As the polar ring contains HI, it is useful to detect dark matter, but
as the host galaxy is lenticular and usually gas-poor, we cannot benefit
from the standard analysis of spirals. Therefore 
the study of the exceptional galaxy NGC 660, the only polar-ring
spiral galaxy known is very important. It has been extensively studied by van Driel et
al. (1995) and van Driel and Combes (1997). The disk has a flat
rotation curve and the polar ring is a rising one, which is rather
puzzling. No conclusion about the flatness was reached, although the
authors noted that
several problems cannot be ignored, such as the fact that the ring is
very massive, so that it cannot be considered to be formed by test particles
tracing the potential, together with the fact that, obviously, the polar ring
velocity and the disk velocity cannot be measured at the same
radius. These objections raised by van Driel and Combes (1997) also
hold for other dark matter studies in polar ring galaxies.

\subsection{Binary galaxies}

Binary stars constitute the best direct method to determine stellar
masses. It is therefore to be expected that binary galaxies should provide
galactic masses. The first and closest example is the double
system formed by M31 and the Milky Way, and we should begin with this one.

\subsubsection{M31 and the Milky Way}

The Local Group contains more than 35 galaxies, most of which are
dwarf ellipticals and irregulars with low mass; this
complicated system may be considered as being formed by two main galaxies, M31 and the
Milky Way, with other dynamically less important satellites belonging
either to one of them or to the pair. 

This picture is derived from
galactic luminosities, but when possible dark matter is taken into
account, it
is not clear at all. M31 has a visible mass of about $4 \times 10^{11}
M_\odot$. The Milky Way, $10^{11} M_\odot$. Next are M33 with $4
\times 10^{10} M_\odot$, LMC with about $2.3 \times 10^{10} M_\odot$,
SMC with $6.3 \times 10^9 M_\odot$, IC10 with $3 \times 10^9 M_\odot$
and other minor members. Note that this list, when
ordered following the total mass, could be changed. For instance, LMC
has a visible mass of $\sim 1/5$ the mass of the Milky Way. As it has been
suggested that irregulars may 
contain more dark matter than bright galaxies, the
total mass of LMC could be as large as, or even more massive than,
that of the
Milky Way. In this case it could no longer be considered our
``satellite''. Let us however retain the more standard viewpoint
and consider that M31 and the Milky Way are dynamically dominant and
form a binary system.

M31 has a line-of sight velocity of $\sim$ -300 km $s^{-1}$, and therefore
it is approaching us. Taking into account our motion of rotation
within the galaxy of about $220 km s^{-1}$, it is easy to deduce that
the speed of M31 with respect to the centre of our Galaxy is about $-125
km s^{-1}$. Both galaxies are approaching one another, with M31 therefore being 
an exception in the general motion of expansion of the
Universe. There are different interpretations of this fact:

a) ``Ships passing in the night''

Besides the expansion velocity following Hubble's law, galaxies have a
peculiar velocity. For instance, our galaxy is moving with respect to the
CMB black body at about 620$km s^{-1}$. Within a cluster, peculiar
motions are also of the order of $600 km s^{-1}$. Even if these high
velocities could be interpreted in other ways, such as bulk motions of
large inhomogeneities or only characteristic of rich clusters, it is
evident that some thermal-like peculiar velocities of this order of
magnitude characterize the velocity dispersion of present galaxies,
once the Hubble flow is subtracted. If we write for the velocity of
a galaxy
\begin{equation}
  v_i = H_0 r_i + V_i
\end{equation}
where $V_i$ is independent of $r_i$, for distances less than $V/H_0$,
Hubble's law becomes imprecise and of little use, peculiar velocities being
larger than expansion velocities. The law is imprecise for distances
shorter than about 10 Mpc and becomes absolutely unsuitable for r$<$1
Mpc. Therefore, a simple interpretation for the approaching motion of
M31 is that it is due to pure initial conditions, and is
unrelated to the mass of the Local Group.

Van der Bergh suggested that our Galaxy and M31 might not form any
coherent system, and that both galaxies ``were passing each other as
ships pass in the night'' (Lynden-Bell, 1983).

b) The ``timing'' argument of Kahn and Woltjer.

The most widely accepted interpretation of the negative velocity of M31 was
first given by Kahn and Woltjer (1959). They assumed that this double
system has negative energy, i.e. it is held together by
gravitational forces. However, considering visible matter only, they
estimated the kinetic energy of the system to be about $1.25 \times
10^{58}$ erg, and the gravitational energy $-6 \times 10^{57}$
erg. Even with an apparent positive energy (unbounded system) they
considered the possibility of large quantities of intergalactic
material in the form of gas, which would render the total energy
negative. This gaseous intergalactic mass was not confirmed by later
observations. Instead, today, the argument of Kahn and Woltjer is
considered as a proof for either the existence of large dark matter
halos surrounding M31 and the Milky Way or (at least) a large common
DM super halo pervading the Local Group. They deduced, with a simple
order of magnitude argument, that the effective mass was larger than
$1.8 \times 10^{12} M_\odot$, about six times larger than the reduced
mass of M31 and the Milky Way. Lynden-Bell (1983) has presented a more
precise description. 

It is interesting to note, also in this historic paper, that Kahn and Woltjer
(1959) considered that the ram pressure produced by this hypothetical
intergalactic gas, due to the motion of both galaxies with respect to
it, was responsible for warps of both galaxies. This hypothesis for
the origin of warps has today been largely forgotten, but it could
explain the coherence in the orientation of the warps of M31, M33 and
the Milky Way shown by Zurita and Battaner (1997). This coherence can only
be explained by the hypothesis of Kahn and Woltjer and by the magnetic
hypothesis (Battaner, Florido and Sanchez-Saavedra, 1990, 1991;
Battaner, 1995; Battaner, Florido, 1997; Battaner and
Jimenez-Vicente, 1998; Battaner et al. 1991; see also Binney, 1991,
and Kuijken, 1997).

Coming back to the ``timing'' argument, let us obtain a similar order
of magnitude, by an argument closer to that presented by Lynden-Bell
(1983). Suppose that the pregalaxies later to become M31 and the Milky
Way were formed at Recombination. Inhomogeneity seeds were previously
developed, but at Recombination, photon decoupling allowed matter to
freely collapse. Identifying Recombination as the epoch of the Local
Group birth, at about $10^6$ years after the Big Bang, is equivalent
to this birth being produced at the very beginning of the Universe, as
$10^6$ years is negligible when compared with 14 Gyr, at present. 

Then the Universe was much smoother, so we can assume a
vanishing initial transverse velocity. The Local Group, i.e. the two
galaxies, were born so close to each other that gravitation was stronger than the
expansion effect, so that we assume that during the period of the
birth of both galaxies, there was a negligible relative velocity
between them, in the line connecting them. Therefore, we assume
that 14 Gyr ago, both galaxies were at rest with respect to each other,
and since then their mutual gravitational attraction has reduced their
separation and is responsible for the 125 km$s^{-1}$
approaching velocity observed today.

The general equations for the orbit in the framework of Newtonian
Mechanics adopt the following parametric form
\begin{equation}
  r = a (1 - \epsilon \cos{\eta})
\end{equation}
\begin{equation}
  \Omega t = \eta - \epsilon \sin{\eta}
\end{equation}
where $r$ is the mutual distance, $t$ the time and $\epsilon$ the
eccentricity, while $\Omega$ and $a$ are constants. The parameter $\eta$ is
called the eccentric anomaly. The sum of the masses of both galaxies,
$M$, is related to these constants, through
\begin{equation}
  GM = \Omega^2 a^3
\end{equation}

If $\epsilon$ were zero, we would have $r=a$ (constant) and $\eta =
\Omega t$, e.g. a circular orbit with a constant velocity. But given
that our initial transverse velocity was assumed to be null, our orbit
cannot be circular, but rather, it will become approximately a
straight line. We thus consider $\epsilon =1$.

Figure 13 presents various possibilities:

the first possibility provides the lowest mass and we will
concentrate on this one. We have
\begin{equation}
  r = a (1- \cos{\eta})
\end{equation}
\begin{equation}
  \Omega t =\eta - \sin{\eta}
\end{equation}
therefore
\begin{equation}
  \dot{r} = a \sin{\eta} \dot{\eta}
\end{equation}
\begin{equation}
  \Omega = (1- \cos{\eta}) \dot{\eta}
\end{equation}

At the birth (approximately, at the Big Bang) we set $t = t_1$. Then,
$\dot{r}_1 = 0$, as we have assumed. $\dot{\eta} \ne 0$, always, as
otherwise (37) would imply $\Omega =0$. Therefore, $\sin{\eta_1} =0$,
which gives either $\eta_1 =0$ or $\eta_1 = \pi$. But $\eta_1 =0$
would imply $r_1 =0$, while we have started with the distance of
the galaxies being a maximum (2$a$). Therefore, $\eta_1
=\pi$. Hence, $r_1 =2a$ (as expected), $\Omega t_1 =\pi$, $\dot{r}_1
=0$, $\Omega = 2 \dot{\eta_1}$.

At the present time, we set $t =t_2$. Then
\begin{equation}
  r_2 = a (1- \cos{\eta_2})
\end{equation}
\begin{equation}
  \Omega t_2 = \eta_2 - \sin{\eta_2}
\end{equation}

Then
\begin{equation}
  \Omega t_2 - \Omega t_1 = \Omega (t_2-t_1) = T \Omega
  = \eta_2 - \sin{\eta_2} - \pi
\end{equation}
because $t_2 - t_1 =14$, if we adopt 1 Gyr as time unity. We
know $r_2 = 650$ (taking 1 kpc as distance unity). We also
know $\dot{r_2} = -125$ (if we adopt 1 kpc/Gyr as unity for the
velocity; 1 km/s $\approx$ 1 kpc/Gyr !)
\begin{equation}
  \dot{r}_2 = a \sin{\eta_2} \dot{\eta}_2
\end{equation}
\begin{equation}
  \Omega = (1-\cos{\eta_2}) \dot{\eta_2}
\end{equation}

Eliminating $\dot{\eta_2}$
\begin{equation}
  \dot{r}_2 = {{a \sin{\eta_2} \Omega} \over {1- \cos{\eta_2}}}
\end{equation}

With (16) and (43)
\begin{equation}
  {{\dot{r}_2} \over {r_2}} = {{\sin{\eta_2} \Omega}\over
   {(1-\cos{\eta_2})^2}}
\end{equation}
and taking the value of $\Omega$ given by (40)
\begin{equation}
   {{\dot{r}_2} \over {r_2}}={{\sin{\eta_2}} \over {(1-\cos{\eta_2})^2}}
          {{\eta_2 -\sin{\eta_2} - \pi} \over {T}}
\end{equation}

Defining, $\varphi_2 = \eta_2 - \pi$
\begin{equation}
  {{\dot{r}_2} \over {r_2}}T = {{\varphi_2 \sin{\varphi_2} + \sin^2{\varphi_2}} \over
        {(1+ \cos{\varphi_2})^2}}
\end{equation}

Taking the numerical values for $r_2$, $\dot{r}_2$ and $T$,
the solution of this equation, approximately, gives $\varphi_2 = 1.59$,
$\eta_2 = 4.73$. Hence (with (40)), $\Omega =  0.18 Gyr^{-1}$. Therefore
\begin{equation}
 \Omega t_2 =\Omega t_1 + 14 \Omega =\pi + 14 \times 0.18 =5.66
\end{equation}
\begin{equation}
 t_2 =31 Gyr
\end{equation}
(Note that the time of the Big Bang is $t_1 =17 Gyr$. We are not
taking the Big Bang as the origin of time!). Then, with (38), we have:
\begin{equation}
  a= 662 kpc
\end{equation}

In our modest calculation, at the beginning both galaxies were $2a =1324$ kpc apart and they were
at rest. Now they are 650 kpc apart (about half the initial distance) and they
are approaching at 125 km/s.

With all these values, we deduce for the mass of the pair of galaxies
\begin{equation}
  M \sim 2 \times 10^{12} M_\odot
\end{equation}
which is clearly much more than the visible mass of the pair of about
$5 \times 10^{11} M_\odot$. Despite the long calculation, the order of
magnitude is just given by $M = V^2 r/G$, where $r$ and $V$ are the
distance and velocity of M31.

c) In the above argument we considered two mass points with mutual
attraction, but the dark matter apparently encountered may be
distributed in a single extended halo. If
the force of gravity acting on the Galaxy were due to this Local Group
super-halo, the equation to be integrated would be
\begin{equation}
  \ddot{r} + {4 \over 3} \pi G\rho r=0
\end{equation}
where $\rho$ is the density of the intergalactic medium, which, for
simplicity we assume to be constant. In this case the angular
velocity of the periodic motion would be 
\begin{equation}
  \Omega = \left( {4 \over 3} \pi G\rho \right)^{1/2}
\end{equation}

We can, as before, obtain detailed values of $\Omega$ and the
initial distance between the new born Milky Way and the centre of the
Local Group, identified with the position of M31. In this case (r=a,
$\dot{r} =0$ at $t=0$; the origin of time is now the Big Bang,
approximately. Now, $a$ is the maximum separation of the Milky Way,
instead of $2a$, as in the previous case). We adopt $r = 650 kpc$,
$\dot{r} = -125 kpc/Mpc$, $T = 14 Gyr$ as before,
\begin{equation}
  r = a \cos{\Omega t}
\end{equation}
\begin{equation}
  \dot{r} = -a \Omega \sin{\Omega t}
\end{equation}
Dividing the formulae
\begin{equation}
  {{\dot{r}} \over r}= -\Omega T
\end{equation}
hence
\begin{equation}
  \Omega = 0.083 Gyr^{-1}
\end{equation}
and 
\begin{equation}
  a = 1633 kpc
\end{equation}

For the density of dark matter in the Local Group, we obtain
\begin{equation}
  \rho = 2.7 \times 10^{-29} g cm^{-3}
\end{equation}

This value is much lower than the minimum value estimated by Kahn and
Woltjer (about $1.6 \times 10^{-28} g cm^{-3}$) and slightly higher than
the critical density to close the Universe ($\sim 10^{-29} g
cm^{-3}$). The common halo hypothesis is not easy to reject.

d) The Local Group, rather than two main galaxies and several
satellites together with some minor members, should be considered as a
primordial inhomogeneity which has only recently collapsed to
form its present galactic members. Like any other inhomogeneity it has
evolved through the radiation dominated epoch with $\delta = \Delta
\rho/\rho \propto R$, decaying transverse velocities and increasing radial
velocities in a moderate collapse. Then
inhomogeneities reached an acoustic epoch, which for masses typical of
the Local Group began at $z = 10^5$ approximately (see later, Fig. 22). After the
Recombination epoch the Local Group pursued its process of collapse with
the relative density contrast increasing as R, where $R$ is the
cosmological scale factor, the transverse
velocities decreasing as $R^{-1}$ and - what is most important for our
purposes- the radial velocities increasing as $R^{1/2}$. After that, the
collapse became non-linear and these variations with the cosmic scale
factor became complicated and faster. As $\delta \gg 1$ we find 
ourselves in the non-linear regimen, but we will consider a linear 
evolution to find typical orders of magnitude. In this picture a naive
 formula relating the
present velocity $V_0$ of an inhomogeneity with present size
$\lambda_0$ and actual relative density contrast is (Battaner, 1996)
\begin{equation}
  V_0 =H_0 \lambda_0 \delta_0
\end{equation}

If the Milky Way and M31 were condensations within the Local Group,
$V_0$ would be identified with the relative velocity between these two
galaxies, with $\lambda_0$ and $\delta_0$ being typical parameters
characterizing the size and the density contrast of the Local Group.

This interpretation of the negative recession velocity of M31 is
fully compatible with the scenario of an approach between the two
galaxies within an expanding universe but somewhat in contrast with
present hierarchical models, in
which small structures form first, which will be accounted for later. 
As the velocities, before
Recombination, do not reach high values (Florido and
Battaner, 1997) we can start our calculations at Recombination. From the
above formula, taking $V_0 \sim 125 km/s$, $H_0 = 60 km/(s Mpc)$ and
$\lambda_0 \sim 0.65 Mpc$, we obtain $\delta_0 \approx 5.5$. Then
\begin{equation}
  \rho_0 = 5.5 ={{\rho_{Local Group} - <\rho>} \over {<\rho>}}
\end{equation}
where $<\rho>$ is the average density in the Universe. Hence, for the
Local Group
\begin{equation}
  \rho_{LG} = (\delta + 1) <\rho>
\end{equation}

Let us adopt for $<\rho> = 0.3 \times 10^{-29} g cm^{-3}$, thus
obtaining
\begin{equation}
  \rho_{LG} \cong 2 \times 10^{-29} g cm^{-3}
\end{equation}

Let us compare the different results. Methods c) and d) give a similar
order of magnitude, about $2.7 \times 10^{-29} g cm^{-3}$. The mass
corresponding to this density depends on the volume. The density
surely decreases outwards. Suppose a moderate equivalent radius of 650
kpc; then the mass of the Local Group would be $4 \times 10^{11}
M_\odot$, which is approximately the visible mass. Or suppose an
equivalent radius of 1 Mpc. In this case, we obtain $1.5 \times
10^{12} M_\odot$, in reasonable agreement with method a).

Not only should the results be compared, but also the basic formulae when
the numerical coefficients close to unity are ignored. Essentially,
methods b) and c) use $M \approx V^2r/G$, where $V$ is the
approaching velocity of M31 and $r$ its distance. Of course, the more
detailed arguments presented provide a more precise result, but which
cannot greatly differ from this value ($2.3 \times 10^{12}
M_\odot$). However, method d) is quite different. The
order-of-magnitude lying behind the calculation is of the type $M
\approx {{Vr^2} \over H_0} <\rho>$. In a critical Universe
$<\rho> = 3 H_0^2/8\pi G$. This method is not intrinsically related to
the other two. The orders of magnitude coincide because, curiously, V/r
is of the order of $H_0$. In most pairs the orbital period is of the
order of $H_0^{-1}$.

Summarizing, unless M31 and the Milky Way are like ``ships passing in the
night'' (a possibility that cannot be totally disregarded), the Local
Group seems to have 4 times more mass than we see as stellar
light. But we don't know where this mass lies, whether in galactic dark
matter halos or in a large common super halo. The difficulties
encountered in the interpretation of the closest binary system are
translated to the interpretation of other binary systems.

\subsubsection{Statistics of binary galaxies}

It is tempting to observe binary galaxies to
obtain galactic masses. Typical periods are, however, so large that
orbits cannot be observed. On
the other hand, typical distances between the two galaxies are much larger
than visible galactic sizes, and
therefore we could, in principle, obtain total masses.

The observations permit the obtention of ``projected
distances'', $\Delta r$, and of ``differences in the velocity
component along the line-of-sight'', $\Delta v$. From both series of
data we must infer a mean $M/L$ ratio. The analysis must be
statistical as no parameter of the individual orbits is known.

The first problem to resolve, and a very serious one, is the selection of a suitable
sample. Chance superpositions must be avoided: including in the sample
two unbounded galaxies, for which the velocity difference is due to
Hubble's flow, could give a very high $\Delta v$ and hence an
overestimation of the mean mass. If the pair is not isolated, the
influence of a third galaxy could produce a misinterpretation of the
results; galaxies are often in small or large clusters and are
rarely found in truly isolated pairs.

Usually, only pairs with a projected separation of less than a given value,
$R$, are accepted in the sample. Binney
and Tremaine (1987) have warned of this danger. Suppose that we take
$R$ as smaller than the mean
true pair distance. Then, the velocity should be mainly perpendicular
to the line-of-sight, and therefore much greater than the projected
velocity along the line of sight; then $\Delta v$ would be
underestimated, as would the galactic masses.

Limit values of $\Delta r$ and of $\Delta v$ become necessary, but
then, we find the results that we expect. Indeed, Sharp (1990) after
comparing the large discrepancies between different workers, even with
similar samples, was very pessimistic about the ability of these
statistical approaches to derive galactic masses.

To interpret statistical distributions of $\Delta r$ and $\Delta v$
in order to obtain $M$, the mass of the two galaxies, it is necessary
to adopt a law for the distribution of the true separation $r$; for
example, $\varphi (r) \propto r^{-\gamma}$, deduced with the two-point
correlation function of galaxies (Peebles, 1974). Observations
indicate that $\gamma \sim 1.8$. This distribution might  not be valid
for close binary systems. It is also necessary to adopt a hypothesis
about the orbits, and more precisely about the value of the parameter
$\beta$, defined as
\begin{equation}
  \beta = 1- {{<v_\theta^2 >} \over {< v_r^2>}}
\end{equation}
where $v_\theta$ and $v_r$ are the azimuthal and radial components of
the velocity. If the orbits are radial, $<v_\theta^2> =0$ and we
should take $\beta =1$. If the orbits are perfectly circular, then $<
v_r^2> =0$ and $\beta =-\infty$. An interesting intermediate
assumption is the condition of isotropy $<v_r^2> = <v_\theta^2>$,
hence  $\beta=0$.

A natural way to study binary samples is the adoption of galaxies as
mass points. In a classical analysis by White et al.
(1983), however, it was demonstrated that the mass point model does not
fit the data. This model predicts a correlation between $\Delta v$ and
$\Delta r$ (clearly $(\Delta v)^2$ should correlate with $(\Delta
r)^{-1}$), which is not found. This negative result is highly
interesting, as it can be interpreted as being due to the existence of greatly extended halos. If
the force, instead of $-GM/r^2$, were of the type $\propto r^{-1}$, the
correlation $[(\Delta v)^2 \leftrightarrow (\Delta r)^{-1}]$ would not
exist. 

This classical paper also claimed other evidence favouring the
existence of massive extended halos.
They found dark-to-luminous mass ratios higher than those found with
rotation curves. This is really to be expected, because mass
determinations from rotation curves are made at a maximum radius lower
than the rim of the halo, while a companion could be far away, at a
distance greater than the sum of both halos. Indeed their results are
compatible with extrapolations of observed rotation curves. These
authors found a relation of the type $M/L \propto
L^{-3/4}$. Therefore, low luminosity galaxies should contain large
amounts of dark matter. Lake and Schommer (1984) confirmed very high
$M/L$ values in a sample of dwarf irregular pairs.

However, Karachentsev (1983, 1985) found no evidence of dark matter in
binary systems, using very large samples, even containing some
galaxies with well observed flat rotation curves.

Honma (1999) found M/L for spiral pairs in the range 12-16, lower than
M/L for ellipticals, confirming previous results by Schweizer
(1987). These values are clearly lower than those 
previously reported.  

Among the large list of workers who have attempted to obtain
proof of dark matter with this method, noticeable are the studies
by van Moorsel (1987), Charlton and Salpeter (1991) and others,
favouring the scenario of a large common dark matter envelope, as we
have seen in the pair formed by M31 and the Milky Way. From the
cosmological point of view, whether dark matter lies in individual or in
common halos is unimportant, but from the point of view of galactic
structure and evolution, the two models are completely different.

As stated by Binney and Tremaine (1987) ``the mass-to-light ratio of
binary galaxies is probably large, but not so large as the ratio of
the mass of papers on this subject to the light they have shed on
it''. Even the concept of binary systems is controversial: with
typical velocities of 100 $km s^{-1}$ and separations of 100 kpc, a
typical value for the orbital period is of the order of Hubble's
time. In most cases, such as in the M31-Milky Way pair, a simple orbit
has not been completed. The Universe is expanding with a typical time of
the orbital period. Nevertheless, despite the large variety of
results, it should be emphasized that the most widely
accepted point of view is that binary galaxies possess large amounts of
dark matter, either in individual halos or in common super halos.

\subsection{Globular clusters and satellites}

Globular clusters and satellite galaxies are stellar components at
large distances from the centre and are therefore ideal probes of the
gravitational potential of a galaxy. The Milky Way 
may constitute, once more, the best example. Actually, globular
clusters are typically at distances of less than 10 kpc, much lower than
typical halo sizes, and are therefore not so suitable. Satellite
galaxies are further away but they are statistically scarce,
probably only nine satellites belonging to our Galaxy. However, satellites
are also observable in other galaxies.

Suppose a Keplerian potential $-GM/r$ and that, therefore, for satellites
with elliptical orbits
\begin{equation}
  r= a (1- e \cos{\eta})
\end{equation}
\begin{equation}
  \omega t = \eta - e \sin{\eta}
\end{equation}
as in 3.5.1. Time derivatives are
\begin{equation}
  \dot{r} = -a e \sin{\eta} \dot{\eta}
\end{equation}
\begin{equation}
  \omega =(1-e \cos{\eta}) \dot{\eta} ={r \over a} \dot{\eta}
\end{equation}

Therefore
\begin{equation}
  \dot{r}^2 r =\omega a^3 e^2 \sin{\eta}^2 \dot{\eta}
\end{equation}

Let us calculate the mean time value of this quantity during a complete
orbit
\begin{equation}
  <\dot{r}^2 r> = {\omega \over {2 \pi}} \int_0^{2\pi} \omega a^3 e^2
  \sin{\eta}^2 {{d\eta} \over {dt}} dt = {{\omega^2} \over {2\pi}}
  a^3 e^2 \int_0^{2\pi} \sin^2{\eta} d\eta =
  {1 \over 2} \omega^2 a^3 e^2 = {1\over 2} GM e^2
\end{equation}

Therefore, the galactic mass $M$ could be obtained from the quantity
$<\dot{r}^2 r>$. If we now consider not a single cluster or satellite but an
assemblage of them in a given time, the mean time value would not differ
from the mean value of $\dot{r}_i^2 r_i$ of the different clusters, at
present. We need to know the eccentricity, which would be different
for the different clusters. The mean-square value for isotropic orbits
is 1/2. Binney and Tremaine (1987), who have presented this argument,
propose
\begin{equation}
  M = {4 \over G} <\dot{r}^2 r>
\end{equation}

This method was applied by Lynden-Bell et al. (1983), who obtained $M
= 3.8 \times 10^{11} M_\odot$ with 9 satellites of the Milky Way
(both Magellanic Clouds, Leo I and II, Fornax, Sculptor, Ursa Minor,
Draco and Carina).

The adoption of a Keplerian potential was not fully justified as 6 out
of the 9 selected satellites are at distances of less than 100 kpc, lower
than a
reasonable size of the dark matter halo. The unknown value of the
eccentricities is also a major source of errors. If the orbits were
radially elongated, with $e=1$, the calculated mass would be closer
to the galactic mass deduced from the visible matter alone. That
orbits could be elongated rather than isotropic is somewhat suggested
by the fact that (with the exceptions of Leo I and Leo II, which are
too far away)
all the satellites of the Milky Way are roughly aligned in a line
connecting ($b  = -45^o$, $l = 270^o$) $\rightarrow$ ($b = 45^o$, $l =
90^o$).

Ashman (1992) and Trimble (1987) have summarized previous work carried
out by Little and Tremaine (1991), Zaritsky et al. (1989), Salucci and
Frenk (1989), Peterson and Latham (1989), Kulessa and Lynden-Bell
(1992) and others. The results obtained by the different authors are
very different, depending on the inclusion or exclusion of some distant
satellites, in particular on the inclusion or exclusion of Leo I. However, a
mass of $10^{12} M_\odot$ and a halo radius of 100 kpc are typical
values.

This type of analysis will be more promising in the future, when proper
motions of the satellites of the Milky Way become
available. Wilkinson and Evans (1999) have incorporated into the
computation the known proper motions of 6 satellites. They obtain a
value of about $2 \times 10^{12} M_\odot$ for the mass of the Milky
Way, with the inclusion or exclusion of Leo I not being so important as 
when only radial motions are considered. The results are model
dependent and these authors have chosen a peculiar one, called the
``truncated flat'' rotation curve, in which the density decreases as
$r^{-2}$ in the inner parts and decreases as $r^{-5}$ in the outer ones.

Tidal radii (see Section 3.3) of globular clusters and satellite
galaxies have also been considered (e.g. Innanen et al. 1983) but these
radii are obtained by an extrapolation of the photometric data, which
introduces a lot of uncertainty. A mass of about $9 \times 10^{11}
M_\odot$ and a halo size of at least 44 kpc were obtained by these
authors.

Satellites of other galaxies have been studied by Zaritsky et
al. (1993), Zaritsky and White (1994), Zaritsky (1997) and others. The
problem has much in common with that of binary galaxies but presents
particular interest because, as satellites are supposed to be low mass
systems they can be considered as test particles orbiting in the total
mass of the primary. The above authors have observed 115 satellites around 69
isolated primary galaxies. They conclude that DM halos do exist and
that they extend to distances of over 400 kpc, actually a
very large figure, but many characteristic facts are
difficult to explain: a) There is a complete lack of correlation
between $\Delta v$ and $\Delta r$, which impedes the obtention of the
mean value of galaxies. b) There is a complete lack of correlation
between $\Delta v$ and the HI widths. It has been mentioned that
Salucci and Persic (1997) and, as will be mentioned later, 
theoretical models do not expect a large correlation between these
two quantities, but not a vanishing one, either. c) Satellites seem to be
preferentially distributed near the plane perpendicular to the rotation
axis of the primary. d) The assemblage of all the satellites seems to present
a ``rotation curve'' around a typical primary without signs of
becoming flat. Note that the possibility of a common DM halo
without an internal DM structure is a picture compatible with these
observations.

\section{Theory}
The theoretical interpretation of rotation curves is one of the goals
of the so called CDM (Cold Dark Matter) hierarchical models of formation and evolution of
galaxies, which are at present the most widely accepted models. Other
explanations, such as the MOND (Modified Newtonian Dynamics)
and the magnetic models, are tentative and will also be mentioned. 

\subsection{The nature of galactic dark matter}

Galaxies are born out of primordial fluctuations with an evolution
probably driven by gravitation as the dominant effect. Gravitation, as
a geometric concept, has the same effect on the different types of
particles. Some forces other than gravitation, 
such as the interaction with photons, dissipative effects,
magnetic fields, etc., could also have an influence and act on the involved
particles differentially, but an overall trend for galaxies and
clusters 
to have a similar composition to the general composition of the
Universe is to be expected.

Our knowledge about the composition of the Universe has changed in recent
times with respect to the classical view, summarized, for instance, by
Schramm (1992). This new conception has been reviewed, for instance, by
Turner (1999a,b). The dominant matter is considered to be cold dark matter (CDM),
consisting of particles moving slowly, so that the CDM energy density
is mainly due to the particle's rest mass, there being a large series of
candidates for CDM particles, but axions and neutralinos being the most
attractive possibilities. 

Big Bang nucleosynthesis studies have been able to accurately determine the baryon
density as $(0.019 \pm 0.0012)h^{-2}$. The cluster baryon density has
also been accurately determined by X-ray and the Sunyaev-Zeldovich effect
to be $f_B = (0.07 \pm 0.007)h^{-3/2}$ and, assuming that rich clusters
provide a fair sample of matter in the Universe, also
$\Omega_B/\Omega_M = f_B$, from which, it follows $\Omega_M = (0.27
\pm 0.05)h^{-1/2}$. The Universe is however flat, $\Omega =1$, with
the CMB spectrum being a sensitive indicator. Therefore $\Omega =1= \Omega_M
+ \Omega_\Lambda$, where $\Omega_\Lambda \sim 0.7$ represents the
contribution of the vacuum energy, or rather, the contribution of the
cosmological term $\Lambda$. With this high value of $\Omega_\Lambda$
the Universe should be in accelerating expansion, which has been
confirmed by the study of high-redshift supernovae, which also suggest
$\Omega_\Lambda \sim 0.7$ (Perlmutter, Turner and White, 1999;
Perlmutter et al. 1999). The stellar or visible matter is estimated to
be $\Omega_V = 0.003-0.006$. All these values can be written in a list
easier to remember, with values compatible with the above figures,
adopting the values of $H_0 = 65 km s^{-1} Mpc^{-1}$; h=0.65:
  \[ \Omega_V = 0.003 \]
  \[ \Omega_B = 0.03 \]
  \[ \Omega_M =0.3 \]
  \[ \Omega = 1 \]
less precise but useful for exploratory fast calculations.

A large cluster should have more or less this composition, including
the halo of
course, even if a halo could contain several baryonic
concentrations or simply none. Therefore, a first direct approach to
the problem suggests that halos are non baryonic, with baryonic matter
being a minor constituent. 

This is also the point of view assumed by most current theoretical
models (this will be considered later, in Section 4.2.2), which follow
the seminal papers by Press and Schechter (1974) and White and Rees
(1978). We advance the comment that, in these models, a dominant collisionless non dissipative cold
dark matter is the main ingredient of halos while baryons, probably
simply gas, constitute the dissipative component, able to cool,
concentrate, fragment and star-producing. Some gas can be retained
mixed in the halo, and therefore halos would be constituted of non-baryonic
matter plus small quantities of gas, its fraction decreasing
with time, while mergers and accretion would provide increasing
quantities to the visible disks and bulges. Therefore, a first
approach suggests that galactic dark matter is mainly non-baryonic, which would
be considered as the standard description. Baryons, and therefore
visible matter, may not have condensed completely within a large DM
halo, and therefore the baryon/DM ratio should be similar in the largest
halos and in the whole Universe, although this ratio could be different in
normal galaxies.

However, other interesting possibilities have also been proposed. The
galactic visible/dark matter fraction depends very much on the type of galaxy,
but a typical value could be 0.1. This is also approximately the
visible/baryon matter fraction in the Universe, which has led some
authors to think that the galactic dark matter is baryonic (e.g. Freeman,
1997) in which case the best candidates would be gas clouds, stellar remnants
or substellar objects. The stellar remnants present some
problems: white dwarfs require unjustified initial mass functions;
neutron stars and black holes would have produced much more metal
enrichment. We cannot account for the many different
possibilities explored. Substellar objects, like brown dwarfs, are an
interesting identification of MACHOs, the compact objects producing
microlensing of foreground stars. Alcock et al. (1993),
Aubourg et al. (1993) and others have suggested that MACHOSs could provide a substantial
amount of the halo dark matter, as much as 50-60\% for masses of about
0.25 $M_\odot$, but the results very much depend on the model assumed 
for the visible and dark matter components, and are still
uncertain. Honma and Kan-ya (1998) argued that if the Milky Way does
not have a flat rotation curve out to 50 kpc, brown dwarfs could account
for the whole halo, and in this case the Milky Way mass is only $1.1
\times 10^{11} M_\odot$.

Let us then briefly comment on the possibility of dark gas clouds, as
defended by Pfenniger and Combes (1994), Pfenniger, Combes and
Martinet (1994) and Pfenniger (1997).  They have proposed that spiral galaxies
evolve from Sd to Sa, i.e. the bulge and the disk both increase and at
the same time the M/L ratio decreases. Sd are gas-richer than Sa. It
is then tempting to conclude that dark matter gradually transforms
into visible matter, i.e. into stars. Then, the dark matter should be
identified with gas. Why, then, cannot we see that gas? Such a scenario could be the
case if molecular clouds possessed a fractal structure from 0.01
to 100 pc. Clouds would be fragmented into smaller, denser and colder
sub-clumps, with the fractal dimension being 1.6-2. Available millimeter
radiotelescopes are unable to detect such very small clouds. This
hypothesis would also explain Bosma's relation between dark matter and gas
(Section 2.3), because dark matter would, in fact, be gas (the
observable HI disk could be the observable atmosphere of the dense
molecular clouds). In this case, the dark matter should have a disk
distribution.

The identification of disk gas as galactic dark matter was first
proposed by Valentijn (1991) and was later analyzed by
Gonz\'alez-Serrano and Valentijn (1991), Lequeux, Allen and Guilloteau
(1993), Pfenniger, Combes and Martinet (1994), Gerhard and Silk (1996)
and others. H$_2$ could be associated to dust, producing a colour
dependence of the radial scale length compatible with large amounts of
H$_2$. Recently, Valentijn and van der Werf (1999) detected
rotational lines of H$_2$ at 28.2 and 17.0 $\mu$m in NGC 891 on board
ISO, which are compatible with the required dark matter. If confirmed,
this experiment would be crucial, demonstrating that a disk baryonic
visible component is responsible for the anomalous rotation curve and
the fragility of apparently solid theories. Confirmation in other
galaxies could be difficult as H$_2$ in NGC 891 seems to be
exceptionally warm (80-90 K).

A disk distribution is, indeed, the most audacious statement of this
scenario. Olling (1996) has deduced that the galaxy NGC 4244
has a flaring that requires a flattened halo. However, this analysis
needs many theoretical assumptions; for example, the condition of
vertical hydrostatic equilibrium requires further justification,
particularly considering that NGC 4244 is a Scd galaxy, with vertical
outflows being more important in late type galaxies. Warps have also been
used to deduce the shape of the halo. Again, Hofner and Sparke (1994)
found that only one galaxy NGC 2903, out of the five studied, had a
flattened halo. In this paper, a particular model of warps is assumed
(Sparke and Casertano, 1988), but there are other alternatives (Binney
1991, 1992). The Sparke and Casertano model seems to fail once the
response of the halo to the precession of the disk is taken into
account (Nelson and Tremaine, 1995; Dubinski and Kuijken,
1995). Kuijken (1997) concludes that ``perhaps the answer lies in the magnetic
generation of warps'' (Battaner, Florido and Sanchez-Saavedra 1990). On
the other hand, if warps are a deformation of that part of the disk
that is
already gravitationally dominated by the halo, the deformation of the
disk would be a consequence of departures from symmetry in the
halo. To isolate disk perturbations embedded in a perfect unperturbed
halo is unrealistic. Many other proposals have been made to
study the shape of the halo, most of which are reviewed in the cited
papers by Olling, and in Ashman (1982), but very different shapes have
been reported (see section 3.4).

There is also the possibility that a visible halo component could have
been observed (Sackett et al. 1994; Rausher et al. 1997) 
but due to the difficulty of working at these faint levels,
this finding has yet to be confirmed.

Many other authors propose that the halo is baryonic, even if new
models of galactic formation and evolution should be developed (de
Paolis et al. 1997). This is in part based on the fact that all dark
matter ``observed'' in galaxies and clusters could be accounted for by
baryonic matter alone. Under the interpretation of de Paolis et al.
(1995) small dense clouds of $H_2$ could also be identified with
dark matter, and even be responsible for microlensing, but instead of
being distributed in the disk, they would lie in a spherical halo.

\subsection{CDM theoretical models}

Theoretical models of galaxies consider their origin and evolution. It
is difficult to review the early history of these theories and identify which
of them have had a decisive influence on our present ideas. Most
contemporary theories about the origin of galaxies are based on three decisive papers
by Press and Schechter (1974), White and Rees (1978) and Peebles
(1982), which will be commented on later, and have in common the
hypothesis that the dark matter is cold (CDM) and that, at a given time,
CDM halos arose through a hierarchy of different sized halos
formed from mergers of smaller halos. At least four steps characterize the
evolution of a galaxy:

a) Small density fluctuations, probably originated by quantum
fluctuations before the epoch of Inflation or at cosmological phase
transitions, grow during the radiation dominated universe and provide
a fluctuation spectrum after the epoch of Recombination.

b) CDM overdensities accrete matter and merge. The hierarchical
formation of greater and greater halos produces the present galactic
and cluster structures.

c) Baryons cool and concentrate at the centre of halos and constitute
the visible  component of galaxies. The explanation of the Hubble
sequence and the origin of rotation of galaxies would be goals of the
study of this phase.

d) Once the basic structure of a galaxy with its different components
has been established, it is necessary to follow its evolution due to
star formation, gas ejected from stars, progressive metal enrichment,
matter flows connecting the intra and extra media, small internal
motions, etc.

\subsubsection{Growth of primordial fluctuations}

Suppose we start our analysis shortly after Annihilation. Then, a primordial
energy density fluctuation spectrum must be assumed. One of the
most simple hypotheses is the spectrum of Harrison and Zeldovich which
is rest mass independent and which arises naturally from the quantum
fluctuation at Inflation, but there are other more exotic
possibilities; indeed, the spectrum has been characterized by some
parameters which are considered free in some numerical calculations. The
subsequent evolution is a consequence of this initial spectrum and of
the nature of the matter, mainly through the equation of state.

Most models do not explicitly consider this first phase. It is
considered that an unknown primordial density fluctuation spectrum is
responsible for an unknown post-Recombination spectrum and this,
therefore, is equivalent to assuming the initial spectrum after
Recombination and this complicated phase is thus avoided. We consider this
procedure somewhat dangerous because even if the initial spectrum is
random some regular structure may be inherited after
Recombination. For example, primordial magnetic fields may be
responsible for very large scale filaments ($\sim$ 100 Mpc) as
discussed later. Moreover, the existence of periodic structures
forming a lattice, actually observed whatever the cause may be, must be
understood  to assess how CDM halos merge at later epochs. These
points will be addressed later.

As in the case of stellar collapses, the basic concept is Jeans'
Mass. We must know which masses are able to collapse and how the
collapse grows as a function of time. Both phenomena depend on the epoch during
the thermal history of the Universe. The basic treatment was developed
by Lifshitz (1946), Zeldovich (1967) and Field  (1974) and has been
clearly incorporated in the well-known book by Weinberg (1972). Some
more recent books also address this analysis (Kolb and Turner, 1990;
Battaner, 1996).

The protogalactic collapse has some differences with respect to 
the protostellar collapse, mainly:

a) Protostellar collapses are considered to be isothermal, because
photons are able to quit the protostellar cloud freely and the
temperature remains constant. It is then obtained for Jeans' Mass, $M_J
\propto \rho^{1/2}$. The fact that $M_J$, the minimum mass able to
collapse, increases when the collapse proceeds produces the
fragmentation of the cloud until the smaller fragments are so dense
that the isothermal regime breaks down. The pre-Recombination collapse
involves clouds made up of CDM particles, baryons and, mainly,
photons. Photon clouds have no way to remain isothermal when they
contract. Adiabatic collapses are to be assumed, which does not lead to
any fragmentation.

b) Contraction within the expansion. During the collapse, the
dimensionless quantity $\delta$, defined as $\delta =
(\rho -<\rho>)/<\rho>$ (where $\rho$ is the inhomogeneity density and
$<\rho>$ its mean value in the Universe), increases, but as $<\rho>$
decreases because of the general expansion, $\rho$ need not necessarily
increase. The collapse is relative. Indeed, present densities in a
galaxy are greater than, but comparable to, densities before the
collapse. As a zeroth-order language, isolation rather than absolute
contraction gives rise to galaxies. The time evolution of
$\delta$, i.e. of the
relative overdensity, provides a simpler description. 
The effect of expansion is not at all negligible, because the
characteristic time of expansion, 1/H, is of the order of the period
of Jeans' wave, $\lambda_J/V_s$, where $\lambda_J$ is Jeans'
wavelength and $V_s$ the speed of sound, with both being variable during
history of the Universe.

From the point of view of the physics involved, pre-Recombination
collapses require a general-relativistic treatment as they are
fluctuations in a very hot medium (photons) and the curvature they
produce is not only non-ignorable but a dominant effect.

Jeans' Mass is calculated to be $M_J \propto R^3$ during the era
between Annihilation of electrons and positrons and the transition
epoch dividing the radiation and matter dominations; $R$ is the
cosmological scale factor. Between this last epoch (i.e. Equality) and
Recombination, Jeans' Mass increases to a constant asymptotic value,
$M_J \approx 4 \times 10^{19} M_\odot$, which is never reached because,
at Recombination, the scenario abruptly changes, with a sudden fall
from about $10^{17} M_\odot$ to about $10^5 M_\odot$. In the
post-Recombination era $M_J \propto R^{-3/2}$. The complete function
$M_J(R)$ is depicted in Fig. 14.

In this picture, we may follow the stability of an inhomogeneity with
a rest mass of $10^{12} M_\odot$, a typical value of the galactic
mass, dark matter included. Its mass is in principle higher than
Jeans' Mass, and therefore we initially find this protogalaxy in a
collapsing phase. The collapse is not so fast, as we will see later,
and is truncated when $R/R_0 \sim 10^{-5}$ approximately. The
proto-galaxy then enters a stable state and  Jeans' wave just
produces acoustic oscillations. There is not much time to oscillate
in this Acoustic era, less than one complete period, because the
Recombination sudden falls, leading our homogeneity to unstable conditions
again. In other words, once baryons are no longer coupled to photons
they are free to collapse.

CDM particles may alter this picture if they have no interaction with
photons, as they are free to collapse when they become dominant. They
then create potential wells where, after Recombination, the baryons 
fall. In this case the Acoustic era would be absent.

In the same way that the study of Jeans' waves provides the value of
typical stellar masses, it would be desirable to obtain typical values
of masses of galaxies and also of clusters and superclusters, because the
analysis mentioned considers any inhomogeneity. A large enough mass
would always collapse, but we could expect at least a minimum value of
collapsed systems.

If the dominant matter particles were baryons, or any other type of
particles interacting with photons, then damping by non-perfect fluid
effects would affect the oscillations in the Acoustic Era, therefore 
preventing small mass inhomogeneities from reaching
Recombination. The mechanism of {\it{photon diffusion}} is of this type. The
fast photons would tend to escape from the overdensity cloud and then
push baryons outwards, via the interaction due to Thomson
scattering. This is equivalent to a viscosity and a heat conduction, which
are expected to be important when the photon mass free time is of the
order of the cloud size. The so called Silk
Mass is calculated is such a way. Numerical estimations provide 
values of the Silk Mass of the
order of $10^{12} M_\odot$, a very significant value. However, we will
see that the model of CDM hierarchical structure formation considers a
different scenario, with high masses only being limited by the finite time
of the Universe. Even if a mechanism similar to photon diffusion
had been at work before Recombination, much smaller masses, much lower
than $10^{12} M_\odot$, would be the components of the initial merging
CDM blocks. The smaller galaxies are
of the order of $10^7-10^8 M_\odot$.

Another mechanism, called {\it{Free Streaming}}, would give a lower limit to
collapsing clouds, if the DM particles were hot.
 Suppose they were neutrinos, for example; in this case, they would escape
from the initial inhomogeneity if this homogeneity were small. When the
expansion proceeds and the temperature of the Universe is low enough,
the neutrino speed becomes small, which limits the distance a neutrino
is able to run. When $kT \sim m_\nu c^2$, where $m_\nu$ is the
neutrino's mass, the neutrino can be considered stopped. Normal
estimations of the free streaming lower limit mass are of the order of
$10^{12} M_\odot$ too, although current ideas about the nature of dark
matter favour CDM.

Once we have considered the question of when an inhomogeneity is
unstable, and therefore when an overdensity region grows and $\delta$
increases, let us  briefly speak about the $\delta (t)$ function, or its
equivalent, $\delta(R)$. Again a pre-Recombination treatment
requires general relativistic tools, Newtonian Mechanics being
adequate after Recombination. However, this later epoch is much more
complicated from the mathematical point of view, because we know that
at present $\delta >1$, which means that the evolution is
non-linear. In the radiation dominated epoch it was $\delta << 1$ and the
standard linear perturbation analysis is a very good approximation.

It has been obtained that growth perturbations increase as  $\delta \propto t$,
therefore $\delta \propto R^2$, during the radiation dominated epoch
before entering the Acoustic era. During this Acoustic era, if it 
really existed, it is apparent that $\delta$ is a constant, or,
rather, periodic. After Recombination, inhomogeneities grow as $\delta
\propto t^{2/3}$, therefore $\delta \propto R$, until $\delta$ is
closer to unity. Then, the simple linear analysis technique is no
longer adequate. Non-linear calculations suggest that first $\delta
\propto R^2$, afterwards $\delta \propto R^3$, but then the
hierarchical models, as commented below,
constitute the most widely accepted technique to study this recent evolution.

Figure 15 plots $\delta (R)$, but is only a rough description
due to the many factors which are at present poorly understood.

\subsubsection{CDM Hierarchical Models}

As mentioned above, current theoretical models of galaxy formation
and evolution are based on historical papers, in particular those
by Press and Schechter (1974) and by White and Rees (1978). Other
pioneering papers (e.g. Gunn, J.E. and Gott, J.R., 1972; Gunn, 1977,  and
references therein) have also contributed to the presently accepted
scenario, which has had a considerable success in explaining a large
variety of galaxy and clustering properties.

Previous studies led by Zeldovich
(e.g. Zeldovich, 1970; Sunyaev and Zeldovich, 1972) considered that
small mass objects formed from nonlinear processes in clusters with a
larger hot dark matter mass.
Press and Schechter analyzed the opposite point of view, according to which
larger mass objects form from the non-linear interaction of smaller
masses, with these being formed before. Some of these ideas were suggested by
Peebles (1965).
Press and Schechter (1974) did not mention dark matter, but a ``gas''
of self-gravitating mass.

Starting from an initial spectrum of perturbations shortly after
Recombination, baryonic aggregates with a small mass condensed, merged to form
larger condensations, which in turn merged and so on. In this way, the
condensation proceeded to larger and larger scales, at later and later
times. They proposed that this merging series very weakly depended on
the spectrum of seed masses initially assumed. ``When the condensation
has proceeded to scales much larger than the seed scale, the {\it{gas}}
should have essentially no memory of its initial scale 
and the condensation process should approach a self-similar
solution''.

One of the most decisive papers in the modern history of Astronomy was
written by White and Rees (1978), which may be considered the
progenitor of nearly all current theoretical models of galaxy
formation. This model adopted the hierarchical clustering scenario
proposed by Press and Schechter (1974) but introduced two new basic
ingredients: dark matter and the cooling of the baryon system to produce
the visible component of galaxies.

First, these authors proposed $\Omega \sim 0.2$, which is certainly close to
present-day estimates. They considered that the baryon to dark matter
fraction in the Universe should not very much differ from that in a
rich cluster, like Coma, from where they adopted M/L =400. This is in
agreement with modern estimates of $150 h \le M/L_B \le 500h$
(Bahcall, Lubin and Dorman, 1995). As $L_B$, the blue radiation energy
density is of the order of $2 \times 10^8 h L_\odot Mpc^{-3}$ (see for
instance, Zucca et al., 1997, for a current value),  
$\Omega \sim 0.2$ is then deduced.

From this, they proposed that approximately 80\% of this matter was
dark matter and of the remainding 20\%, half was still uncondensed
baryons and the other half constituted the luminous component.

In this scenario, small halos formed first through a merging process; 
the first generation of halos
produced a new generation and so on. New
generations in the hierarchy are therefore born later and are more
massive. The process is interrupted by the finite time of the
Universe, and therefore no clustering is to be expected at a large enough
scale. The smaller scale virialized systems merged into an amorphous
whole, mainly constituted of dark matter but also of gas, as a minor
component. When the gas cooled it fell into the centre of the DM halo
and there became sufficiently concentrated to produce stellar
collapses, which rendered it visible.

``When a halo is disrupted in a larger system the luminous galaxy in
its core can preserve its identity because dissipation has made it
more concentrated than the surrounding dark material'' wrote White and
Rees. Therefore small galaxies could be reminiscent of the first
generation halos. As they formed earlier, dwarf galaxies could have a small
mass and a higher density. When these small halos with a luminous core
merged to produce a larger halo with a larger luminous core, the small
baryonic concentration would not be destroyed and should be
identifiable as orbiting the large galaxy. The familiar observation of
a large galaxy, such as ours, surrounded by many  dwarf galaxies
would then be explained in a very natural way.

Note that this  elegant idea could be in conflict with
current interpretations of the observations, which seem to indicate
that dwarf galaxies have their own halo and they are not so ``dwarf''
as they may possess specially massive dark halos. We will return 
to this point when dealing with the magnetic hypothesis.

In more detail, the fate of the gas would depend on other
factors. When two or more smaller halos merge, there is an intense
heating, produced by shocks during the violent relaxation that
accompanies the formation of the halo. The gas could be heated until
it reaches a pressure-supported state. At a temperature of about $10^4$
K it would be ionized and able to cool radiatively, via
bremsstrahlung, recombination, and so on. The cooling process would
settle the concentrated gas into the centre and produce stars,
which become a visible component. The pressure-supported gas contraction
would be quasi-static. But this  slow concentration could be abruptly
truncated by a new merging. Therefore, a visible baryonic component
would be formed only when the radiative cooling time 
is less than the typical dynamic or merging time.

Therefore ``the luminous material that condensed in their centers may
nevertheless have survived to the present day in identifiable stellar
systems'' (White and Rees, 1978). For instance, this would be  not only the case of
satellite dwarf galaxies but also of large galaxies within a rich
cluster like Coma. Such large clusters would possess a very large common
halo, rather than small individual halos. By merging, because of the
violent relaxation, halos virialize very fast and lose any internal
structure other than the baryonic cores.

\subsubsection{Recent developments}

After these important papers in which the basic scenario was outlined
and the general assumptions justified, it was necessary to develop
more detailed models, mainly numerical and N-body simulations due to the
high complexity of the various physical processes involved.

A schematic list of models and reviews is now given, which should be
completed with references therein: Cole (1991), White and Frenk
(1991), Navarro and White (1993), Kauffmann, Guidernoni and White
(1994), Cole et al. (1994), Navarro, Frenk and White (1996, 1997),
Lacey and Cole (1993), Avila-Reese, Firmani and Hernandez (1998),
Avila-Reese et al. (1999), Sensui, Funato and Makino (1999), 
Salvador-Sole, Solanes and Manrique
(1998), Baugh et al. (1999), Subramanian, Cen and Ostriker (1999),
Steinmetz (1999), van den Bosch (1999) and a large series of papers,
reflecting the importance of the topic.

Models can be classified as ``semi-analytical'' (in which some
processes are given a simplified treatment assuming simple recipes, based on
either previous numerical calculation or on theoretical ideas),
numerical simulations (e.g. hydrodynamical simulations, collision-less
simulations) N-body simulations (the most widely used) and even
analytical. Some hybrid models are difficult to classify in this
scheme.

It is first necessary to adopt a cosmological model, the most
popular one being the ``standard'' CDM (with $\Omega =1$, $h=0.5$, for
instance) or the $\Lambda$CDM (more in consonance with current
values, $\Omega =0.3$, $\Lambda =0.7$, h= 0.65). A primordial
fluctuation spectrum must often be adopted, usually a power law $P(k)
\propto k^n$, with $n$ ranging from 0 to -1.5 (for example), where $k$
is the wave number. Another important parameter used by most models is
$\sigma_8$. In general, the variance $\sigma$ is defined as
$<\delta^2>^{1/2}$; then $\sigma_8$ is the present value for a
scale-length of 8 Mpc. This parameter is adopted ``a priori'' taking
into account the present large-scale structure, rather than 
considering a real free parameter. Usual values adopted are $\sigma_8 \sim
0.6$ for the standard CDM and $\sigma_8 \sim 1$ for the $\Lambda$CDM.

Other parameters characterize the calculation methods. For instance
the initial redshift, the number of particles in N-body simulations
and the box in Mpc$^3$ in which the calculations are performed. The so
called ``Virgo consortium'' (Jenkins et al. 1997) is able to handle $256^3$
particles and a large volume of the order of 60 Mpc. Parameters
controlling the resolution of the simulation and the efficiency with
which gas cools have a higher influence on the results (Kay et al.
1999).

These models not only deal with the formation of halos, but also with the
ability of gas to form stars, with matter and energy outputs, mainly due
to supernova explosions, the evolution of the baryonic component, the
explanation of the Hubble Sequence, how spirals merge to produce
spirals and so on. From our point of view, the rotation of galaxies
strongly depends on the structure of the halos, which is determined in
the first stage of the computations. The latest evolution of visible
galaxies is, paradoxically, the most difficult to understand and to 
model. For instance, the Initial Mass Function (IMF) is largely
unknown and yet is decisive in galactic evolution.

The hierarchical process of merging, the formation and internal structure
of dark matter halos is said to be the best known process. This could
be due, in part, to the relative simplicity of the process, but
also to the evident fact that it is easier to make predictions about
the unobservable. In general, even if some observable facts remain
insufficiently explained, these families of theoretical models
provide a very satisfactory basis to interpret any evolutionary and
morphological problem.

\subsubsection{General  remarks}

The basic scenario cannot be accepted without discussion. In the CDM
model, in which small structures form first, it is predicted that at a
large enough scale no structure should be encountered and that the
density distribution should be completely random. This random
distribution should be found at scales larger than about 30
Mpc. However, this might not be the case, as there is a large body
of evidence suggesting a regular large structure forming  a
lattice at a larger scale. Broadhurst et al. (1990) found 10 periodic
peaks separated by about 128 $h^{-1}$Mpc in a pencil beam survey,
which cannot be due to chance. Einasto et al. (1994) 
presented very clear evidence of a regular network with
superclusters residing in chains separated by voids of diameters
100$h^{-1}$Mpc. Such a regular lattice has also been confirmed by
other authors (Tucker et al. 1997, Landy et al. 1996, Einasto et al. 1997
and others). Tully et al. (1992) compared the structure with a three
dimensional chess-board. Similar regularity has been found in the
distribution of QSO absorption-line systems (Quashnock et al. 1996) and
in the CMB spectrum (Atrio-Barandela et al. 1997). The Tartu Group has
been specially active in demonstrating this large-scale structure
(Toomet et al. 1999).

Typical sizes of the lattice elements would be about 100-150
$h^{-1}$Mpc, but the regularity in the alignment of these elements
can be detected for much greater distances. In the Tully et al. (1992)
supercluster distribution a straight line consisting of a chain of
superclusters can be identified, from Tucana to Ursa Major, or even to
Draco, in other words, a straight-line chain 700 $h^{-1}$Mpc long. We
will return to this point when discussing the cosmological magnetic
field. 

These observations have been rejected by many authors. The regularity
found by Broadhurst et al. (1990) has not been found in other
directions, but if a lattice is formed by filaments and voids, that is
precisely what should be expected. Only in particular selected beams
would a periodicity be detected. In the power spectrum of the Point
Source Catalogue redshift survey (Sutherland et al. 1999) no
periodicities, spikes or preferred directions were
found. There was only the marginal evidence of a ``step'' in the power
spectrum at $k \sim 0.08 h Mpc^{-1}$, but this was just a $2 \sigma$
effect that the authors considered a statistical fluctuation.

The possible crystal large scale structure is therefore under debate
at present. Correlation function analysis is probably not
appropriate to study a lattice of filaments and sheets, which would be somewhat
deformable, elastic-like, due, for instance, to the gravity action
caused by  the largest superclusters. Other statistical methods to
detect ``foam lattices'' should be developed. The
evidence of a crystal-like structure for scales larger than about 100
Mpc is overwhelming.

Even if this possible observational fact were in complete
contradiction with the hierarchical CDM models, strictly speaking, in
practice, it would not invalidate them. It could be that, within a large
structure ($\sim$ 100 Mpc), the models would be available at a much shorter
scale ($\le$ 30 Mpc).  Another explanation for the large scale should be
sought, but the smaller scale models could remain valid. 
Current theoretical models,  instead of an
initial random distribution of $\delta$, would start with a very
large wide filament-sheet lattice as the initial condition. In practice, the
existence of a large scale crystal is not incompatible with CDM
models.

In addition, the hypothesis of a fractal universe without upper 
limits (Sylos-Labini, Monturi and  Pietronero 1998) should be borne in
mind.

Two quasi-philosophical criticisms can always be made of numerical
models. If they assume a hypothesis and adjust a number of free
parameters to agree with observations, the conclusion is that if the
hypothesis is true, the set of free parameters proposed is correct,
but the hypothesis has not been proved to be true. In the best case,
the hypothesis is just compatible with observations. Furthermore,
for N-body simulations, if we find
results matching observational facts, we know that the physics used
is able to explain these facts, but we are still unaware of
the in-depth explanation.

\subsubsection{Some successes and failures}

Following Frenk et al. (1997) some basic facts such as the abundance of
DM halos, their merging history and their internal structure are
reasonably well understood. Semi-analytical models satisfactorily
explain the luminosity function, the number counts and colours, the
evolution of the Hubble sequence, the morphological type, the history
of star formation, etc. However, no model has succeeded in
producing a faint end slope of the galaxy luminosity function flatter
than $\alpha \approx 1.5$, whilst the observations indicate $\alpha
\approx 1$.

Reasonable results are also obtained for the morphologies of
galaxies. Galaxies with disks and bulges are directly obtained.
They may merge producing an elliptical galaxy. There is
increasing evidence that ellipticals arise from the mergings of
spirals. Apart from
the theoretical models, Pfenninger (1997) vigorously argues in  favour
of the
evolution from the so called ``late'' to the so called ``early''
galactic types. There is general agreement on this, though an
elliptical may then develop a new disk. However, the gaseous
disks obtained are too small, due to the loss of angular momentum to the halo
when they form. Note that in the early interesting
models by Larson (1974) where no dark matter was considered, just gas
and stars, and the calculations were not obtained from a cosmological
scenario, no combination of free parameters was able to produce large
enough disks. After many years the formation of disks is still a poorly
known process.

One of the most remarkable successes is the ability of semi-analytical
models to match the counts of faint galaxies as a function of
magnitude, redshift and morphology, particularly when standard CDM
cosmology is adopted. Frenk et al. (1997) wrote ``This agreement
is the most striking indication so far that the models contain some
element of truth''. The redshift distribution of galaxies with
magnitudes in the range 22.5$<$B$<$24.0 is reproduced in Fig. 16. This
agreement was only obtained when the data from Cowie et al. (1996) and
the Miller-Scalo IMF become available.

The Lyman Break galaxies (Steidel et al. 1996) probably constitute the
first generation of star forming galaxies. They are therefore a
challenge for theoretical models, especially considering that their
adjustable parameters are set taking into account z=0
properties. Lyman Break galaxies have little emission in the U filter,
because the redshifted Lyman Break with rest-wavelength at 912\AA  lies
redwards from the wavelength range of the U filter,
while redder filters are still unaffected and transmit a higher
signal. This is what should be expected for galaxies with $z \ge3$ and
what Keck has spectroscopically confirmed. This discovery was
disconcerting for hierarchical CDM models as they had predicted that galaxy
formation was a very recent phenomenon (with z$\approx$1). It seems,
however, that the use of the Miller-Scalo IMF instead of the earlier
Scalo IMF again provides an agreement with the observations of the Lyman
Break galaxies. Furthermore, a readjustment of the $\sigma_8$ parameter is
required. Agreement is obtained for both the standard CDM and the
$\Lambda$CDM models, both of which predict a similar early star formation
history. Some of the present galaxies could have had a Lyman Break
progenitor, specially the brightest galaxies. A bright spiral galaxy
could be the descendant of a single fairly massive Lyman Break object;
a bright elliptical could be the descendant of two less massive 
ones, merging at more recent
epochs (Baugh et al., 1998; Frenk et al. 1997; Gobernato et al. 1998). 
As studied by Baugh et
al. (1999), a remarkable success of the theory is that models can be
adjusted so that the agreement with the z=0 galaxy clustering (Postman
et al 1998 and others) also agrees with the clustering data at z=3 given by
Adelberger et al. (1998). Figure 17 plots the theoretically
obtained present B luminosity.

Salvador Sol\'e et al. (1998) took into account that continuous
accretion between mergers and tiny mass captures have a very different
effect from notable mergers. Between abrupt major mergers, the central parts
of halos grow steadily and their virial radius continues to expand.

\subsubsection{CDM Models, halo structure and rotation of spirals}

CDM models predict a halo structure which is responsible for
the rotation curve of the spiral galaxies. Halo structures and
rotation curves are therefore closely connected problems. Let us
assume that halos are spherical.

Gunn and Gott (1972) concluded that the gravitational collapse could
lead to the formation of virialized halos with almost isothermal profiles. A
tempting assumption for the halo density distribution is therefore
the so called non-singular ``isothermal'' sphere
\begin{equation}
  \rho(R) ={{\rho_0} \over {1+ \left({R \over {R_c}} \right)^2}}
\end{equation}
defined with two parameters: the central density, $\rho_0$, and the
core radius, $R_c$. In this isothermal profile $(d\rho/dR)(R=0) =0$ and
$\rho(R=0) = \rho_o$ is finite, two desirable properties for
a density profile. Out to $R \sim R_c$, the density remains more or
less flat, i.e.,  there is a ``core'' of radius $R_c$. Note that, in
this case, the ``circular velocity'', $V$, is defined as
\begin{equation}
  V^2(R) ={{GM(R)} \over R}
\end{equation}
where $M(R)$ is the mass in a sphere of radius $R$. It can then be
deduced that the circular velocity is given by
\begin{equation}
  V^2(R) = 4 \pi G\rho_0 R_c^2 \left[ 1-{R \over R_c} \arctan{R \over
  R_c} \right]
\end{equation}
which is an increasing function of $R$, asymptotically reaching
$V_{max} = V(R =\infty)$ given by 
\begin{equation}
  V_{max} = \sqrt{4 \pi G\rho_0 R_c^2}
\end{equation}
which might be an undesirable property (an asymptotically Keplerian curve would
be preferable).

Other types of halos have been reviewed by Bertschinger (1998)
and others. Recently Navarro, Frenk and White (1996, 1997) deduced from their CDM
models that halos should be described by the so called ``universal''
or NFW profiles
\begin{equation}
  {{\rho(R)} \over {\rho_{crit}}} = {{\delta_c} \over {{R \over R_s}
  \left( 1+{R\over R_s} \right)^2}}
\end{equation}
where $\rho_{crit}$ is the density of the critical Einstein-de Sitter
Universe
\begin{equation}
  \rho_{crit} = {{3H^2} \over {8 \pi G}}
\end{equation}
and $\delta_c$, the characteristic contrast density (dimensionless) and $R_s$,
the scale radius, are the parameters of the profile. It is singular,
$\rho(R=0) = \infty$, which is certainly an ``unpleasant'' property,
even if the mass $M(0)$ converges. The NFW profile was called
``universal'' because the authors found it in a large variety of halo
masses, spanning 4 orders of magnitude, from individual galaxies to 
cluster halos, as well as for a large variety of cosmological
scenarios. Some authors (e.g. Avila-Reese, Firmani and Hernandez,
1998) deduced that the density profile depends on the environment,
with the NFW appropriate only for isolated halos. The circular
velocity of this NFW halo can be calculated by
\begin{equation}
  \left( {{V_c(r)} \over {V_{200}}} \right)^2 ={1 \over x}
  {{\ln(1+cx) - {{cx} \over {1+cx}}} \over 
  {\ln(1+c) - {c \over {1+c}} }}
\end{equation}
where $V_{200}$ is the circular velocity at $R_{200}$, called the
``Virial'' radius. This virial radius is that radius for which $<\rho>
= 200 \rho_{crit}$, where $<\rho>$ is the mean density in a sphere of
radius $R_{200}$. Cole and Lacey (1996) showed that this radius
approximately separates the virialized and infall regions. The
parameter $c$, called the concentration, is defined as
\begin{equation}
  c = {{R_{200}} \over R_s}
\end{equation}
and is dimensionless. $x$ is simply $R/R_{200}$.

As $V_{200}^2 =GM_{200}/R_{200} = G (200 \times 3H^2/8\pi
G)(4\pi(R_{200})^3/3)/R_{200} = 100 H_0^2 R_{200}^2$, we have $V_{200}
= 10 H_0 R_{200}$, or
\begin{equation}
  V_{200} = R_{200} h
\end{equation}
if we measure $V_{200}$ in km$s^{-1}$ and $R_{200}$ in kpc. Also:
\begin{equation}
  M_{200} = (200 \rho_{crit}) {4 \over 3} \pi R^3_{200} = 100 {H^2 \over
  G} R_{200}^3
\end{equation}
therefore $M_{200} \propto R_{200}^3 \propto V_{200}^3$. If the
luminous mass were proportional to the halo mass, $M_{200}$, and if
$V_{opt}$ were related to $V_{200}$ ($V_{opt}$ is the disk
velocity at $R_{opt}$, the optical radius), then a relation similar to
the observational Tully-Fisher relation would be obtained. In conventional
astronomical units
\begin{equation}
  M_{200} = 2.33 \times 10^5 V_{200}^3 M_\odot
\end{equation}
where $V_{200}$ is to be expressed in km/s. The Tully-Fisher relation
clearly establishes a relation between the luminosity and some power of
the optical rotational velocity, $L \propto V_{opt}^3$ or $L \propto
V_{opt}^4$. (The exponent depends on the wavelength of the
observations. The higher value of 4 is for the infrared. See van der Bosch
(1999) for a recent critical review).

The obtention of the Tully-Fisher relation from the outcoming halo
density distribution presents some problems. The slope obtained and 
the scattering of the points agree with the observations, but the 
theoretical curve is displaced with the observational
curve (Frenk et al. 1997). Or equivalently, it is possible to vary the
free parameters to match the Tully-Fisher relation but then the
amplitude of the galaxy luminosity function is not matched. This is at
present a failure of theoretical models. The number of predicted dark
halos is excessive. Navarro, Frenk and White (1996) suggested several
possibilities: other cosmological parameters, the existence of a large
number of halos with no visible component, the non-detection of many
existent low surface brightness galaxies, etc. 

Even if the observational Tully-Fisher relation is basically 
reproduced, there is still no convincing explanation for the numerical
outputs. As Navarro (1998) remarks: ``Our analysis has
made use of this surprisingly tight relation between disk luminosity
and rotation speed but provides no firm clues to elucidate its
origin''. 

There is a general argument based on Shu (1982) that does not explain the
Tully-Fisher relation but does introduce some light. If the total
luminosity of a galaxy is roughly $L \propto \Sigma_0 R_{opt}^2$
(where $\Sigma_0$ is the central surface brightness, notably the same
for all spirals, and $R_{opt}$ the optical radius) and if the total mass
within the optical radius is roughly ${\cal M}\propto R_{opt}
V_{opt}^2$ (where $V_{opt}$ is the asymptotic observed velocity) and
if ${\cal M}/L$ is roughly a constant, then $R_{opt} \propto
V_{opt}^2/\Sigma_0$. If $\Sigma_0$ is really a constant $R_{opt}
\propto V_{opt}^2$ and therefore $L \propto V_{opt}^4$. This argument
relies on the constancy of $\Sigma_0$, which so far has no explanation. 

The NFW circular velocity reaches a maximum at $R \approx 2R_s = 2
R_{200}/c$ and declines beyond that radius. NFW density profiles are
two-parametric and it is possible to choose $V_{200}$ and $c$ to
characterize the halos, or the equivalent set of characteristic density and halo
mass. A very exciting result of these theoretical models is that the two free
parameters show a clear correlation. The reason behind this is that the
halo density reflects, and is proportional to, the true density when
the halo was formed, with the initial small halos being denser because they
formed earlier, when the density of the expanding Universe was higher.
But we also know that, due to the hierarchical halo
formation, more massive halos were born later. Then,
the existence of decreasing functions $\rho_c(t)$ and $M_{200}(t)$
implies a correlation between $\rho_c(t)$ and $M_{200}(t)$. Therefore,
in practice, rotation curves are intrinsically one-parametric. 
Figure 18 plots the circular velocity curves for different
values of the concentration, $c$. Low values of the concentration
parameters denote slowly rising curves, i.e. small ancient galaxies.

We saw above that the ``maximum disk'' hypothesis provides
halo circular velocities. Do NFW halo profiles fit these profiles? Are
disks really maxima? Do isothermal profiles provide a better fit than NFW
profiles? A comparison of the theoretically predicted halos and the
observations is necessary to answer these questions and to test the
models. This comparison was made by Navarro (1998), who adopted about
100 disk galaxies from published observations and tried to deduce the
NFW free parameter in each case. This analysis was also made under the
isothermal halo assumption.

For this task, we should not forget that galaxies also have visible
components, i.e. that spirals possess a disk and a bulge. For the
surface brightness of the disk,  Navarro (1998) assumed, as usual,
an exponential disk
\begin{equation}
  \Sigma (R) = \Sigma_0 e^{-R/R_d}
\end{equation}
where $\Sigma_0 = \Sigma (R=0)$ and $R_d$, the disk radial scale
length, are two parameters. Exponential optical disks constitute a
reasonable zeroth-order description, but there is perhaps a misuse in
the literature, as above discussed.  It is however an
appropriate and almost a necessary assumption in studies similar to that of
Navarro.

For the surface brightness of the bulge it is assumed that
\begin{equation}
  \Sigma_b(R) = \Sigma_{b0} e^{-(R/R_b)^4}
\end{equation}
where $\Sigma_{b0}$ and $R_b$ are two parameters. Navarro assumed the
same M/L ratio in the bulge and in the disk, and taht this ratio was
another free parameter.

It is also usually assumed that the halo responds ``adiabatically'' to
the growth of the disk (Barnes and White, 1984; Flores et al. 1993),
which was also  assumed by Navarro, Frenk and White (1996, 1997). This
means a variation of the NFW profile within the disk
region. Certainly, the formation of the disk must somewhat modify the
halo density profile, probably in a way that is very difficult to model. This
assumption was introduced in order to avoid disagreements with
observations of the rotation of dwarf galaxies and therefore, as a
correction of an initial theoretical failure. In fact, the rotation
curve of dwarf galaxies indicates that the halo circular velocity rises
almost linearly, which would mean a constant density (i.e. a halo core)
in clear contradiction with the NFW profiles. As mentioned above, this
hypothesis of the adiabatic response of the halo to the formation of
the disk also alleviates the problem of the halo-disk conspiracy.

The results indicate that theoretical models must introduce a
higher degree of sophistication, because even isothermal profiles give
similar or
better results, specially for low surface brightness galaxies. Moreover,
observational rotation curves are very often well fitted by halos
with no disk and no bulge! This type of fitting is
meaningless, but Cosmology could benefit from it. The 
concentration parameter, $c$, obtained after this peculiar halo-only fitting
provides an upper limit which can be compared with the theoretical
predictions on $c$. The standard CDM model in general predicts higher
concentrations than the upper limits obtained. Therefore, if the
theoretical models are considered an efficient basis to interpret the
observations, then cosmological models with small $\Omega(\sim 0.3)$
and large $\Lambda (\sim 0.7)$ are favoured.

A puzzling observation within the DM interpretation of rotation curves
is the absence of correlation between the asymptotic velocity of disks
and the orbital velocity in binary systems. If the asymptotic velocity,
$V_{rot}$, is found in a region dynamically dominated by the halo, and
the orbital velocity, $V_{orb}$, of a galaxy considered the secondary
would, clearly, reflect the total mass of the primary, $V_{rot}$ and
$V_{orb}$ should correlate. Navarro (1998) seeks the explanation
in his Fig. 19 (top-right: log$V_{200}$ versus log$V_{rot}$).
 If we observe the solid line in this figure, we see
that disks with $V_{rot}\le 150 kms^{-1}$ have $V_{200} > V_{rot}$ and
disks with $V_{rot} > 150 kms^{-1}$ are all predicted to have similar
halo velocities, $V_{200} \sim 200 kms^{-1}$. Then, disk-dominated
galaxies would be surrounded by halos of approximately the same
mass.

This explanation of rotation curves remains incomplete. 
Samples do not contain
dwarf galaxies. The Navarro sample is probably also lacking early
bulge-rich galaxies. In Fig. 26 there are two other curves,
the M/L =h curve and the varying M/L curve, which are not unreasonable
(for instance, $V_{rot} = V_{200}$ implies very small
variations of M/L) but which, however, do not imply the same
conclusion. Even if the solid line shows us a change in slope, the
increasing function does not firmly predict the complete absence of
correlation. The discussion of Navarro concerning this point is very
illustrative but the puzzling behaviour of binary galaxies is not completely
cleared up. The existence of  an upper limit of the halo mass needs
further justification.

Furthermore, Navarro and Steimetz (1999) find it difficult to reconcile the
theory with data for the Milky Way and with the Tully-Fisher relation;
they
consider that substantial revision of the theoretical models is needed.

As  a general conclusion, observational rotation curves are not
incompatible with NFW halos, but the confrontation seems somewhat
discouraging. We have examined the work of Navarro
(1998) in some detail because it is probably the most serious and complete study
linking the observed properties and the model outputs for rotation 
curves. However, either
the observations do not constitute a proof of the CDM models, or dynamic
ingredients other than halo and disk density profiles are necessary to
study the rotation of spirals. Considering the success of the models
in explaining and predicting other observational facts, we would
suggest this second possibility as more plausible. In particular, we
will later argue that ignoring magnetic fields in the interpretation
of the rotation curves could be unrealistic. 

\subsection{MOND}

The interpretation of the rotation curve of spiral galaxies is based
on the assumption that Newtonian Dynamics is valid. This assumption is
not accepted by the so called ``Modified Newtonian Dynamics'' (MOND)
developed and discussed by Milgrom (1983a,b,c), Sanders(1990) and others. An
implicit rule in our approach to Cosmology is that our physical laws are
valid everywhere, unless they lead us to unacceptable conclusions. This law is
therefore apparently violated. But Newtonian Dynamics was established
after considering nearby astronomical phenomena, and we are
allowed to modify it now, when we are aware of distant large-scale
phenomena, unknown in Newton's time. Moreover,
MOND aims not only to explain the rotation curve of spirals, but
to propose an alternative theory of Gravitation and/or Dynamics. With
respect to galaxies, the introduction of MOND provides very remarkable
fits, rendering this theory a very interesting alternative.

As an introduction, let us consider the outermost disk, where the
galactic mass can be considered a central point producing a
gravitational potential that predicts a
Keplerian decrease. But instead of the standard form for the acceleration
of gravity, $g = GM/r^2$, assume that it is expressed as $K(M)/r$,
where $K(M)$ is a constant depending only on the galactic mass. If
this force is matched by the centrifugal force, $\theta^2/r$, we
readily obtain $\theta^2 = K(M)$, therefore obtaining that $\theta$ is
a constant and thus the rotation curve paradox is automatically solved,
without the need for dark matter, at least for bright galaxies.

What would be the dependence of $K$ on $M$? A good assumption in the
absence of dark matter could be that the $M/L$ ratio is independent of
$R$. Therefore, if $K(M) = xM^{1/2}$ with $x$ being any constant, 
then we would directly obtain $M
\propto \theta^4$, and therefore, $L \propto \theta^4$, which is the
Tully-Fisher relation. Thus, we have already solved the two basic
problems of the rotation of spirals: flat rotation curves and the
Tully-Fisher relation. Therefore, $g$ would depend on $M^{1/2}$, or equivalently, on
$(GM)^{1/2}$. Hence, $g$=constant$(GM)^{1/2}R^{-1}$. The constant
would have the dimensions of the square root of an acceleration and would
be a universal constant. Let us call this acceleration
$a_0$. Finally, $g=(GMa_0)^{1/2}R^{-1}$. The constant $a_0$ was
introduced by Milgrom (1983a).

Let us estimate its value. If this kind of gravity is supported by
rotation, $\theta^4 = GMa_0$. Taking $\theta \sim 200 km s^{-1}$ and
$M= 10^{11} M_\odot$, as typical values in a spiral (without dark
matter), we obtain $a_0 \sim 1.2 \times 10^{-8} cm s^{-2}$. More
precise estimations provide $2 \times 10^{-8} cm s^{-2}$ (Milgrom,
1983b, for $H_0 = 50km s^{-1} Mpc^{-1}$) or $1.2 \times 10^{-8} cm
s^{-2}$ (Begeman et al. 1991, with $H_0 =75$). 

But we know that $g=GM/R^2$ for familiar dynamical systems. It could
be that the standard expression is valid for small $R$, and the tested
expression $(GMa_0)^{1/2} R^{-1}$ is valid for large $R$. In
the whole region $g$ could obey the sum of the two
\begin{equation}
  g= {{GM}/R^2} + {{(GMa_0)^{1/2}} \over R} 
\end{equation}
the first term would predominate at low $R$, and
the second at large $R$. This expression was first proposed by Sanders
(1990).

The transition region would be characterized by a similar order of
magnitude of the two terms, i.e. for $R$(transition) around
$\sqrt{GM/a_0}$. For $M$ of the order of a galactic mass, the
transition would take place at around 10 kpc (but a galaxy is not a
central point mass), thus suggesting that at large radii, the new
gravitational term would predominate, which solves both the flat
rotation curve and the Tully-Fisher problems.

This equation  is just a first simple example to see how our problems
can be solved by modifying Newton's Universal Law of
Gravitation. There is a second way, consisting in accepting this law,
but modifying Newton's Second Law, $\vec{F}=m\vec{a}$. Which procedure
is the best? Apparently, the first one is preferable, as the whole of
Newtonian Dynamics remains valid and we just modify a law that was proposed
by Newton from the observations, but without claiming it to be a
fundamental principle. In the same way, Einstein's Field Equations
could be modified without rejecting General Relativity. Therefore,
modifying the gravitational law would be a much softer procedure than
modifying the Second Law. However, for gravitational purposes, both
procedures are equivalent.

Let us reconsider the problem from a more general point of view.

In MOND, Milgrom (1983a) proposed, instead of the Second Law, that
\begin{equation}
  \vec{F} = m\mu \left({a \over a_0}\right) \vec{a} 
\end{equation}
with $\mu(x)$ being a function to be determined, of which we only know
\begin{equation}
  \mu(x \gg 1) =1 
\end{equation}
\begin{equation}
  \mu (x \ll 1) =x 
\end{equation}
i.e., for low accelerations, much less than $a_0$, the Second Law
would be substituted by $\vec{F} = m(a/a_0) \vec{a}$, being the force
proportional to the squared acceleration. In this way, in a galaxy,
with $a = \theta^2/R$ we would have $GM/R^2 = (\theta^4/R^2)a_0$ again
solving both the flat rotation problem and the Tully-Fisher relation.

Alternatively, the modification of the gravitational law could be
expressed in a general form as
\begin{equation}
  \vec{g} =a_0 I^{-1}(g_N/a_0) \vec{e}_N
\end{equation}
where $I$ is an unknown function, and therefore its inverse, $I^{-1}$, is
also an unknown function; $g_N$ is the standard Newtonian gravitational
acceleration and $\vec{e}_N$ is a unit vector with the standard
direction of $\vec{g}_N$. 

The directions of all vectors are the same, and therefore we can denote our
derivations without vector arrows. If $g_N$ is the classical Newton
gravitational acceleration we rewrite (85) as
\begin{equation}
  {g_N \over a_0} = \mu \left({a \over a_0}\right) {a \over a_0}
\end{equation}

The arguments of both functions $\mu$ and $I^{-1}$ are any variable,
but we will keep the notation $g_N/a_0=u$ and $a/a_0 =x$, therefore,
instead of (89)
\begin{equation}
  u= x \mu(x)
\end{equation}
and, instead of (88)
\begin{equation}
  {g \over a_0} = I^{-1} (u)
\end{equation}
(91) is equivalent to modifying the gravitational force while retaining
    Newton's Second Law, $g=a$, therefore
\begin{equation}
  {g \over a_0} = I^{-1}(u) = {a \over a_0} =x
\end{equation}
hence
\begin{equation}
  u = I(x)
\end{equation}
With (90)
\begin{equation}
  I(x) = x\mu(x)
\end{equation}
If $x \ll 1$, $\mu(x) = x$, hence $I(x) = x^2$, $I^{-1}(u) = u^{1/2}$.

If  $x \gg 1$, $\mu(x) = 1$, hence $I(x) = x$, $I^{-1}(u) = u$.

We have stated that the modification of Newton's Second Law, while retaining
the expression of the gravitational force, is equivalent to the
modification of the gravitational force and retaining Newton's Second
Law. We will show this with two examples, based on the
above expressions.

The first example considers the modification of Newton's Second Law,
in the way first proposed by Milgrom (1983b), corresponding to
\begin{equation}
  \mu(x) = {x \over {\sqrt{1+x^2}}}
\end{equation}

This expression is interesting as it is the simplest form of accomplishing
our asymptotic conditions (86) and (87). With (95) Newton's Second Law
would be replaced by
\begin{equation}
  \vec{F} = m \vec{a} {{a \over a_0} \over {\sqrt{1+ \left({a \over
  a_0}\right)^2}}}
\end{equation}
which for $a \gg a_0$, effectively reduces to $\vec{F} = m \vec{a}$.

Then our question becomes what is the equivalent transformation of the
gravitational force, i.e. producing the same dynamical
effects as the simple proposal of Milgrom (1983b) in (95)? We just
need to find $I$, and then $I^{-1}$ can be inserted in (88). With (94)
\begin{equation}
  I(x) ={x^2 \over {\sqrt{1+x^2}}}
\end{equation}

With (93)
\begin{equation}
  u = {x^2 \over {\sqrt{1+x^2}}}
\end{equation}
hence
\begin{equation}
  x= \left( {{u^2 + u\sqrt{u^2+4}} \over 2} \right)^{1/2}
\end{equation}

The signus (-) provides an non physical solution. With (92)
\begin{equation}
  I^{-1}(u) = \left({{u^2 + u\sqrt{u^2+4}} \over 2} \right)^{1/2}
\end{equation}
Observe that if $u \gg 1$, then $I^{-1}(u) =u$, which means that in
regions where Newton's classical gravitational force is high ($g_N
\gg a_0$), the gravitational force coincides with the classical
one. But, if $u \ll 1$, $I^{-1}(u) = u^{1/2}$ and $g= (a_0
g_N)^{1/2}$, it is rather different. The complete expression of $g$ would
become 
\begin{equation}
  g= a_0 \left( {{(g_N/a_0)^2 + (g_N/a_0)\sqrt{(g_N/a_0)^2 +4}} \over
  2} \right)^{1/2}
\end{equation}
an interesting formula because it is obtained without modification of
the three general laws of Newtonian Mechanics, explains flat rotation
curves without dark matter and explains the Tully-Fisher relation.

As a second example, suppose the inverse problem. We start with a
modification of Newtonian gravitational force and want to know 
the equivalent modification of Newton's Second Law. We must first
propose a new form of the gravitational force. We suppose that (86) is
the correct expression as we have seen that it also explains flat
curves and the Tully-Fisher relation. First, we rewrite (86) as
\begin{equation}
  {g \over a_0} = {g_N \over a_0} + \left({g_N \over a_0}
  \right)^{1/2}
\end{equation}

With (88) and $u = g_N/a_0$, this is written
\begin{equation}
  I^{-1}(u) = u+ u^{1/2}
\end{equation}
or
\begin{equation}
  I(u+ u^{1/2}) = u
\end{equation}
With (93)
\begin{equation}
  x = u + u^{1/2}
\end{equation}
or
\begin{equation}
  u = {{1+2x \pm \sqrt{1+4x}} \over 2}
\end{equation}
With (93) again
\begin{equation}
  I(x) = {{1+2x \pm \sqrt{1+4x}} \over 2}
\end{equation}
With (94)
\begin{equation}
  \mu(x) = {{1+2x \pm \sqrt{1+4x}} \over {2x}}
\end{equation}

If $x \gg 1$, $\mu(x) =1$, which matches (86). If $x \ll 1$, $\mu(x)
={{1 \pm 1} \over {2x}}$. If we take the signus (+), we have $\mu(x) =
x^{-1}$ which is not correct, following (87). We then take the signus
(-); then we apparently obtain $\mu(x) =0$. But we should then
expand $\sqrt{1+4x} \sim 1+2x -2x^2$ and therefore $\mu(x) = x$ in
agreement with (87). The signus (-) therefore gives us the physical
solution. The complete modified Second Law would then be
\begin{equation}
  \vec{F} = m\vec{a} \left({{1+2{a \over a_0} - \sqrt{1+4{a \over
  a_0}}} \over {2{a\over a_0}}} \right) 
\end{equation}
which could be a general expression, Newton's Second Law
could be just an approximation for high accelerations. 

\subsubsection{MOND applied to different astrophysical systems}

Milgrom (1983a) showed that this theory of gravitation could explain
the advance of Mercury's perihelion, first interpreted by Leverier as
due to Vulcano, hypothetical intramercurial planet, later
``observed'' by Lescarbault, and finally explained by General
Relativity. Another interesting application concerns the distance of
Oort's cometary cloud, as commented below. Milgrom (1983b) applied
MOND to the problem of the vertical distribution of stars in relation
to the velocity dispersion.

Of course, MOND was successful when applied to galaxies, as it was
originally intended to explain rotation curves. It is not remarkable 
that MOND explains
rotation curves, but what is really remarkable is that a very large
variety of galaxies can be fitted under this hypothesis, with only one
parameter, i.e. the M/L ratio of the bulge.

This task of fitting real data was undertaken by Begeman, 
Broeils and Sanders (1991) and
later continued by Sanders (1996) and Sanders and Verheijen
(1998). The method consists basically of obtaining $\vec{g}_N$ by classical
procedures and then considering equation (84) to fit the
results. With the sole exception of NGC 2841, the results were very
good, excellent in some cases, even better than the multiparameter
fitting considering a dark matter halo. About 80 spiral galaxies with
a large variety of luminosities and types were compatible with MOND.

The success was particularly interesting in the case of Low Surface
Brightness galaxies. These galaxies can be considered to have a low
surface density, too, in the absence of dark matter. Accelerations are
therefore so low that the whole galaxy can be considered within the MOND
regime. Milgrom (1983a,b,c) even deduced that positive slopes in the
rotation curve could be expected, which was later confirmed by
Casertano and van Gorkom (1991). These galaxies were studied by de
Block and McGaugh (1998), who found reasonable and constant $M/L$ ratios,
when the use of classical Newtonian Dynamics provides $M/L$ ratios
ranging from 10 to 75 (van der Hulst et al., 1993).

As mentioned above, dwarf spheroidal galaxies are also interpreted as being
characterized by very large $M/L$ ratios, in the range 10-100 (Mateo,
1994; Vogt et al., 1995). Gerhard (1994) applied MOND to 7 dwarf
spheroidals, without finding any agreement. This negative result was
confirmed by Gerhard and Spergel (1992), finding unacceptable
differences in the $M/L$ ratios required, but Milgrom (1995)
reanalyzed these 7 galaxies and obtained a reasonable agreement between MOND
and the observations.

De Block and McGaugh (1998) studied 15 galaxies
with low surface brightness, finding a low dispersion in the $M/L$
ratios. Sanders and Verheijen (1998) carried out the analysis in
the infrared K' band, where extinction and recent star formation
effects do not alter the photometric profiles, and obtained values in
good agreement with those predicted by MOND. 

Rodrigo-Blanco
and P\'erez-Mercader (1998) have directly considered the
modification of the Newtonian acceleration of gravity from the
rotation curve of 9 galaxies, also without the need of dark matter.

Van den Bosch and Dalcanton (2000) compared the results obtained with
semi-analytical models and with MOND. This search was undertaken
because in their opinion ``the dark matter scenario is certainly
starting to lose its appealing character'' due to the mix of baryons,
CDM and HDM needed, as well as a non-zero cosmological constant;
therefore, other alternatives should be seriously reconsidered. These
authors found that both theories can explain rotation
curves almost equally well, even if MOND needs a similar amount of fine-tuning.

Milgrom (1983c) also studied a large variety of systems with dark
matter problems, such as binary galaxies, small clusters, rich
clusters and in particular, Virgo. All these systems are characterized by
low acceleration, and no great quantities of dark matter were
required. The and White (1988) considered optical and x-ray
observations to check MOND in Coma, finding that models without
dark-matter are compatible. Sanders (1999) finds a non-negligible
difference between the dynamic and the luminous mass, this ratio being
about 2, less than that obtained by standard Newtonian Dynamics but
far from unity. He attributes this discrepancy to the fact that accelerations in
the innermost part of  clusters are not much lower than $a_0$
($\approx 0.5a_0$). If this application is correct, it could be due
to the non-detection of considerable luminous matter at the centre of
rich clusters.

Milgrom (1983a) also discussed the determination of mass in clusters
by the gravitational lensing method. Qin, Wu and Zou (1995) concluded
that no dark matter was needed under the MOND interpretation. The 
Faber-Jackson (1976)
relation, $L\propto \sigma^4$, where $\sigma$ is the velocity
dispersion in an elliptical galaxy, is also explained under Milgrom's
hypothesis.

More recently, Milgrom (1997) has studied the filaments that
characterize the large-scale structure of the Universe, comparing the
M/L ratio obtained with that of Eisenstein, Loeb and Turner (1997) who
found $M/L \sim 450 h$ in solar units in a filament in the
Perseus-Piscis supercluster, in contrast with that of Milgrom of only
$M/L \sim 19$, again requiring little or no dark matter.

$a_0$, with a value of about $2\times 10^{-8} cms^{-2}$, is very close
to that of $H_0c$ of the order of $6.5 \times 10^{-8} cm s^{-2}$, which
suggests that MOND could have some implications in Cosmology. Milgrom
(1983a,b) suggested that $a_0$ could be connected with the
cosmological constant. In any case, the matter contained in the
Universe could be considerably less. Felten (1984) developed a
MOND-Cosmology, proposing that the homogeneous and isotropic universe
would not be possible for small scales. Sanders (1998) continued
the discussion, finding another law for the growth of the cosmological
scale factor. This cosmology also provides a suggestive scenario for
the development of large scale structures.

\subsubsection{Final comments about MOND}

A modification of the physical laws should be attempted when all
classical hypotheses lead the wrong way. However, such modifications
are not uncommon in Physics. In particular, in Astrophysics, we could
remember the so called Steady-State Cosmology, which added a term of matter
creation. MOND is particularly attractive because, with a single
adjustable parameter, and a very limited range of allowable values, it
is able to explain the basic facts of galactic
dynamics very satisfactorily.

Equations (96) and (109) are not as simple as Newton's
equation. Simplicity, beauty and symmetry are apparently non-scientific
aesthetic concepts, but have often inspired scientific
discoveries and should not be completely absent when a renewal of
the fundamental laws in Physics is proposed. Indeed, simplicity is an
ingredient in MOND. Equation (95), for example,
was proposed by Milgrom because of its simplicity. Nevertheless, MOND
equations could be not simple but nevertheless true.

However, Newtonian Dynamics, irrespective of its historic origin, is
at present an approximation deduced from General Relativity and therefore
it enjoys the protection afforded by this wholly accepted theory. So, proposing
corrections to Newtonian Dynamics means rejecting General Relativity,
one of the most perfect physical theories. Unless MOND acquires a
similar justification by General Relativity, it would remain difficult  
to be accepted, and to date no such derivation has been
reported (Sanders, 1998). For such a task, the modification of
Einstein's Field Equations would be more acceptable than a
reformulation of the full theory of Relativity.

\subsection{The magnetic hypothesis}

Following this hypothesis (Nelson, 1988; Battaner et al. 1992, Battaner
and Florido, 1995; Battaner, Lesch and Florido, 1999; see also Binney, 1992) the rotation
curve of spiral galaxies may be explained by the action of magnetic
fields in the disk. If this hypothesis is correct, the cosmological
implications would be very important.

In this review about the magnetic scenario we will take
into account the following questions:
\begin{itemize}
{\item Are magnetic fields ignorable in the dynamics of the
outer disk?}
{\item What kind of magnetic fields do explain the flat rotation
curve?}  
{\item What mechanisms may produce these magnetic fields?}
{\item What is the ultimate origin of cosmic magnetic fields?}
{\item What is the overall picture of magnetic fields in
cosmology?}
\end{itemize}

\subsubsection{Are magnetic fields ignorable?}

At large radii, gravity decreases as $R^{-2}$. In contrast, magnetic
fields evolve locally due to gas motions. As in the case of the Sun,
at large enough radii, magnetic fields may become more important than
gravity, or even dominant.

It is at present evident that $10 \mu G$ fields exist in the inner
disks. It is increasingly evident that 1$\mu G$ fields exist in the
intergalactic medium, this point being addressed later. It is
therefore to be expected that in the region in between -the outer
disk- magnetic fields larger than $1 \mu G$ exist, even if they have
not been observed. 

Then, at some radii the magnetic energy density should reach the order
of magnitude of the rotation energy density
\begin{equation}
  {1 \over 2} \rho \theta^2 \sim {B^2 \over {8 \pi}}
\end{equation}
or, equivalently, the Alfven and the rotation speeds should have the same
order of magnitude. 

This equality of both energy densities will take place for a magnetic
field strength that depends on the gas density and on the rotation
velocity. Rotation velocities typically range from 50 to 200 $km
s^{-1}$. Typical values of $n$, the number density of atoms, can be
estimated as in our galaxy. Burton (1976) showed a plot in which $n
\approx 0.2 atoms cm^{-3}$ at 10 kpc and $n \approx 0.01 atoms cm^{-3}$ at
20 kpc. With an exponential decrease $n$ should be about $ 2 \times
10^{-3} atoms cm^{-3}$ at 25 kpc. Burton (1992) gives similar values
reaching $5 \times 10^{-4} at cm^{-3}$ at about 30 kpc. For other
galaxies it should be kept in mind that we measure the surface density
in atoms $cm^{-2}$ and to obtain $n$, the flaring of the layer must
be taken into account. For instance, for the Milky Way, the FWHM
thickness is about 300 pc at the Sun distance and at $R \sim$20 kpc,
it is higher than 1 kpc (Burton, 1992). HI disks appear to have a
cut-off at about $10^{19} atoms cm^{-2}$ (Haynes and Broeils,
1997). Van Gorkom (1992) found this cut-off at $4 \times 10^{18} atoms
cm^{-2}$. For a thickness of about 1 kpc, this corresponds to $n =
10^{-3} atoms cm^{-3}$. This cut-off is probably due to the ionization
of the intergalactic UV radiation, not to a cut-off in the hydrogen
itself. However, in our Galaxy, we observe number densities lower than
this. We should take $n = 0.3 - 3\times 10^{-4} atoms cm^{-3}$ in the
region of more-or-less flat rotation curves.

The following table gives the magnetic field strength required to
produce the same kinetic and magnetic energy densities.

Magnetic fields $\sim 0.1$ this strength should already have a
measurable influence; a magnetic field of this strength would even be a
dominant effect. To interpret this table it is therefore to be
emphasized that with the strengths given, even gravitation would be
negligible.

\vskip 1cm

\begin{tabular}{cccc}
 $\theta$ &  200  &   100  &  50  \\
 n      &       &        &      \\
 $3\times 10^{-1}$  &  35  &  18  &  8  \\
 $3\times 10^{-2}$  &  11  &  6   &  3  \\
 $3\times 10^{-3}$  &  3   &  1   &  0.7 \\
 $3\times 10^{-4}$  &  1   &  0.6 &  0.3 
\end{tabular} 

\vskip 1cm

$n$ in atoms$cm^{-3}$, $\theta$ in $km s^{-1}$, $B$ in $\mu$G.

For instance, in our galaxy, the magnetic field would be negligible at
the Sun distance, important at 20 kpc and dominant at the rim.

Moreover, we can compare the gravitational attraction and the magnetic 
force. For an order of
magnitude calculation let us adopt the point mass model,
$\rho GM/R^2$, where $M$ is the galactic mass (without dark matter) and
the magnetic force is, as we will see later, of the order of
$(B^2/R)(1/8\pi)$. For a galaxy like the Milky Way we obtain similar
values as before. A non-negligible magnetic field should be of the
order of $6 \mu G$ at R = 10 kpc, of $1 \mu G$ at R= 20 kpc and of
$0.4 \mu G$ at R= 0.4 kpc. 

For dwarf late-type galaxies, which are usually considered to need
higher dark matter ratios, the magnetic fields required are higher but
nevertheless worryingly large. Using the same estimation formula $\rho
GM/R^2 \sim B^2/(8\pi R)$ we have approximately
\begin{equation}
  B^2 \approx 10^{-8} {{\Gamma \Sigma} \over {RH}}
\end{equation}
where $\Gamma$ is the visible M/L ratio in solar units, $\Sigma$ the
typical surface brightness in $L_\odot pc^{-2}$, $R$ the radius in kpc
and $H$ the scale height in kpc. For typical values, $\Gamma =1$,
$\Sigma =0.3$, $R =5$, $H=0.5$ (from Swaters, 1999) we obtain that a
strength of $B
\sim 3.5 \times 10^{-5} G$ would produce a force as important as
gravitation in the whole galaxy and that $\sim$1/10 this value $\sim 4
\times 10^{-6}$G would be non-negligible.  The orders of magnitude
obtained by the equality of kinetic and magnetic energy are again
similar.

Following the analysis of Vall\'ee (1994), the hypothesis of
magnetic-driven rotation curves is unsustainable. The
magnetic field strengths required in the model by Battaner et al.
(1992) were too high, by at least a factor of 2, as compared to the
weaker magnetic field strengths observed. Battaner et al. (1992)
indeed required high magnetic fields, of about 6 $\mu$G at the
rim. However, after the publication of the study by Vall\'ee (1994), Battaner and Florido
(1995) recalculated the strength required, by means of a two-dimension
model including escape and flaring, obtaining much lower values, of
the order of 1$\mu$G. These values are not incompatible with
observations, for instance reviewed by Vall\'ee (1997), in his
exhaustive analysis of cosmic magnetic fields at all scales, and in
particular in spiral galaxies. In this review, he collates a number of
measurements obtained by other  authors and by himself. On the other
hand, there are not many measurements available for the outermost
region of the disk, as discussed later.

The figures in the above tables are worrying. It can be concluded that {\it{interpreting
rotation curves, while ignoring the influence of magnetic fields may be
completely unrealistic}}. It is therefore remarkable that a fact that
may be so
far-reaching concerning our cosmological beliefs has been object to such scarce 
attention. 

The magnetic hypothesis takes this fact into consideration and tries
to determine whether magnetic fields alone, without requiring any dark matter,
and without modifying our physical laws, are able to explain the
observed flat and fast rotation curves. The existence of dark matter
cannot be completely excluded, but here we explore the extreme case
with no DM at all.

\paragraph{Extragalactic magnetic fields}

Observations are probably still too scarce to reveal the magnitude and
distribution of extragalactic magnetic fields. Kronberg (1994) has extensively
reviewed all available observations based on synchrotron radiation and
its Faraday rotation, and has proposed several properties of which 
the following are outstanding:

a) Typical values of intergalactic magnetic field strengths are in the
range 1-3 $\mu G$. These are larger than previously thought, so their
influence on a large variety of phenomena must be revised.

b) These values are found nearly everywhere and are noticeably independent
of the density of the zone observed. They are found in cluster cores,
in clusters and in regions between clusters (e.g. between Coma and
A1367, Kim et al. 1989). The first measure of magnetic fields in the
intracluster medium was reported by Vall\'ee et al. (1987), finding
$\sim 2 \mu$G in A 2319. Feretti et al. (1999) have obtained a field
strength of between 5 and 10 $\mu$G in Abell 119.
In superclusters, high value strengths have been reported
(e.g. Vall\'ee, 1990, finding about 2 $\mu$G in the Virgo Supercluster
for the ordered component of the field).

Kronberg has speculated about a ubiquitous
magnetic field. In some particular objects, such as radiosources,
magnetic fields can be much higher, but this 1$\mu G$ background field
seems to be ubiquitous.  The value
of 3$\mu G$ is particularly interesting since then the magnetic energy
density equals that of CMB, thus suggesting an equipartition of both
energies. Note that both energy densities decrease as
$R^{-4}$ (being $R$ the cosmic scale factor), but this equipartition,
if it exists at all, cannot be primordial, as argued below. However, 
magnetic fields have never
been reported in the large-scale $\sim$100 Mpc sized voids. Only
Vall\'ee (1991) has searched for an excess rotation measure in the Bootes
Void, estimating that the magnetic field strength was less than
0.1$\mu$G.

c) Magnetic fields of this magnitude were also present in quasar
absorption line clouds, usually interpreted as pregalactic
systems. Therefore pregalactic clouds were  magnetized as much as
present galaxies. Field strengths of this order have also been measured at
redshifts 0.395 and 0.461 (Kronberg, Perry and Zukowski, 1992;
Perley and Taylor, 1991).

We favour another global picture that is fully compatible with
observations but slightly different to Kronberg's view of ubiquitous
1$\mu$G field strength. This global picture is also based on our own
theoretical work, which will be commented later. We assume that magnetic fields
vanish, or have very small strengths in the large-scale voids, in
agreement with Vall\'ee (1991),  i.e. in
most of the volume of the Universe, but are much higher
in the filaments of matter ($\sim$100 Mpc long, $\sim$10 Mpc thick)
characterizing the large scale structure. Therefore, magnetic fields
would be neither ubiquitous nor in energy equipartition with the
CMB, but, in any case, they are high, about 1-3$\mu$G, in the medium
surrounding nearly all galaxies. The medium around galaxies should have 1 $\mu$G
strengths because this is the value at the particular sites where galaxies
lie.

More recently, the review by Eilek (1999) confirms the existence of
$\mu$G field strength in clusters (even in cluster halos), being much
higher at the centre ($B_\parallel \sim 50 \mu$G in M87, for
instance). 

\paragraph{Magnetic fields in the outermost region of galactic
disks}

Measurements carried out in this zone have not been reported. By
roughly interpolating between the large 10$\mu$G fields in the inner
disk and the lower than 1$\mu$G fields  outside the galaxy, we cannot
exclude fields $\ge 1 \mu G$ in the outermost disk.

Objections to the existence of $1 \mu G$ fields at large radii could
be raised, with the argument that no detectable synchrotron emission has been
reported. However, the non-detection of synchrotron emission cannot be
interpreted as the absence of magnetic fields. Kronberg (1995) wrote that
``synchrotron radiation can tell us only that magnetic field is
present, but not measure its strength''.

Despite this pessimistic point of view, let us make some simple
estimations.
When the relativistic electrons responsible for the synchrotron emission
have an energy distribution given by $N dE = N_0 E^{-\gamma}dE$, with
$N_0$ and $\gamma$ being constants, then the synchrotron intensity can be
calculated with (e.g. Pacholczyk, 1970; Ruzmaikin, Shukurov and
Sokoloff, 1988)
\begin{equation}
  I \propto N_0 \nu^{{1-\gamma} \over 2} B_\perp^{{1+\gamma}\over 2}
\end{equation}
where $B_\perp$ is the component of $\vec{B}$ perpendicular to 
the line-of-sight. The
calculation of $B_\perp$ once $I$ is measured, is difficult because
$N_o$ is unknown. The spectrum of the synchrotron continuum itself
$[I, \nu]$ permits the easy obtention of $\gamma$, but not of $N_0$,
meaning the number density of relativistic
electrons is unknown. To surmount this difficulty the most usual assumption is
that of equipartition.

Equipartition is equivalent to the assumption of equal values of the turbulent
and magnetic energy densities and that the energy density is the
minimum for a given magnetic field, in which case (Ruzmaikin, Shukurov
and Sokoloff, 1988)
\begin{equation}
  B^{7/2} \propto {{\cal L} \over V} \propto q \propto I
\end{equation} 
where ${\cal L}$ is the luminosity of an emitting cloud, $V$ the
volume and $q$ the flux.

We will later show that magnetic fields with a gradient slightly less
than $B \propto R^{-1}$ can  produce a flat rotation curve. If
for an estimation we take $B \propto R^{-1}$, then
\begin{equation}
  I \propto R^{-7/2}
\end{equation}
i.e. $I$ decreases much more rapidly than $B$ does (Lisenfeld, 2000). Therefore, we
would not observe synchrotron emission where the magnetic field presents
significant values.

The coefficient in (113), $I \propto B^{7/2}$, is not perfectly known
because it depends on the ratio of protons to electrons in cosmic
rays, which has a value in the range 1-100, but following current estimates
(Lisenfeld et al. 1996) for a typical VLA beam of 15 arcsec$^2$,
2.6 $\mu$Jy would correspond to 1$\mu$G. However, the confusion limit,
or minimum detectable flux at, say, 1.5 GHz is about 20 $\mu$Jy,
noticeably larger than the expected 2.6$\mu$Jy.

Some works take the equation (112) with a hypothesis about $N_0$. If
relativistic electrons are born in type-II Supernova explosions, which
in turn are produced in regions of star formation, i.e. in sites with
high gas density, and if relativistic electrons are not able to travel
far from the birth region, then $N_0 \propto \rho$, could be an
interesting, simple and acceptable assumption. But in this case, the radial 
decrease of $I$
would be much faster; much faster even than the exponential (with
typical radial scale length about 3 kpc). The reduction of $\rho$
because of the external flaring would give a still faster truncation
of the synchrotron continuum. If we assume that type-I Supernovae also
contribute to producing relativistic electrons, the truncation of $I$
will be even faster, as a result of the stellar truncation typical in
all disks. In the Milky Way it takes place at about 12 kpc (Porcel,
Battaner and Jim\'enez-Vicente, 1997).

Moreover, there is another argument to show that the absence of
synchrotron radiation does not imply the absence of magnetic fields. It is
observed that the synchrotron spectrum suddenly steepens for large
radii. This feature takes place, for instance, in NGC 891 (Hummel et
al, 1991; Dahlem, Dettmar and Hummel 1994) for $\ge$6 kpc. If the
slope of the [$\log{I}$, $\log{\nu}$] curve, usually called $\gamma$,
is high, the number of very high energy electrons is relatively
low. It is known (Lisenfeld et al. 1996) that these very high
energy electrons have less penetration capacity, i.e. they cannot
travel far from their sources. The simplest form of interpreting
the increase of $\gamma$ at those radii when the synchrotron becomes
undetectable is a truncation of the relativistic electron
sources. It is then probable that, the absence of cosmic electrons, rather than
the absence of magnetic fields, is responsible for the low synchrotron
intensity in the outermost disk.

\subsubsection{The magnetic model}

In this Section we try to determine what kind and magnitude of
magnetic fields are necessary to explain the rotation of the outermost
disk. These magnetic fields could introduce some instabilities
in the disk, related to flaring, winds and escape, which are also
examined.

\paragraph{The one-dimension model}

This exploratory model was developed by Battaner et al. (1992)
following the magnetic hypothesis previously proposed by Nelson
(1988). The two models are not equivalent. For instance, Nelson's magnetic
field must have a non-vanishing pitch angle (i.e. the angle between
the direction of the field -assumed to be contained in the galactic
plane- and the azimuthal direction). Battaner et al. (1992) consider pure
azimuthal (toroidal) fields or, rather, the azimuthal component of the
field producing the required force.

In the radial component of the equation of motion, it is necessary to
include magnetic forces, which are of the form ${1 \over {4 \pi}}
\vec{B} \cdot \nabla \vec{B} - {1 \over {8\pi}} \nabla B^2$
(e.g. Battaner (1996)). The first term in cylindrical coordinates
(R, $\varphi$, z) will be
\begin{equation}
  {1 \over {4 \pi}} \vec{B} \cdot \nabla \vec{B} =
  {1 \over {4\pi}} (B_R, B_\varphi, B_z) \left[ \begin{array}{ccc} 
    {{\partial B_R} \over {\partial R}} & {{\partial B_\varphi}\over
    {\partial R}} & {{\partial B_z} \over {\partial R}} \\
    {1 \over R}{{\partial B_R} \over {\partial \varphi}}-
    {{B_\varphi}\over R} & {1 \over R}{{\partial B_\varphi}\over
    {\partial \varphi}} + {B_R \over R} & {1 \over R}{{\partial
    B_z}\over {\partial \varphi}} \\
    {{\partial B_R} \over {\partial z}} & {{\partial B_\varphi}\over
    {\partial z}} & {{\partial B_z} \over {\partial z}}
    \end{array} \right]
\end{equation}

if the field is azimuthal, ($B_R = B_z =0$), under azimuthal symmetry,
($\partial/\partial \varphi=0$), the radial component of this force is
simply $- B_\varphi^2 /R$.

The magnetic pressure gradient force $-{1 \over {8 \pi}} \nabla B^2$
would have a radial component simply given by $-(1/8\pi) \partial
B_\varphi^2/\partial R$, and therefore the radial component of the
magnetic force would be
\begin{equation}
  F_{m,R} = -{1\over {4\pi}}{{B_\varphi^2} \over R}
            -{1 \over {8\pi}} {\partial \over {\partial R}}
               B_\varphi^2 =
            -{1 \over {8 \pi}}{1 \over {R^2}}{{\partial (R^2
		B_\varphi^2)} \over {\partial R}}
\end{equation}
By including this force in the radial component of the equation of
motion, we obtain
\begin{equation}
  \rho \left[ -{{\partial {\cal F}} \over {\partial R}} +
  {{\theta^2} \over R} \right] -
  {{\partial p} \over {\partial R}} -
  {1 \over {8 \pi R^2}}{{\partial (R^2 B_\varphi^2)} \over {\partial
  R}} = 0
\end{equation}
in which steady-state conditions and vanishing
viscosity are assumed. ${\cal F}$ is the gravitational potential. Here, the
pressure gradient force in the radial direction is usually ignored.

From this equation, with current estimates for the different terms,
but excluding any dark matter halo, it is easy to integrate numerically
and obtain $B_\varphi (R)$. This was done by Battaner et al.
(1992).

For didactic purposes only, it could be interesting to consider an
ideal analytical calculation, assuming, for such large radii,
that gravitation itself could be considered negligible. In such a
case
\begin{equation}
  B_\varphi^2 = {{B_0^2 R_0^2} \over R^2} +
  8 \pi \theta^2 \int_{r_0}^R \rho R dR
\end{equation}
where $B_0 = B_\varphi (R=R_0)$ and $R_0$ is the radius where the
integration begins. As $\rho$ decreases exponentially (or faster, due
to flaring, truncation, etc.) the last integral converges. For 
large enough $R$ the first term would become dominant and we obtain that
$B_\varphi$ should become asymptotically
\begin{equation}
  B_\varphi (R \rightarrow \infty) \rightarrow B_\varphi^* (R) \propto
  R^{-1} 
\end{equation}
This profile, $B_\varphi^*(R)$, does not produce any force, neither
inwards nor outwards, and will be called critical. The real profile
should have a slope lower than the critical one, to produce the fast
rotation, as the magnetic pressure gradient force is probably an
outward force. On the other hand, the magnetic tension $B^2_\varphi/R$
is always inward and does not depend on the gradient, ${{\partial
B_\varphi} \over {\partial R}}$.

An intuitive reasoning underlying these equations, about how a magnetic
tension produces an inward force, is that the term
$1/4\pi \vec{B} \cdot \nabla \vec{B}$ pushes the gas along the field
lines. In a ring where the magnetic field lines are circular and
contained in the ring, a radial inward force will be produced. This
force will also be present in the disk composed of many rings and if
it is higher than that produced by the magnetic pressure, a net inward
force would be added to gravity, which can only be compensated with
an enhanced centrifugal force. Therefore the disk must rotate more
rapidly.

In the exploratory model by Battaner et al. (1992), the calculated
strengths are close to the critical (or asymptotic)
profile $B_\varphi^* (R) \propto R^{-1}$ for very large radii. In this
basic model a strength of about 6 $\mu$G at $R = 25 kpc$ was obtained,
which is indeed very high. In more recent models, which will commented
later, much lower values of $B_\varphi$ are obtained, even less than
1$\mu$G . The authors considered that the predicted synchrotron radiation
was not in conflict with observations reported by Beck (1982) and
that stability problems in the disk
could arise if the strength reached such high values.

To explore this problem, Cuddeford and Binney (1993) developed a
single model, with which they demonstrated that a disk with such a large
magnetic field would produce excessive flaring. They assumed in this
model that the magnetic pressure was $\alpha$ times the density, and
this constant, $\alpha$, was considered independent of $z$, but it was
allowed to have a dependence on $R$. There exist several observations
that prove that $B^2/8\pi$ decreases with $z$ much more slowly than
does the
density, mainly in galaxies with a radio-halo (Ruzmaikin,
Shukurov \& Sokoloff, 1988; Hummel et al. 1989; Hummel, Beck \& Dahlem,
1991; Breitschwerdt, McKenzie \& V$\ddot{o}$lk, 1991; Wielebinski,
1993; Han \& Qiao, 1994 ...) and so the large flaring calculated by
Cuddeford and  Binney was clearly overestimated. Nevertheless, this
paper was very illustrative in showing that the vertical component of
the motion equation must unavoidably be integrated with the
radial one, to assess the problem properly. If the magnetic field
strength capable of driving the rotation curve is too high the disk may
become thicker and flare. Moreover, Vall\'ee (1997) considers that 
the strength required would unacceptably expand the HI disk.

But even when adopting $B^2/8\pi = \alpha \rho$ with $\alpha(r)$ being
independent of $z$ (i.e. the condition assumed by Cuddeford and
Binney, 1993), the flaring of the disk would not be so large as
estimated by these authors. If the disk were too thick the gas far
from the plane would escape if it were slightly perturbed by
very small vertical winds. As such winds are probably present in
spiral galaxies, gravitation at such a high $z$ would be too weak to
retain gas moving outward.

An expanded disk cannot be retained. Clouds very far from the plane would
be blown away by instabilities producing vertical winds and strong
disk-corona interaction.

To demonstrate this and solve the problem raised by Cuddeford and
Binney (1993) and Vall\'ee (1997), Battaner and Florido (1995) developed a second
model in which they adopted the most disadvantageous magnetic vertical
profile, according to the $B^2\propto \rho$ condition, but
considered vertical winds and escape of gas, in the z-direction, to a
galactic corona. Let us include a summary of this model.

\paragraph{The two-dimension model. Flaring and escape}

In this model by Battaner and Florido (1995) the vertical 
component of the equation of motion is written as
\begin{equation}
  \rho v_z {{\partial v_z}\over {\partial z}} + \left( {{v_A^2}\over
  2} + \beta \right) {{\partial \rho}\over {\partial z}} +
  \rho G M R^{-3} z =0
\end{equation}
where $v_z(R,z)$ is the vertical velocity, $v_A(R)$ is the Alfven
velocity considered to be a constant in the vertical direction
(because of the assumption $B^2 \propto \rho$, 
as $v_A =B/\sqrt{4\pi \rho}$); also $\beta =p(R,z)/\rho(R,z)$ is
assumed to be constant (isothermal condition, or rather, constant
cloud-to-cloud velocity dispersion) and the vertical component of the
gravitational force is simplified to that due to a central point
mass; this assumption is in part justified because it is assumed that
no dark matter halo is present and because we are considering just the
outermost part of the disk. Many of these simplifying conditions are
not necessary and, indeed, are only justified in a model with
exploratory aims.

Therefore the motion in the vertical direction was considered to be
the result of four terms: the inertial term, the magnetic, the
gravitational and the pressure gradient forces. Horizontal velocities
were considered, but the term $v_R \partial v_z/\partial z$ was
neglected.

The equation of continuity restricts the possibilities of the vertical
flux. It was assumed that
\begin{equation}
  \rho (R,z) v_z(R,z) = \rho(R,0) v_z(R,0)
\end{equation}

To calculate $v_z(R,0)$ at $z=0$ in the plane (or slightly above the
plane, to avoid the problem arising from the symmetry in the galactic
plane; $v_z(R, \Delta z) = -v_z(R, -\Delta z)$, for  very
small $\Delta z$, this would imply $v_z(R,0) =0$) some assumptions
must be made; i.e.
adopting a physical mechanism responsible for the vertical flux, which
could be produced by supernova explosions. In this case it would be
preferable to adopt a hypothesis of the type $v_z (R,0) \propto \rho$,
if for large time scales supernovae are born where the gas is
denser. Battaner and Florido, instead, considered that Parker
instabilities were the main origin of the vertical flux and assumed
for the flux at $z=0$:
\begin{equation}
  \rho(R,0) v_z(R,0) = k {{B^2(R,0)}\over {8\pi}}
\end{equation}
the value of $k$ being considered a free parameter. This condition is
equivalent to
\begin{equation}
  v(R,0) = k {{v_A^2(R)} \over 2}
\end{equation}

Instead of $\rho$, it is preferable to use
\begin{equation}
  y(R,z) ={{\rho(R,z)} \over {\rho(R,0)}}
\end{equation}
because the calculation of $\rho(R,0)$ and the profile $y(R,z)$ were
made using different equations. Using Alfven's velocity the radial
component of the equation of motion is now written as
\begin{equation}
  {1\over 2} {{\partial}\over {\partial R}} v_A^2(R) +
  {1 \over 2} v_A^2(R) \left[{1 \over {\rho(R,0)}}
  {{\partial \rho(R,0)}\over {\partial R}} + {2 \over R} \right] -
  {{v_H^2 (R)} \over R} =0
\end{equation}

The variable $v_H(R)$ is defined by
\begin{equation}
  {{v_H^2(R)} \over R} = {{\theta^2(R)} \over R} -
  {{\partial {\cal F}(R)} \over {\partial R}}
\end{equation}
which is what in more conventional theories is called the ``halo
velocity'' as a way of introducing the halo potential. This quantity
$v_H(R)$ was used because it can be found directly in the literature,
but no dark matter halo was introduced. No inertial terms, neither
$v_z \partial v_R/\partial z$ nor $v_R\partial v_R/\partial R$, were
considered to be important and the gradient pressure force was again
considered negligible.

The surface density was adopted from the literature, and therefore
$\int_{-\infty}^{\infty} \rho(R,z) dz$    was kept constant. To
compensate for the escape, it was assumed that the horizontal flux from
the central region was so easily established that it was  able
to supply the necessary escaped mass at all radii. The function $v_R(r)$ was
however found to be negligible. When the vertical velocities reached a
value higher than the typical velocity dispersion, say 8 km/s, gas
clouds were considered to have escaped from the disk, but not necessarily from
the galaxy.

Though not affecting the numerical computations, it is interesting to
obtain two functions of interest in the interpretation. One of
these is the ``Flaring Function'', $Z(R)$ defined by
\begin{equation}
  \int_0^{Z(R)} \rho(R,z) dz ={1\over 2} \int_0^\infty \rho(R,z) dz
\end{equation}
and the other is the total mass loss rate
\begin{equation}
  \dot{M} = 2\int_0^\infty \rho(R,0) v_z(R,0) 2\pi R dR = 
            {1 \over 2} \int_{R_0}^\infty kB^2(R) R dR
\end{equation}
where $R_0$ is the adopted inner boundary radius. $Z(R)$ is important
because if the disk is highly magnetized, $Z(R)$ can become
unacceptably large. If $\dot{M}$ is calculated to be too large, the
whole galaxy could evaporate.

A numerical integration of these equations was carried out by Battaner
and Florido (1995) taking M31 as a representative galaxy, with a
convergent procedure that we do not reproduce here, but just show the
results (see Fig. 20).

The free parameter $k$ should have a value of between $10^{-9}$ and
$10^{-8} cm^{-1} s$, with $k = 3 \times 10^{-9}$ being the value giving the
most reliable results. The mass loss rate, $\dot{M}$ was found to
be 0.054, 0.16 and 0.55 $M_\odot yr^{-1}$ for $k = 10^{-9}$, $3 \times
10^{-9}$ and $10^{-8} cm^{-1} s$; this is still rather low, lower in
any case than the typical value given by fountain models (15 $M_\odot
yr^{-1}$; Kahn, 1994). Part of this gas that escaped from the disk would
eventually fall back into the disk. Even if this were not the case and
in the worst situation in which all the gas escaping from the disk 
escaped from the galaxy, the total mass loss during the whole history
of the galaxy (assuming the flux to be constant in time) would be of the order of
0.16$M_\odot yr^{-1}\times 10^{10} yr \sim 1.6 \times 10^9 M_\odot$,
an acceptable value.
                
Even with the simplifying conditions assumed, a
coherent general scenario is obtained:

a) The Alfven speed increases outwards is always lower than the
rotational velocity but has a common order of magnitude. $B$ and
$\rho$ decrease but $B^2/\rho$ increases.

b) The effect of flaring and escape reduces the magnetic field required
to drive the rotation of the outer disk. In the previous simple model,
$6 \mu G$ at 30 kpc was obtained. Now it is  a full order of
magnitude less. This is a very exciting figure, as it confirms that
moderate magnetic field strengths can have a decisive influence on the rotation
curve.

c) The flaring seems to be high but this is in reasonable agreement with
observations. For low radii, the adoption of the central point mass
potential is not appropriate. For instance, Z(R) is too high for radii
less than about 17 kpc. But in our Galaxy, where precise  data exist for
very large radii, a value of $Z \sim 6$ kpc at R= 20 kpc from the
plots provided by Diplas and Savage (1991) is reproduced by the
model fairly well.

d) The densities are reasonable. At 30 kpc, values of the order of $1.6
\times 10^{-28} g cm^{-3}$ ($10^{-4}$atoms cm$^{-3}$) were obtained,
and at 25 km, $5 \times 10^{-28} g cm^{-3}$. In any
case these values are compatible with the observed surface density, as
this function was adopted from the observations.

e) Velocities (vertical at the base of the galactic plane and radial)
are small, of the order of a few km/s, which are nearly undetectable and
therefore do not introduce problems of disagreement with any
measurement. Consider that 2 km/s at $z \sim 0$ may produce 10 km/s at
$z \sim 8 kpc$ and $R \sim 20 kpc$, due to
continuity. (The flux would be conserved, so the decrease in density
accelerates the vertical speed). Velocities of this order of magnitude
are observed even in a quiet disk (see for instance Jim\'enez-Vicente
et al. 1999).

\paragraph{Some indirect arguments against and favouring the
magnetic hypothesis}

Apart from the objection by Cuddeford and Binney (1993) to the
one-dimension model above
mentioned, which was solved in the two-dimension model,  
Persic and Salucci (1993) considered the magnetic hypothesis as ``neither
necessary nor sufficient''. Setting aside the question of how a theory
can be ``not necessary'', they argued that galaxy pairs  need
dark matter, but we have seen that galaxy pairs admit other
interpretations, in particular that of common halos.

S\'anchez-Salcedo (1996) has considered the possibility that a
relation found by Bosma (1978, 1981, 1993) between HI and dark matter
density could be explained under the magnetic hypothesis. On
qualitative grounds this would be reasonable, because the higher the
gas density, the higher the magnetic strength that could be amplified.

There is a possible connection between the truncation of stellar disks
and the magnetic hypothesis for the rotation curves. Once stars are
born, the centripetal magnetic force, previously acting on the
progenitor gas cloud, is suddenly interrupted and stars move to
larger orbit radii or escape. This escape would be responsible for the
truncation of stellar disks, which is a common feature in
spirals. 

Vall\'ee (1994, 1997) also addresses this point. He considered that
newly formed stars would acquire ballistic velocities of the order of the
rotation velocity of the parent gaseous cloud. Stars with this velocity have
not been observed. If stars were rotating in Keplerian orbits -Vall\'ee
argues- they should decelerate. The answer to this problem raised by
Vall\'ee (1997) lies in the fact that many new-born stars could escape. Others
would simply migrate to more energetic orbits. A numerical model of
high velocity new-born stars, escape in the radial direction and
truncation of the stellar disk is currently being constructed.

Pfenniger, Combes and Martinet (1994) and Jopikii and Levy (1993)
argued that, following the Virial theorem, magnetic fields should have
an expansive effect, in contrast with the magnetic centripetal force
deduced by Battaner et al. (1992). This is a peculiar argument, as the
Virial theorem is deduced from the equation of motion, which was the
equation integrated by Battaner et al. (1992). This could constitute a
paradox in the one-dimension model. The solution should be found in
the two-dimension model (Battaner and Florido, 1995), where
it was shown that there is an escape of gas in the vertical direction. In
other words, magnetic fields have a contracting effect in the radial
direction, but an expansive one in the direction perpendicular to the disk;
hence, the net effect of magnetic fields could be expansive. We have
also suggested that a large fraction of new-born stars could escape in
the radial direction, which is also an expansive dynamical
effect. It should be borne in mind that the two-dimension model (1995) was published after
the study of Pfenniger, Combes and Martinet (1994).

The inclusion of magnetic fields by Nelson (1988), Battaner et
al. (1992) and others have -in the opinion of Pfenniger, Combes and
Martinet- ``the implicit hope that by complicating the physics new
alternatives can emerge''. However, magnetic fields were introduced in
Physics several centuries ago, while the inward force
due to the magnetic tension in a magnetized disk is a conclusion of
really elementary physics.

\subsubsection{Mechanisms producing magnetic fields in the outermost disk}

Magnetic fields may explain rotation curves if a) there is a
sub-critical strength gradient and b) they have a sufficient order of
magnitude. The next step is to deduce
the existence of these fields theoretically and to identify the
mechanisms that produce them.
   
A first simple model with this objective was presented by Battaner,
Lesch and Florido (1999), in which a mechanism is responsible for a
critical slope, $B_\varphi \sim B_\varphi^* \propto R^{-1}$. A highly
convective disk in the vertical direction maintains a highly turbulent
magnetic diffusivity, establishing a connection and equilibrium between
extragalactic and galactic fields. The origin of galactic fields is
extragalactic and they are amplified and ordered by differential
rotation. The problem of the origin of magnetic fields is then shifted
to the intergalactic medium, a topic that will be addressed in the next
section.

With this model, we depart from the classical approach, basically consisting
of the $\alpha \Omega$ dynamo or similar models. We can allow ourselves
this liberty because the classical dynamo theory (summarized,
for instance, in the review by Wielebinski and Krause, 1993) has been
subject to severe criticism and does not offer a clear scenario. The
standard dynamo approach does not take into account the back reaction
of the turbulence on the amplified magnetic field, which is very
strong at small scales (Kulsrud, 1986; Kulsrud and Anderson,
1992). Another important shortcoming of the standard dynamo theory lies in
the following fact: The $\alpha \Omega$ dynamo exponentially amplifies
a preexisting seed field up to the present, with strengths of the order
of 1-10 $\mu$G. The field is amplified e-times in each rotation. Suppose
that the galaxy has rotated about 20 times since its birth. Then, the
field has been amplified by a factor of about $e^{20} \simeq 5 \times
10^8$. Therefore, the initial strength would have been about
$10^{-15}$G. This is in contradiction with the $\mu$G fields measured in
3C295 (with z = 0.395) (Kronberg, Perry and Zukowski, 1992). Moreover
Perley and Taylor (1991) detected such large fields at
z=0.461. Absorption Line Systems of quasar spectra, usually
interpreted as pregalactic structures, also have $\mu$G fields
(Kronberg and Perry, 1982; Watson and Perry, 1991). Observations of
Lyman-$\alpha$ clouds at $z \sim 2$ also show $\sim 3 \mu G$-fields
(Wolfe, Lanzetta and Oren, 1992), similar to other highly redshifted
disks (Wolfe, 1988; Kronberg et al., 1992). If new-born galaxies were
so highly magnetized, the $\alpha \Omega$ dynamo would have amplified
these initial fields to a present value of about 500 G, in
astonishing disagreement with observations. Even if, before reaching
this value, some saturation mechanism had appeared, the
classical dynamo is incompatible with pregalactic
$\mu$G-strengths. Therefore, the topic is now free for speculation
and the search for alternative scenarios.

This argument not only invalidates the classical dynamo theory but
also many hypotheses about the origin of primordial magnetic fields
that were conceived as providing $\sim 10^{-15}$G at the epoch of galaxy
formation. Galaxies were probably formed out of an already strongly
magnetized medium, with an equivalent-to-present $\sim 1 \mu G$ field,
the same order of magnitude as the present intergalactic medium field.

\paragraph{Turbulent difussion and the galactic magnetic field}

In this Section let us summarize the work by Battaner, Lesch and
Florido (1999). The similarities of the strengths in the
interstellar, the extragalactic and the pregalactic media suggest a
fast and efficient connection between them. In this work, it is proposed
that this connection is the result of a highly turbulent magnetic
diffusion in the vertical direction.

It is an observational fact that convective phenomena are very active in disks. 
Galaxies sometimes exhibit ``boiling'' disks, with
NGC 253 being a good example (Sofue, Wakamatsu and Malin, 1994), where dark
filaments, lanes, arcs and other micro-structures are features
revealing a very complex convective region. This fact is in part
explained by ``fountain'' models (Shapiro and Field, 1976;
Breitschwert et al., 1991; Kahn, 1994; Breitschwerdt and Komossa,
1999). 
Of course, these turbulent
motions constitute a transporting of magnetic fields, as a result of the
condition of frozen-in lines. Because of this
transporting, extra and intergalactic fields merge.

Suppose first that no dynamo is acting on the galactic gas. The
equation of induction will tell us (e.g. Ruzmaikin, Shukurov and
Sokoloff, 1988; Battaner, 1996)
\begin{equation}
  {{\partial B_R} \over {\partial t}} = \beta {\partial \over
  {\partial z}} \left( {{\partial B_R} \over {\partial z}}-
  {{\partial B_z} \over {\partial R}}\right)
\end{equation}
\begin{equation}
  {{\partial B_\varphi} \over {\partial t}} = B_R {{\partial \theta}
  \over {\partial R}} + B_z {{\partial \theta} \over {\partial z}} -
  \theta {{B_R} \over R} + \beta \left( {{\partial^2 B_\varphi} \over
  {\partial z^2}} + {\partial \over {\partial R}} \left( {{\partial
  B_\varphi} \over {\partial R}} + {{B_\varphi} \over R} \right)
  \right)
\end{equation}
\begin{equation}
  {{\partial B_z} \over {\partial t}} = -\beta \left( {\partial \over
  {\partial R}} \left( {{\partial B_R} \over {\partial z}} -
  {{\partial B_z} \over {\partial R}} \right) + {1 \over R} \left(
  {{\partial B_R} \over {\partial z}} - {{\partial B_z} \over
  {\partial R}} \right) \right)
\end{equation}
where $B_R$, $B_\varphi$ and $B_z$ are the magnetic field strength
components and $\beta$ is the coefficient of turbulent magnetic
diffusion. Cylindrical symmetry has been assumed. The usual expression
to calculate $\beta$ is
\begin{equation}
  \beta = {1 \over 3} vl
\end{equation}
where $l$ is a typical length of the larger convective cells, say $l
\approx 1 kpc$, and $v$ is a typical convection velocity corresponding
to the larger scale turbulence, say 20 km s$^{-1}$. Hence $\beta$ is
of the order of $2 \times 10^{27} cm^2 s^{-1} \approx 6 kpc^2
Gyr^{-1}$. In comparison, $\beta$ is taken as being of the order of $10^{26}
cm^2 s^{-1}$ in the inner disk), of $5
\times 10^{27} cm^2 s^{-1}$ in the galactic corona 
 and of $8 \times 10^{29} cm^2 s^{-1}$ in the
intergalactic medium in a cluster (Ruzmaikin, Sokoloff and Shukurov,
1989; Sokoloff and Shukurov, 1990; Ruzmaikin, Shukurov and Sokoloff, 1988).

The characteristic diffusion time is calculated with $l^2/\beta$,
therefore having a typical value of 0.2 Gyr, very little compared with
the lifetime of the galaxy. Extragalactic magnetic fields would have spatial
variations at scales much larger than a galaxy. The field strength
can be assumed to be constant outside the galaxy, as a boundary
condition. This external steady state magnetic field could have
produced an initial penetration of magnetic fields which would have
been subsequently ordered by differential rotation, resulting in a
predominantly toroidal field. Or rather, the disk was born out of already
magnetized material, then was magnetized at birth and maintains a
permanent interchange with the magnetized environment, because of the
high magnetic diffusivity.

However, the magnetic field is assumed to be homogeneous outside and
toroidal inside. A configuration that continuously transforms a
constant into a toroidal field was proposed by Battaner and
Jim\'enez-Vicente (1998), but here we need to adopt convenient
boundary conditions taken at a large enough height.

All three components -$B_x$, $B_y$, $B_z$- are constant in the extragalactic medium. But not
all penetrate and diffuse inwards equally. For instance,
there is no difficulty for $B_z$ to penetrate, because it is not
perturbed by rotation. And if the transport is so effective we
could even assume that $B_z$ is a constant in the whole outer disk
considered, equal to the extragalactic value of $B_z$. We then assume
as a reasonable mathematical assumption that $B_z$=cte everywhere in
the integration region.

It is more difficult for the other components to penetrate (or exit). For
instance, $B_R$ penetrates into the disk at a given time and point (R,
$\varphi$); the rotation would transport the frozen-in magnetic field
lines into the azimuthally opposite position (R, $\varphi + \pi$) in
half a rotation period. The direction of the penetrated field
vector there would be opposite to the vector transported from the
opposite azimuth. The two vectors would meet with the
same modulus and opposite direction and would destroy one another
through the reconnection of field lines. It is therefore
tempting, in a first simplified model, to assume that $B_R =0$, at
the boundaries. We may even adopt $B_R =0$ everywhere inside the
disk. 

With respect to $B_\varphi$, we have a similar situation. $B_\varphi$
when penetrating at (R, $\varphi$) would be frozen-in transported to (R,
$\varphi  +\pi$) in half a rotation and then interact with the field
penetrated there. Reconnection would then act and we could reasonably
adopt $B_\varphi =0$ at the boundaries. But $B_\varphi$ is easily
amplified by rotation and can be generated from $B_z$, which is
non-vanishing; therefore we cannot assume $B_\varphi =0$ everywhere;
rather it is $B_\varphi (R,z)$ that we want to calculate. We also assume
steady-state conditions, $\partial/\partial t =0$. The equations then
become greatly simplified which also permits a simplified
interpretation of what is essential in the process, much more
understandable than a lengthy numerical calculation. In the above
equations, we set $\partial/\partial t =0$, $B_z$=constant, $B_R =0$
and obtain
\begin{equation}
  0 \equiv 0
\end{equation}
\begin{equation}
  0 = B_z {{\partial \theta} \over {\partial z}} + \beta \left(
  {{\partial^2 B_\varphi} \over {\partial z^2}} + {\partial \over
  {\partial R}} \left( {{\partial B_\varphi} \over {\partial R}} +
  {{B_\varphi} \over R} \right) \right)
\end{equation}
\begin{equation}
  0=0
\end{equation}

The first and third are tautologic, telling us that we could have
deduced much of what was assumed (as Battaner, Lesch and Florido did),
but that is unimportant. We now see that our assumptions do not lead
to incoherent results.

The second equation would provide us with $B_\varphi$ if $\theta(z)$ were
known. Let us further assume $\partial \theta/\partial z =0$, which is
not unrealistic given the relatively low thickness of the disk. In order
to find a fast solution (which is not necessary, but just didactic)
let us assume that $\partial^2 B_\varphi/\partial z^2$ is
negligible (it cannot be zero, as $B_\varphi$ must be zero at the
 boundary). In fact,
some galaxies have a radio halo (e.g. NGC 253, Beck et al. 1994; NGC
891, Dahlem, Lisenfeld and Golla, 1995 in
other spirals). The decrease of magnetic field strength with $z$ is
observed to be very slow, even in galaxies with no radio halo
(Ruzmaikin et al., 1988; Wielebinski, 1993) and also in the Milky Way
(Han and Qiao, 1994). Then for small $\mid z \mid$ we simply obtain
\begin{equation}
  {\partial \over {\partial R}} \left( {{\partial B_\varphi} \over
  {\partial R}} + {B_\varphi \over R} \right) =0
\end{equation}
therefore
\begin{equation}
  B_\varphi \propto {1 \over R}
\end{equation}
which is precisely the critical profile. Once we see how the critical
profile is supported with this mechanism, it is expected that other
more realistic calculations would be able to provide sub-critical
profiles, capable therefore of producing inward magnetic forces.

The symmetries of the magnetic field predicted in this simple model
agree with those obtained with Faraday rotation by Han et al. (1997) in
our own galaxy. This model does not need a dynamo but provides a
large-scale structure with much in common with the so called AO
mode. We also predict an antisymmetry of the azimuthal field in both
hemispheres for $\mid l \mid < 90^o$. This AO dynamo mode has also been
observed in other galaxies, but in view of the symmetry
similarities with our predictions, these galaxies could be interpreted
in terms of the mechanisms sought by Battaner, Lesch and Florido
(1999).

\paragraph{A dynamo-like mechanism}

Though we favour the previous model based on the action of 
turbulent magnetic diffusion, let us show that there is also a special
kind of dynamo or amplification mechanism that, in contrast 
with the standard $\alpha \Omega$,
quickly reaches steady state conditions, and that also produces the
critical profile of the magnetic field strength.

Let us continue considering azimuthal symmetry. In addition to $\beta$
terms, let us consider $\alpha$ terms, i.e. the effect based on a
mean value of the quantity $<\vec{v} \cdot \nabla \times \vec{v} >$,
non vanishing in turbulence velocity fields, because the Coriolis
force  introduces order into the chaotic turbulence. With this
symmetry the induction equation becomes
\begin{equation}
  {{\partial B_R} \over {\partial t}} = -\alpha {{\partial B_\varphi}
  \over {\partial z}} + \beta {\partial \over {\partial z}} \left(
  {{\partial B_R} \over {\partial z}} - {{\partial B_z} \over
  {\partial R}} \right)
\end{equation}
\begin{equation}
  {{\partial B_\varphi} \over {\partial t}} = B_r {{\partial \theta}
  \over {\partial R}} + B_z {{\partial \theta} \over {\partial z}} -
  \theta {B_R \over R} + \alpha \left( {{\partial B_r} \over {\partial
  z}} - {{\partial B_z} \over {\partial R}} \right) + \beta \left(
  {{\partial^2 B_\varphi} \over {\partial z^2}} + {\partial \over
  {\partial R}} \left( {{\partial B_\varphi} \over {\partial R}} +
  {B_R \over R}\right) \right)
\end{equation}
\begin{equation}
  {{\partial B_z} \over {\partial t}} = \alpha \left( {{\partial
  B_\varphi} \over {\partial R}} + {B_\varphi \over R} \right) - \beta
  \left( \left( {\partial \over {\partial R}} \left( {{\partial B_R}
  \over {\partial z}} - {{\partial B_z} \over {\partial R}} \right)
  \right) + {1 \over R} \left( {{\partial B_R} \over {\partial z}} -
  {{\partial B_R} \over {\partial R}} \right) \right)
\end{equation}

If, as in the previous Section, we set $B_R =0$, $B_z$ constant, and
assume steady state conditions, this equation system reduces to
\begin{equation}
  0= \alpha {{\partial B_\varphi} \over {\partial z}}
\end{equation}
\begin{equation}
  0 = B_z {{\partial \theta} \over {\partial z}} + \beta \left(
  {{\partial^2 B_\varphi} \over {\partial z^2}} + {\partial \over
  {\partial R}} \left( {{\partial B_\varphi} \over {\partial R}} +
  {B_R \over R} \right) \right)
\end{equation}
\begin{equation}
  0 = \alpha \left( {{\partial B_\varphi} \over {\partial R}} +
  {B_\varphi \over R} \right) 
\end{equation}

The first equation confirms our previous assumption. The third
equation is very illustrative, as, even if $\beta =0$, in the absence of
significant turbulent magnetic diffusion, we again obtain the critical
profile $B_\varphi = B_\varphi^* \propto {1 \over R}$. As the first
one informs us that $\partial B_\varphi/\partial z$, then the second
just tells us that this solution is compatible with either $B_z
=0$, or no vertical differential rotation ${{\partial \theta} \over
{\partial z}} =0$. It is important to note that we have
obtained the critical profile both without $\beta$ and including $\beta$.
In contrast with other dynamos, this mechanism can reach a steady
state.

\subsubsection{The origin of cosmic magnetic fields}

We now consider the origin and evolution of cosmological
magnetic fields. Our aims are: a) To classify the long
list of theories which have been proposed; b) To  provide
astrophysical arguments constraining these theories, reducing their
number, if possible; c) To propose or adopt a coherent overall history
of cosmological magnetic fields; d) To determine whether this general scenario
provides a reasonable basis for the magnetic hypothesis of rotation
curves.

An important argument severely reducing the long list of candidate
theories has been given by Lesch and Birk (1998), in
which they prove that small scale magnetic fields cannot
survive the highly resistive pre-recombination era, characterized by
frequent electron-photon interactions. Therefore, cosmic magnetic
fields, were either generated after Recombination or had
coherence cells larger than the horizon before Recombination. In this
latter case, diffusion of magnetic fields cannot proceed over distances
larger than the horizon, and these cells could become a subhorizon in
the post-Recombination epoch, which is more favourable for the existence of
magnetic fields.

Let us then place the limit between the large and small scales at
about 10 Mpc, because this is the minimum length that was superhorizon
before (about) Recombination, or more precisely before ``Equality'' (the
transition epoch between the radiation and matter
domination). Therefore, magnetic field coherence cells longer than
(today) $\sim$ 10 Mpc were not set in motion by diffusion in the resistive
era before (about) Equality. Battaner, Florido and Jim\'enez-Vicente
(1997) and Florido and Battaner (1997) observed a clear transition in the
evolution of magnetic fields for scales
\begin{equation}
  \lambda > {1 \over {mn_0}} \sqrt{{3 \sigma T_0^4} \over {8 \pi c G}} 
\end{equation}
where $mn_0$ is the present cold matter density (baryonic or not; dark
or not), $\sigma$ is the Stephan-Boltzmann constant and $T_0$ the CMB
temperature. This length is equivalent to just a few Mpc.

This transition is very important for our purposes as large scale
fields are not influenced by diffusion or any other micro-physical
effect during the radiation dominated era.

Reviews of cosmological magnetic fields have been written by Rees
(1987), Coles (1992), Enquist (1978), Olesen (1997), Vall\'ee (1997) and
others. Closely related to this topic are the works by Zweibel and
Heiles (1997) and by Lesch and Chiba (1997).

In the absence of loss and production amplification mechanisms,
frozen-in magnetic field lines will evolve due to the flux of Hubble
alone as
\begin{equation}
  \vec{B}_0 = \vec{B} a^2
\end{equation}
where $a$ is the cosmological scale factor, taking its present value as
unity, $\vec{B}$ the magnetic field strength when the Universe was
$a$ times smaller than today and $\vec{B}_0$ the present
strength. This equation is in general not true, because the frozen-in
condition is not guaranteed at all epochs, and because production,
loss and amplification processes other than those due to the Hubble
flow could really have taken place. However, we will adopt this
equation, as a re-definition of $\vec{B}_0$. The equivalent-to-present
magnetic field strength, $\vec{B}_o(t)$, is the strength that would be
observed today corresponding to the real $\vec{B}_0(t)$ when 
the Universe was $a(t)$
times smaller, as a result of frozen-in lines in the Hubble flow, in the
absence of an amplifying or destroying mechanism other than the expansion
itself, even
if it does not coincide with the present one at all. As the effect
of expansion is always important, this definition of the
equivalent-to-present magnetic field strength permits a useful
comparison of strengths during different epochs.

On the other hand, this expression can be more general (Battaner,
Florido and Jim\'enez-Vicente, 1997) and holds under the condition of
$\vec{B}$ constituting a small perturbation of the Robertson-Walker
metrics. A pure U(1) gauge theory with the standard Lagrangian is
conformally invariant (unlike a minimally coupled field), from which
it follows that $\vec{B}$ always decreases in the expansion following
this equation, even in the absence of charge carriers. The Inflation epoch
may be an exception for the reasons given below.

Following Battaner and Lesch (2000), the different theories about the
origin of cosmic magnetic fields can be classified into four main
groups, characterized by the epoch of formation:

a) during Inflation

b) in cosmological phase transitions

c) during the Radiation Dominated era

d) after Recombination

\paragraph{Inflation magnetic fields}

The mean magnetic field of the Universe is zero, if we adopt the
Cosmological Isotropy Principle, and therefore it would be random at the
larger scales. At smaller scales, there can be coherence cells with
different sizes, characterized by a mean field. 
Coherence cells are usually associated with objects or density
inhomogeneities . It could therefore be expected that anisotropies in
the CMB, which are associated with density inhomogeneities, might
correspond to large magnetic coherence cells. Some anisotropies are
larger than $2^o$, i.e. they are larger than the horizon at
Recombination. These larger inhomogeneities were generated before or
during Inflation. In the same way, Inflation provides the most natural
explanation of field inhomogeneities, as it permits a causal
connection between two points with a separation that was until fairly
recently (until Equality, approximately) smaller than the horizon.

Turner and Widrow (1988) were pioneers in analyzing this idea. A cloud
with a present diameter $\lambda$, had in the past a size $a
\lambda$. However, the horizon evolved independently of $a$ during
the first phase of Inflation, and was then proportional to $a^{3/2}$
during the so called reheating phase, to $a^2$ during the
radiation dominated era, and to $a^{3/2}$
after Equality (when the radiative and the matter energy
densities became equal). Therefore, a cloud could be
subhorizon before or during Inflation, become superhorizon thereafter
and again be subhorizon at present. This could explain how 
 points in the cloud at distances further than the horizon
before the present CMB anisotropy are causally connected. In the same way,
magnetic coherence cells, causally connected at Inflation, could have
become superhorizon early and reentered the horizon recently.

Coherence could even be due to electromagnetic waves, as the
oscillating electric and magnetic fields, when the wavelength became
subhorizon, would have appeared as static fields. Only recently,
after around Equality, would conductivity have destroyed large scale
electric fields and controlled large scale magnetic fields.

The problem inherent in this theory is that $a(t)$ was exponential
during Inflation, increasing by a factor of $10^{21}$. This would imply a
decrease in $B$ by a factor of $10^{42}$ (perhaps much more), if $Ba^2$
were a constant, i.e. if the U(1) gauge theory were conformally
invariant. This dilution of magnetic fields must be avoided by some
mechanism. Turner and Widrow considered that the conformal invariance
of electromagnetism is broken through gravitational coupling of the
photon. In this case, the electron would have a mass of only about
$10^{-33}$eV, and therefore be undetectable. Turner and Widrow predicted $B_0
\sim 5 \times 10^{-10}$G at scales of about 1 Mpc, which is really
interesting.

Other authors avoided the complete dilution of primordial fields with
other mechanisms. Ratra (1992) considered the coupling of the scalar
field responsible for inflation (the inflaton) and the Maxwell field,
obtaining $B_0 \sim 10^{-9}$G at scales of 5 Mpc. Garretson, Field and
Carroll (1992) invoked a pseudo-Goldstone-boson coupled to
electromagnetism ($B_0 < 10^{-21}$G at $\sim$1 Mpc). Dolgov (1993)
proposed the breaking of conformal invariance through the ``phase
anomaly''. Dolgov and Silk (1993) considered a spontaneous break of
the gauge symmetry of electromagnetism that produced electrical
currents with non-vanishing curl.

Davis and Dimopoulos (1995) considered a magnetogenesis at the GUT
phase transition, but their theory is included here because this
transition could have taken place during Inflation ($10^{-11}$G at
galactic scales).

Rather interestingly, when considering the Planck era, the Superstring
theory leads to an inflationary pre-Big-Bang scenario which supports
some of the theories explained before (Veneziano, 1991; Gasperini and
Veneziano, 1993a,b; Gasperini, Giovannini and Veneziano 1995a, b;
Lemoine and Lemoine, 1995; etc.) rendering derivations from what were
assumptions. In this scenario, the electromagnetic field is deduced to
be coupled not only to the metric but also to the dilaton
background. COBE anisotropies are  the result of electromagnetic
vacuum fluctuations, involving scales of the order of comoving 100 Mpc,
today. For values of some arbitrary parameters, these models
provide large enough values of intergalactic fields, even in the
absence of galactic dynamos. They are in fact able to explain a
possible equipartition of energy between the CMB radiation and
magnetic fields. This pre-Big-Bang scenario is really promising as an
explanation of primordial magnetic fields and their connection with CMB.

\paragraph{Phase transition magnetic fields}

During the history of the Universe several cosmological phase
transitions have taken place. The best studied are: QCD (250
MeV), electroweak ($10^2$ GeV) and GUT ($10^{16}$GeV). For a
comparison, consider that the present epoch is characterized by $3
\times 10^{-4}$ eV, the matter-radiation decoupling by 1 eV and the
nucleosynthesis epoch by 1 MeV. Typical values of the horizon scale
correspond to a present-day 0.2 pc for the QCD transition, $3 \times 10^{14}$
cm for the electroweak transition and 1 m for GUT.

Hogan (1983) proposed first order phase transitions as a
potential magnetogenetic mechanism. Phase transitions would not have taken
place simultaneously throughout the Universe, but in causal
bubbles. At the rim of these bubbles high gradients of temperature or
any other parameter characterizing the phase transition (for
instance, the Higgs vacuum expectation value) would be
established. These high gradients could produce a thermoelectric
mechanism akin to the Biermann battery. Magnetic
reconnection would stitch the magnetic field lines of different bubbles
and the magnetic field lines would execute a Brownian walk.

There exist a large variety of works and ideas. The electroweak
transition has been considered as a source of magnetic fields by
Vachaspati (1989), Enqvist and Olesen (1993), Davidson
(1996), Grasso and Rubinstein (1995) and others. The QCD phase
transition as a magnetogenesis mechanism has been studied by Quashnock,
Loeb and Spergel (1989), Cheng and Olinto (1994) and others. Magnetic
fields generated at the GUT phase transition have been analyzed by
Davies and Dimopoulos (1995), Brandenberger
et al. (1992). Other interesting theories related to Cosmological phase
transitions have been proposed by Vachaspati and Vilenkin (1991),
Kibble and Vilenkin (1995), Baym, Boedeker and McLerran (1996) and
others.

From the equations for magnetic fields  produced at a given phase
transition, and the spectra at different length scales given by
Vachaspati (1989) it is deduced (Battaner and Lesch, 1999) that $B$,
the equivalent-to-present magnetic strength, only depends on $T_0^2$
(where $T_0$ is the present CMB temperature) and that the spectrum
$B_0(\lambda)$ only depends on $T_0/\lambda$ (or on $T_0^{3/2}
\lambda^{-1/2}$ with a correction proposed by Enquist and Olesen (1993)). In
any case the present spectrum of magnetic strength for different
scales $B_0(\lambda)$ is independent of the specific phase transition.
There is one compensation: earlier phase transitions produced larger
magnetic fields but they have had longer to be weakened by
expansion. We will comment later that the values of $B_0$ can
differ greatly from present magnetic field strengths.

\paragraph{Magnetic fields generated by turbulence during the
radiation dominated era}

There are two classical papers (Matsuda, Sato and Takeda, 1971;
Harrison, 1973) in which magnetic fields were considered to be
generated by turbulence in the radiation dominated universe. There is
also a close relation between vorticity and magnetic fields, $\vec{B} =
-(mc/e) \vec{\omega}$ (where $\vec{\omega}$ is the vorticity) which 
was  deduced by Batchelor (1950)
and considered again by Kulsrud et al. (1997) as an extension of a
previous study by Biermann. The deduction is based on the surprising
similarity between the vorticity and the induction equations. 
With similar initial conditions, both magnitudes, vorticity and
magnetic fields, should evolve similarly. Viscosity has a
different behaviour, because there is a saturation of vorticity, but
the above equation could be used in other astrophysical problems. In the
epoch of radiation domination, however, we will present arguments
against this, so that this equation probably does not hold.

\paragraph{Magnetic Fields generated after Recombination}

When dealing with cosmic magnetic fields in present day astrophysical
problems, it is customary to fully assume hypotheses that are
accepted in magnetohydrodynamics. It is in fact assumed that magnetic
fields can be modified, amplified and even be subject to diffusion or
reconnection, but that they cannot be created. However, the three
above mentioned mechanisms are able to create magnetic fields out of
nothing. There are also more classical mechanisms of net creation of
magnetic fields, with the well known Biermann's battery providing a clear
example (Biermann, 1950; Biermann and Schleuter, 1951). Another
battery mechanism was proposed by Mishustin and Ruzmaikin (1973), in
which the CMB radiation differentially interacts via Compton
scattering with protons and electrons, thus establishing a weak electric
field and weak electrical currents that in turn are able to originate
weak magnetic fields.

In a protogalactic cloud, the conditions are similar to those needed
for the classical Biermann's battery, mainly a combination of
gravitational field with differential rotation. Lesch and Chiba (1995)
showed that
magnetic field strengths in the range $10^{-13}-10^{-16}$G can be
produced at early stages in the protogalactic cloud. This seed may be
exponentially amplified by non-axisymmetric instabilities during the
disk formation epoch, so that magnetic fields of the order of 1$\mu$G
can be reached in less than 1-2 Gyr, as observed in recently born
galactic systems (Chiba and Lesch, 1994).
Kulsrud et al. (1997) and Howard and Kulsrud (1996) also demonstrated
that protogalactic magnetic fields can be created without any seed
after Recombination.

Kronberg, Lesch and Hopp (1998) have proposed that superwinds of dwarf
galaxies of the M82-type, which eject great quantities of matter and
magnetic fields, have effectively seeded the intergalactic medium with
magnetic fields, in a first generation of ($z \sim 10$) galaxies. The
seeding would have been accomplished by $z \sim 6$. Under this
hypothesis, pre-Recombination fields would not be required, at least
to understand galactic fields.

\paragraph{Comments on the different theories}

Let us first discuss some problems inherent to theories based on
phase transitions. Phase transition generated magnetic fields have
small scales. For instance, the most recent one, the QCD transition,
at $\sim$200 MeV, predicts correlation lengths of $10^{-11}$cm, which
grow to 10 cm at present. The horizon at the QCD phase transition
was $10^{-6}$cm, equivalent to 0.2 pc at present. Other phase
transitions also predict small scales. The electroweak phase
transition took place when the horizon was at only a few centimeters,
corresponding to about 1 AU at present. For early phase transitions
the expected scale is even worse.

These fields undoubtedly created in phase transitions
probably have no connection with present magnetic fields, because
there are two
mechanisms that can destroy this kind of small scale fields.

First, the subsequent radiation-dominated universe was highly
resistive, because of the frequent encounters between electrons and
photons. This has been shown by Lesch and Birk (1998). The low
conductivity implies magnetic diffusion. These authors gave a
diffusion time equivalent to
\begin{equation}
  \tau_{diff} = 10^{44} z^{-6} \lambda^2
\end{equation}
where $\tau_{diff}$ is measured in seconds and $\lambda$, the
coherence size, in cm. This time very much depends on redshift, with
the initial times being the most destructive, probably just after
Annihilation, because of the increase in the photon number density. If
we set $\tau_{diff} = \tau_{rec}$ (Recombination epoch) and $z =
z_{ann}$ (Annihilation redshift) we will obtain the minimum scale able
to survive from Annihilation to Recombination
\begin{equation}
  \lambda = 5 \times 10^{-16} z^3_{ann}
\end{equation}

This $\lambda$ will grow to its present comoving size
\begin{equation}
  \lambda_0 = 5 \times 10^{-16}z_{ann}^4
\end{equation}
For $z_{ann} \sim 2 \times 10^9$, we conclude that the minimum scale
able to survive was about 3 kpc, much higher than the scale
predicted by the magnetogenesis mechanisms based on cosmological phase
transitions.

It is understandable that magnetic fields, and density and radiation
inhomogeneities are associated during the radiation dominated
Universe. Therefore if matter or radiation overdensity regions, at a
certain scale, are
destroyed or damped, the same end should be expected for magnetic
fields of this
scale. It is known (Silk, 1968; Weinberg, 1968) that masses smaller than
the Silk mass are damped in the Acoustic epoch, before
Recombination. Jedamzik, Katalinic and Olinto (1996) also concluded
that MHD modes are completely damped by photon diffusion up to the
Silk mass and convert magnetic energy into heat. Damping would also be
very important during the neutrino decoupling era. Therefore, small
scale fields could have been eliminated before the radiation era.

Therefore, small scale fields, even if they were created in phase
transitions, cannot survive the radiation dominated era. They have two
enemies: magnetic diffusion and, probably, photon diffusion.

However, we must mention the important work by Brandenburg,
Enqvist and Olesen (1996) in which they propose that inverse cascade
effects in relativistic turbulence in the expanding medium produce
large scales. Then, inverse cascade would save the small scale phase
transition magnetic fields. But the existence
of a turbulence during this epoch is controversial (Rees, 1987), or at
least it would have had a very peculiar behaviour. In fact $\delta \rho/\rho$
cannot evolve in a random way. If $\delta \rho/\rho$ is small but
positive, it will always increase and remain positive if the cloud
mass is higher than the Jeans mass, because of gravitational
collapse. The Jeans mass is very low, particularly just after Annihilation,
of the order of 1 $M_\odot$, and therefore gravitational collapses, rather
than true turbulence, dominated the evolution of initial
inhomogeneities. Perturbations of the metric tensor are essential in
this era. Even if the inhomogeneities do not grow very fast (as $a^2$)
they cannot be neglected. On the other hand, turbulent motions, if they
really existed, could not affect scales larger than the horizon, and
therefore scales larger than 1 Mpc cannot be produced. In fact, Brandenburg,
Enqvist and Olesen predict much lower amplification factors,
given the initial very small scales to be enlarged.

These arguments seem to exclude phase transitions as mechanisms
providing magnetic fields connected to present fields. Moreover, the
model proposed by Harrison (1973) even if historically interesting, and
emphasizing the effect of the horizon on the turbulence regime, did
not include General Relativity effects, which are not ignorable at all. 

Therefore, despite the possibilities of an extended analysis
of inverse cascade effects, we favour the hypothesis that
large scale fields (larger than the horizon in the radiation dominated
era) were produced at Inflation, as deduced in the scenario of string
cosmology (e.g. Gasperini and Veneziano 1993a,b). Small scale fields,
such as those in galaxies, have two
possible origins: a) The large scale inflation magnetic fields 
were amplified after Recombination as a result of contractions in the
process of forming superclusters, clusters and galaxies after
Recombination; b) They were generated without seeding by battery
mechanisms in the process of galaxy formation.

Irrespective of the exact time and mechanism of magnetogenesis, the
effect of preexisting magnetic fields on the birth and structural
properties of galaxies has been considered in the
literature. Piddington (1969) tried to explain the present
morphology of different types of galaxies from the angle between the
angular momentum and the magnetic field strength. Wasserman (1978)
considered that the formation of galaxies was decided by preexisting
magnetic fields and that these were even able to provide the galactic
angular momentum. Kim, Olinto and Rosner (1996) extended this work to
the non-linear regime.

It is difficult to decide between the two possibilities for the origin
of small scale magnetic fields, and therefore to decide what is the
origin of galactic magnetic fields. We prefer to argue in favour of the
inflationary origin, for the following two reasons, one
theoretical and the other observational:

a) We will see that magnetic fields of the order of $B_0 \sim
10^{-9}-10^{-8}$G may be present in the $\sim$ 100 Mpc long filaments
characterizing the large scale structure of the Universe. 
These structures probably consisted of filamentary concentrations of photons,
baryons and possibly other kinds of dark matter, but the energy
density was smooth and continuous within a filament. After
Recombination, baryons and dark-matter particles begun to form clumpy
structures of a different order (superclusters, clusters, galaxies), and
the contractions involved produced an amplification, until 
the present value of about $10^{-6}$G was reached. The simulations carried out by
Dolag, Bartelmann and Lesch (1999) indicate that initial magnetic
field strengths of $10^{-9}$ G at z=15 provide an amplification of
three orders of magnitude in cluster cores. Therefore if $B_0$ was
$10^{-9}-10^{-8}$G in filaments at Recombination, the subsequent
contractions that undoubtedly took place can account 
for this amplification very easily, only involving two or three orders of
magnitude.

b) If magnetic fields are generated via battery processes similar to
Biermann's, in the galactic formation itself, then magnetic
fields would only be present inside galaxies or in their close
vicinity. However, as mentioned above, strong magnetic fields have been
observed in the intracluster and in the intercluster media.

\subsubsection{Large scale structure and magnetic fields}

Coles (1992) suggested that the failure of CDM models to explain large
scale structures could be satisfactorily surmounted if magnetic fields
were taken into account. Large scale structures are characterized by a
noticeable regularity and periodicity (Broadhurst et al., 1990; Einasto
et al. 1994; Tucker et al. 1997; Landy et al. 1996; Quashnock et al. 1996;
Atrio-Barandela et al. 1997; Tully et al. 1992; Einasto et al. 1997
a,b,c; Retzlaff, 1998; Tadros et al. 1998, Toomet et al., 1999, and 
others) suggesting a network of filaments.

Battaner, Florido and Jimenez-Vicente (1997), Florido and Battaner (1997),
Battaner, Florido and Garcia-Ruiz (1997), Battaner and Florido (1998) and
Battaner (1998) have theoretically analyzed the influence of magnetic
fields on the large scale structure along the radiation dominated universe
and their conclusions may be summarized as follows:

a) Preexisting magnetic fields are able to produce anisotropic density
inhomogeneities in the photon fluid and local metric perturbations. In
particular, they are able to produce filamentary structures in the
distribution of the energy density.

b) Particularly interesting are those filaments larger than about
$\sim$10 Mpc, because they have no problems with magnetic diffusion (as
mentioned above), because their evolution is more predictable and
because they can be observed today. In fact these radiative and
gravitational potential filaments were the sites where baryons, or any
other dark matter component, collapsed at Recombination, forming the
illuminated supercluster filaments that are observed today as elements of the
large scale structure. Non linear effects have very much distorted the
pre-Recombination structures, as well as the larger ones, though to a much lesser
extent, as $\delta \rho/\rho$ remains low. Therefore, these
pre-Recombination radiative filaments should be identifiable today.

c) The orders of magnitude of these magnetic fields are equivalent to
present $B_0 \sim 10^{-8}-10^{-9}$G. If they were much lower, they
would have no influence on the larger scale structure. If they were much higher, the
formation of the galaxy would have taken place much earlier.

d) The filament network, if magnetic in origin, must be subject to
some magnetic restrictions. The simplest lattice matching these
restrictions is an ``egg-carton'' network, formed by octahedra
joining at their vertexes. This ``egg-carton'' universe would have
larger amounts of matter along the edges of the octahedra, which would
be the sites of the superclusters. Outside the filaments there would be
large voids, devoid not only of baryons but also of magnetic
fields (Fig. 21). Magnetic field lines would be concentrated in
the filaments, with their directions being coincident with those of the
filaments.

These theoretical speculations are compatible with present
observations of the large scale structure as delineated by the
distribution of superclusters. It is easy to identify at least four of
these giant octahedra in real data, which comprise observational support
for the egg-carton universe. Two of them, those which are closest and
therefore most unambiguously identified, are reproduced in Fig.
22. Nearly all the important superclusters in the catalogue by Einasto et
al (1997), as well as nearly all the important voids in the catalogue by
Einasto et al. (1994) can be located within the octahedron
structure. This web is slightly distorted by the presence of the very
massive Piscis-Cetus supercluster in one of the filaments.

The magnetic origin of structures at very large scales alleviates the
old problem encountered by CDM models which predict too little
structure at large scales (see, for instance, the reviews by Bertschinger,
1998 and Ostriker, 1993).

A fractal nature could be compatible with the octahedron web, in
agreement with the identification of fractals by Lindner et al. (1996)
from the observational point of view. There could be sub-octahedra
within octahedra, at least in a limited range of length scales. The
simplest possibility is reproduced in Fig. 23 in which 7 small
octahedra contacting at their vertexes have their egg-carton structure
within a large octahedron, the ratio of large/small
octahedron size being equal to 3. The fractal dimension becomes quantified,
with 1.77 and 2 being the most plausible values. The scale of the fractal
structure would range from 150 Mpc, i.e. slightly lower than the
deepest surveys, down to about 10 Mpc (in agreement with Lindner et
al), as shorter scale magnetic fields would have been destroyed
by the resistive radiation dominated universe. Whether the fractal
egg-carton structure continues indefinitely for larger scales as
suggested by Sylos Labini et al. (1998) and others, remains an open question, but
Battaner (1998) proposed this structure under the adoption of the
Homogeneity Cosmological Principle at large enough scales.

The absence of a relation between Faraday rotation and redshift of
quasars indicates that a widespread cosmological aligned magnetic field
must be $B_0 < 10^{-11}$G (Lesch and Chiba, 1997; Kronberg, 1994; Rees
and Reinhardt, 1972; Kronberg and Simard-Normandin, 1976; Vall\'ee,
1983, 1990). However, the distribution of large scale magnetic fields
is probably very far from homogeneous. Not only $<\vec{B}>$=0, but $<B^2>$, even if
not vanishing, is far from homogeneous. Instead, we are interested in
the peak values to be found in the matter filaments, in which case 
this limit should be increased by a large
factor, even if it is very low in voids, i.e. in the largest
fraction of the volume of the Universe. 

There are other upper limits that should be increased by this
factor too, if we are interested in the field strength within the
filaments: for instance, $B_0 < 10^{-7}$G, taking into account the
$^4He$ abundance (Greenstein, 1969; Zeldovich and Novikov, 1975;
Barrow, 1976; Cheng, Schramm and Truran, 1994; Matese and O'Connell,
1970; Grasso and Rubinstein, 1995, 1996 and others), $B_0 < 4 \times
10^{-9}$G, taking into account the neutrino spin flip (though very
much depending on the mass of all neutrinos) (Shapiro and Wasserman,
1981; Enqvist et al., 1993), and $B_0 < 4 \times 10^{-9}$G, based on
the CMB isotropy (Lesch and Chiba, 1997; Barrow, 1976; Barrow,
Ferreira and Silk, 1997).

Observations of the distribution and orientation of warps of galactic
disks, under the interpretation that these warps are produced by
intergalactic magnetic fields (Battaner et al. 1991; Battaner, Florido
and Sanchez-Saavedra, 1990; Zurita and Battaner, 1997) show coherence
regions of about 25 Mpc, though Vall\'ee (1991) did not confirm this
coherent orientation. In any case, the volume of observed galaxies is too
small compared with the scales we are now considering.

The improved sensitivity of experiments intended to measure the CMB would
permit us to gather direct information about large scale magnetic fields
(Magueijo, 1994; Kosowsky and Loeb, 1997; Adams et al., 1996, and
others).Ultrahigh energy cosmic rays can also observe valuable
information (e.g. Lee, Olinto and Siegl, 1995; Stanev et al., 1995;
Stanev, 1997 and others), as well as the delay in
the arrival of the energetic TeV-$\gamma$-rays with respect to the
low-energy-$\gamma$-rays (Plaga, 1995).

\subsubsection{A tentative history of cosmological magnetic
fields}

Despite the large number of theories about the birth, evolution and
present distribution of magnetic fields, some general picture seems to
emerge and could be summarized as follows.

Magnetic fields were created at Inflation, as predicted and
explained in the superstring theory, when the horizon was nearly
independent of the cosmological scale factor. Small scale fields were
washed out during the resistive radiation dominated universe, but
large scale fields, larger than the horizon during the large time
interval between Inflation and Recombination, escaped from magnetic
diffusion and reentered as subhorizon scale fields.

Along the radiation dominated universe, magnetic flux tubes produced
metric perturbations that generated filamentary concentrations of
photons and other matter (including dark) components. Small scale
radiative filaments, if they were actually formed, were dissipated by
photon diffusion mechanisms for masses lower than the Silk
mass. Similarly, small scale fields originated by phase transition
were dissipated by magnetic diffusion just after Annihilation (and
probably also by photon diffusion in the so called Acoustic era, just
before Recombination). Large scale radiative baryonic filaments,
i.e. larger than the horizon along the Radiative era, survived and reached
the Recombination epoch. By
the decoupling of photons, dark matter and baryons remained concentrated in the
filaments. Matter filaments inheriting the properties of primordial
magnetic structures formed a quasi-crystal network mainly consisting
of octahedra contacting at their vertexes, reminiscent of an egg-carton
topology. Magnetic fields were concentrated into the filaments and conserved
their direction. Parameters defining these filaments would be (in
equivalent-to-present units): length: $\sim$100 Mpc; width $\sim$10
Mpc; strength $10^{-9}-10^{-8}$G, at the Recombination epoch, but also
existing fractal substructures.

After Recombination, non-linear contractions leading to superclusters,
clusters and galaxies corrupted and deformed the initial sharper
filaments, becoming clumpy but conserving the large scale
alignment. These collapses amplified the magnetic field strength from
$\sim 3 \times 10^{-9}$G to $\sim 10^{-6}$G, and galaxies therefore
formed out of a microgauss magnetized medium. From the early stages,
magnetic fields played an important role in the dynamics of galaxies,
mainly in the outermost disk, where they became toroidally ordered,
initiated a fast rotation, introduced instabilities into the disk and
produced an escape of gas; they were also in part ejected together with the
gas.

\section{Common halos}

Most galaxies are in more or less large clusters. In the analytic, hierarchical
CDM scenario, halos are the result of the merging of smaller, less
massive, denser, previously formed halos. Once the new large halo is
formed, violent relaxation erases any internal substructure, and therefore
halos within halos should not be expected from this type of model. As an exception,
visible galaxies may survive the merging process, and therefore we might
expect to find several visible galaxies in a halo. High resolution
N-body simulations have, however, been able to resolve some
sub-structures, or subhalos, within dark matter halos (Colin et al.
1999; Benson et al. 1999 and others) even if tidal disruption, spatial exclusion
of subhalos, dynamical friction and other effects complicate the
global picture. In view of these difficulties and given the number of
free parameters inherent to these calculations, let us keep the
classical scenario in which: a) no subhalos exist within a halo; b) several
visible galaxies may reside in the same halo; c) a halo can have no
visible galaxy; d) no visible galaxy can exist outside a halo. This
picture is fully compatible with the essentials of hierarchical CDM
models, even if the above mentioned particular models keep track of 
subhalos. The purpose of
this argument is to comment on the possible picture in which a large
percentage of spiral galaxies are embedded in common halos, instead
of each having their own.

Common halos could be present in clusters and associations at all
levels, from binary systems to rich clusters. The hypothesis of common
halos is not new (see the review by Ashman 1992, for
instance). Let us briefly consider the different systems:

a) {\bf Dwarf irregular satellites around a bright galaxy}. Following
White \& Rees (1978), when the halos of the first small galaxies are
disrupted to form bigger units, the residual gas may again be able to
cool and collapse to form a larger central galaxy. The model naturally
predicts the existence of small satellites around big
galaxies. Therefore, this and some subsequent models implicitly assume
that the satellites have no halo of their own, but are instead in the halo of the
bright galaxy, even if  observations seem to indicate that these satellites
are DM rich (e.g. Ashman 1992).

b) {\bf Binary galaxies}. In the well-known paper by Kahn \&
Woltjer (1959), it was considered, as an alternative interpretation, that
the unseen mass was forming a common envelope. van Moorsel (1987), observing
HI in binaries, suggested that the data were consistent
with a common dark matter envelope surrounding the pair
system. Charlton \& Salpeter (1991) concluded that extremely extended
halos, with radii of around 1 Mpc, were present in their sample of
binaries. This could also be the case of the M31-Milky Way pair; both
could lie in a common halo that has arisen from the mergers of the early
smaller halos of the two galaxies.

c) {\bf The Local Group}. If the 35 galaxies, or more, members of the
Local Group conserved their own halos, there could be a problem with
available volume. To calculate the filling factor, i.e. the volume of
the halos of the 35 galaxies divided by the total volume of the Local
Group, we face the problem that we do not actually
know the individual volumes. But for an exploratory calculation, we may
assume that all halos have the same volume, irrespective of type and
luminosity. To justify this assumption, let us consider that $R_{200}$
is the halo size. From its definition (the radius enclosing a sphere
with mean density 200 times the critical density) it is easily deduced that
$R_{200} \propto M_{200}^{1/3}$ for all galaxies. A relation should
exist between $M_{200}$ (the mass of a sphere with radius $R_{200}$,
which can be taken as the mass of the halo) and the luminosity $L$, an
observational quantity. Salucci \& Persic (1997) give $M_{200}
\propto L^{0.5}$, in which case $R_{200} \propto L^{0.17}$,
i.e. $R_{200}$ is ``nearly'' independent of the luminosity. White et
al. (1983) and Ashman (1992) propose $M_{200}/L \propto L^{-3/4}$, in
which case $R_{200} \propto L^{0.08}$; in this case, the exponent
(0.08) is so small that the adoption of constant $R_{200}$ for all
galaxies is a good first approximation. Let us take $R_{200} \sim
250$kpc as a typical value. Let us adopt the zero-velocity surface
radius (1.18 Mpc; van dem Bergh 1999) as the radius of the Local
Group halo. The filling factor obtained is then 0.33. This figure is
so high that individual halos would be in contact, and eventually
merge. Therefore, a picture more in consonance with the theory is that
there is only one large previously formed common halo. This rough
calculation just considers the most optimistic situation. The filling
factor would be higher if the Local Group were non spherical as
suggested by Karachentsev (1996) and if there were many more galaxies
belonging to the Local Group. Discoveries of new members have recently been
reported and many low surface brightness galaxies would not have been
detected if they had not been in the close vecinity of the Milky Way. Moreover,
consider that  in a
sphere of 500 kpc around the Milky Way there are 11 galaxies and
around M31 there are 15 galaxies. Under the assumption that each
galaxy has its own halo of about 200 kpc, we obtain filling factors much
higher than unity. Another observation suggesting that the Local Group has a common
halo is the observation that the high-velocity clouds have their
kinematic centre in the Local Group barycentre (L\'opez-Corredoira,
Beckman \& Casuso 1999).

d) {\bf Small compact groups}. The evidence and necessity of common
halos is specially clear in the case of Hicson Compact Groups (HCG;
Hickson 1982; Mamon 1995). HCG contain few galaxies, four or slightly
more, and are very compact, with the intergalactic distance and the whole
apparent size of the group being much smaller than the size of typical
halos. From the dynamical point of view, given the small velocity
dispersion, the system would collapse in less than $10^9$ years, after
which the members would merge and form a large elliptical (Barnes
1989; Diaferio, Geller \& Ramella 1994 and others) but in fact they are
noticeably stable and there are few signs of interaction and
merging. These facts led Athanassoula, Makino \& Bosma (1996) to assume a
massive, not excessively concentrated common DM halo. G\'omez-Flechoso
\& Dom\'{\i}nguez-Tenreiro (1997) included a
common DM halo in their N-body simulations 
to stabilize the groups. Common envelope material is
found in X-rays (Ponman \& Bertram 1993) and atomic hydrogen
(Verdes-Montenegro et al. 1997 and references therein). In HCG 49,
Verdes-Montenegro et al. (1999) showed that the HI common envelope is rotating with
a highly symmetrical pattern, following a large-scale potential that
is not due
to any particular galaxy member. Perea et al. (1999) have studied faint
satellite galaxies at large distances from the members but belonging
to the HCG. They found that the common halo is about four times
more massive than the galaxy members. We therefore conclude that a
great deal of
evidence clearly indicates that HCG are embedded in large common
halos.

e) {\bf Rich clusters}. White \& Rees (1978), Navarro, Frenk \&
White (1996) and many other theoretical models had as objectives the
obtention of halos with different sizes, with rich clusters being the
largest considered. Clearly, rich clusters could be the best example
of visible galaxies moving in a large DM cluster, from the point of
view of hierarchical CDM scenarios.

Therefore, the hypothesis of a common halo, as opposed to individual halos, is
compatible with observations of galaxy pairs, almost essential for
groups like the Local Group, compelling for compact groups and tempting
for rich clusters. It is also qualitatively coherent with the scenario
assumed by hierarchical CDM. In analytic and semianalytic models a
large common halo is assumed to be virialized; the violent relaxation
following the successive merging processes would destroy any DM
substructure, though visible galaxies could remain indigest. Then
isolated spirals would not possess a DM halo. Some numerical
calculations have not found any subhalos within halos (e.g. Katz \& White
1993; Summers, Davis \& Evrard 1995) giving rise to the so-called ``overmerging''
 problem.

Benson et al. (1999) obtain many small halos containing no
visible galaxy, which could be due to feedback from supernovae, which
prevents efficient galaxy formation. Though they obtain that the mass-to-light relation
has a minimum for about $10^{12} M_\odot$, the number of visible
galaxies in a halo greatly increases with halo mass (at least for
their $\Lambda$CDM model) indicating that large halos are indeed
common halos of many galaxies. The number of visible galaxies with
blue absolute magnitude brighter than about -19.5 per halo is
statistically lower than unity, but this number probably increases
when fainter galaxies are considered in the results of this model. The
Local Group has only two galaxies brighter than -19.5.

Moore et al. (1999) and others find a DM substructure in DM halos. Following
this calculation, the Milky Way would have about 500 satellites with
mass $\ge 10^8 M_\odot$, and therefore mechanisms avoiding stellar
formation within so many small halos would be required, which has been
discussed by Moore et al. and references therein. Internal
mechanisms do not seem to be responsible: if gas is lost by star-bursts and
winds in a first stage of star formation, it should be explained why
galaxies outside clusters have rotationally supported disks. Moreover,
the strongest star-bursts observed in nearby dwarf galaxies
are insufficient. These authors also discuss the difficulties inherent
in
forming and maintaining disks in the presence of large amounts of
substructure, as disk and small halo interactions will frequently heat
disks and produce ellipticals.

Galaxies could have a very different behaviour depending on their
position in a cluster. Whitemore, Forbes and Rubin (1988) found a
relation between the gradient of rotation curves and the location in
the cluster. Verheijen (1977) found an anomalous behaviour of rotation
curves of galaxies belonging to the Ursa Major cluster. Rubin,
Waterman and Kenney (1999) have found many galaxies with kinematic
disturbances in the Virgo cluster, but tidal effects and accretion
events can explain the observed disturbances. Individual dynamic
studies of disks in clusters are difficult to interpret.

We could conclude that the existence of common halos and the
non-existence of individual sub-halos are suggested both
by the observations and by the theory. Following this picture then,
spiral galaxies would have no halo, but rather  move inside
halos, orbiting off-centre in less dense and more homogeneous DM
environments.

Truly isolated spiral galaxies would have their own halo but these are exceptional. 
van dem Bergh (1999) estimated that half of all galaxies in the
Universe are situated in small clusters such as the Local Group. Soneira
\& Peebles (1977) estimated at 15\% the number of isolated
galaxies. Tully \& Fisher (1978) even claimed that there is no
evidence for a significant number of field galaxies.

The situation would be similar for late
type irregulars and for dwarf spheroidal galaxies. However,
these conclusions would not serve for the DM content of elliptical
galaxies. In a rich cluster, if we assume that the centres of the DM halo
and of the galaxy distribution coincide, at least the giant cD
ellipticals at the centre would have large quantities of DM, with the
cD galaxy and the DM also being coincident. Giant ellipticals, like M87,
have been considered to possess very large amounts of dark matter
since Fabricant, Lecar \& Gorenstein (1980) and Binney \& Cowie (1981) analyzed
their X-ray emission. Indeed these authors noticed that these large
 quantities of DM encountered could belong to the cluster itself rather than to the
galaxy. Huchra \& Brodie (1987) showed that the dynamics of globular
clusters around M87 supported the huge mass found from X-ray
observations, of the order of $10^{13} M_\odot$. It is unclear whether
this conclusion about cD galaxies would also apply for normal
ellipticals. In some cases, the debris from a merger of spirals could
fall into the halo centre. The difference between spirals (and
irregulars) and cD ellipticals would be that the former lie well
outside the halo's centre while the latter coincide with it.

Therefore, we can summarize the present crossroads of the problem of
rotation curves of spiral galaxies by emphasizing that, if we accept the
hypothesis of common virialized halos, with no substructure, for all 
types of clustered visible
galaxies, then there are only
two alternatives:

Either hierarchical CDM models are wrong, for instance,  DM is
  baryonic (e.g. de Paolis et al. 1997; Pfenniger \& Combes 1994), in
  which case we would need a theory of galaxy formation.

Or they are basically valid, in which case, another
 explanation of the rotation curve is needed. For instance, forces
 other than gravitation could be involved, so that models of galaxy
 formation would have no ``responsibility'' in explaining the rotation
 curve. We should take into account the magnetic hypothesis (Nelson 1988; Battaner et al. 1992, Battaner \& Florido
1995, Battaner, Lesch \& Florido 1998). Given the success of current
theoretical CDM hierarchical models in other related topics, we favour this latter
possibility.

\section{Conclusions}

a) Standard interpretation of rotation curves.

There is a general consensus about the history of galaxy formation and
the establishment of the rotation curve of spirals. This standard
history could be summarized as follows:

At Inflation, quantum mechanical fluctuations were generated and then
survived until the epoch of Recombination. The Universe then became
CDM dominated with a baryonic component as a minor constituent. By
gravitational collapse the primordial fluctuation that evolved gave rise to
small DM halos. Adjacent halos merged to produce larger halos and this
merging process has continued until the present. A complete hierarchy
of CDM halos has been produced, those produced later being larger and
the size being limited by the finite time of the Universe.

Once a new halo is formed by merging, violent relaxation effects
destroy part of the previous DM substructure, but not completely,
leaving some CDM subhalos still identifiable. After that, baryons cool and
concentrate at the CDM centre at any size of the hierarchy. Baryon
concentrates then form galaxies and shine. Larger CDM halos, produced
later, also produce larger visible galaxies.

Hierarchical CDM models predict universal halo density distribution
profiles, the so called NFW profiles, irrespective of their size and
position in the hierarchy, following as $R^{-1}$ in the inner region
and as $R^{-3}$ in the outer one. The CDM halo density profiles decide
the rotation curve of the visible galaxies which are small bright
indicators of large and massive CDM halos.

Subhalos within a halo are possible and would correspond to the
existence of small ancient visible satellite galaxies orbiting around a
large galaxy or to  normal galaxies in a large cluster. Some CDM
subhalos were destroyed in the merging process but their visible
baryonic aggregates, because of their high density, were able to
survive. Visible galaxies also merge. The merging of two or more
former disk galaxies produces a larger elliptical.

Some aspects in this short account of the long history are better 
known than others. If we restrict ourselves to the
rotation of spiral galaxies, there are some problems that remain
unclear or are insufficiently explained. We prefer now to select problems
rather than to emphasize successes. Among the outstanding problems let us
highlight the following:

- Theorists themselves are unhappy with the rotation curves obtained;
  in particular the Tully-Fisher relation is unsatisfactorily
  explained.

- It is not clear if the visible galaxies should now be at the centres
  of their ``own'' halos or if they lie off-centre in common
  halos shared with other galaxies. The degree of CDM substructure
  predicted by different authors varies.
  In the first pioneering steps it was
  suggested that dwarf satellites would be in the halo of the larger
  primary galaxy, having no smaller halos of their own. Observations,
  however, and new theoretical model developments, indicate that 
  dwarf satellites not only have their own
  halos, but also that the dark/visible matter ratios are much larger.

- The rotation curves observed can be fitted to the so called
  Universal Rotation Curves, but their density profiles
  do not coincide at all with the theoretical density profiles. The
  universal rotation curves have a core, i.e. a region in which the
  density is more or less constant or slightly decreasing as $R$,
  and in the periphery they decrease as $R^{-3/2}$.

- The so called ``halo-disk conspiracy'', i.e. why the disk
  and halo dominated regions have a featureless flat transition, is not
  completely answered. The fact that some galaxies have rising or
  declining curves does not explain the conspiracy problem in those galaxies that do
  possess a flat curve. The adiabatic compression of the inner
  halo due to the disk formation establishes a halo-disk relation that
  alleviates the conspiracy, but much work is still needed to model
  this interaction.

- Both universal rotation curves and universal halo density
  distributions shed some light on the absence of correlation
  between the orbital velocities of satellite galaxies and rotation
  velocity of the primary galaxies (or equivalently their luminosities, as
  Tully-Fisher relates both quantities). But both velocities
  should be determined by the halo and should partially correlate.

The so called Bosma relation, even if some authors have their doubts
about its validity, indicates a relation between the circular
velocities of both the dark halo and the gas. This relation is not only a
general one, but in some galaxies, small scale changes are present in
the rotation curve and in the gas distribution. This seems
paradoxical, particularly if we consider that the dark matter is more or
less spherically distributed and the gas lies in a disk.

Clearly, other authors would discuss other points that are still
obscure and
others would have preferred to focus on the agreements with
observational facts, which are certainly encouraging and suggest
that the basic scenario has been firmly established. There are, 
however, many other possible histories:
we have seen that some authors consider dark
matter to be baryonic and even that it is in the disk. These theories
have considerable merit, especially because they must be developed against the
general flow of ideas, and because they explain some observational
facts in a simpler way.

Let us just outline another alternative history, rather different from
the above mentioned standard one, if the word ``standard'' can be
properly assigned to any of the present scenarios.

b) The magnetic interpretation of the rotation curve.

At inflation, quantum mechanical fluctuations were generated not only
in the energy density distribution but also created a non uniform
distribution of magnetic fields. Magnetic flux tubes, probably
interconnected with other tubes forming a network, conserved their
shapes during the radiation dominated era, but the strength 
decreased as the expansion proceeded. Finite conductivity effects
destroyed the small scale magnetic field structures, but those that
were large
enough, larger than about 50 Mpc (comoving) survived, producing
filaments in the energy density distribution -probably also sheets-
which after Recombination became dark matter filaments (with baryons
as a minor constituent), more than 100 Mpc long.

The scenario provided by hierarchical CDM models, assumed in the
``standard'' history previously summarized, is fully assumed here too,
with the only exception that mergers and non-linear evolution took
place inside the large density filaments and not in the voids
in between. The heating produced by shocks in the merging events also
affected the magnetic fields, which became disordered and
amplified. Individual halos belonging to a visible galaxy at its
centre were formed in the first generations of halos, but subsequent
mergers produced common halos shared by satellites or groups of
galaxies. Pairs, satellites, poor clusters and rich clusters developed
their superhalos with no dark matter substructure. Only exceptional
truly isolated visible counterparts would retain their own halo
against merging or were the result of merging of the visible
galaxies in a previous common halo.

The fact that hierarchical mergers only took place within the
filaments and not in the large voids, assumed here to be of primordial
origin, could alleviate a problem encountered in CDM hierarchical
models: that of overproduction of halos. Furthermore, these models predict too
little structure at scales larger than about 40 Mpc, a problem that was
detected in the pioneering simulations.

The disk was formed out of magnetized gas and maintained a magnetic
pressure equilibrium with the region outside the visible galaxy,
frozen-in in the low-density uncondensed gas lying in the common
halo. Equilibrium was possible
because of the frequent outbursts of disk material and magnetic fields
due to violent star formation events, as observed for instance in M82
and in other galaxies like ours. This magnetic field acquired a
toroidal distribution, due to the differential rotation, which was
able to generate a centripetal force which produced a higher rotation
in the periphery of disks.

Magnetic fields responsible for the high rotation velocity also
rendered the disk thicker, facilitating the fountain effect and
escape. Magnetic fields would act inwards in the radial direction and
outwards in the vertical direction. The escape from the disk and even
from the galaxy would be a more important effect in dwarf irregulars,
which indeed present larger outbursts of material; they are gas rich,
and therefore have greater ability to amplify magnetic fields. Under this
interpretation, irregulars are not DM rich galaxies, but magnetic
field rich galaxies, because they are gas rich galaxies.

One question naturally arises in this scenario. What is the DM content of
the other galaxies, particularly in ellipticals? Suppose a rich
cluster which has a giant cD elliptical at the centre. It is 
evident that cD galaxies are then also at the centre of the cluster
superhalo and therefore, they are beated in a region with large
amounts of dark matter. Therefore, cD galaxies should have large
amounts of DM, as seems to be the case in M87. It is less clear
what is the situation with normal ellipticals. Therefore, we merely propose that  giant
ellipticals at the centre of large halos would have large quantities
of DM; spirals, lying far from the giant common cluster halo
centre would not posses dark matter halos. Even if they  were
embedded in DM, this would be more
homogeneously distributed around the spiral, and hence it would not have
such a decisive influence on the rotation curve. The DM halo potential
could produce warps as a tidal effect.

In the ``standard'' history, we discussed some current problems and
disagreements between theory and observations. In this alternative
scenario, we comment on their advantages. In this paper we
have reviewed a model that numerically accounts for the basic facts. From
a qualitative point of view, without any precise developments, let us
also consider:

- Galaxies with more gas would produce, and be subject to, higher magnetic fields,
  and precisely these galaxies were considered to have rotation curves
  with a greater discrepancy from the curve expected from the gravitation produced
  by disk and bulge.

- The Bosma relation, establishing a connection between gas and
  DM (which in this picture should be expressed as a relation between
  gas and magnetic fields) would be obvious.

- There would be no conspiracy problems, as the magnetic fields and
  gravitation forces ratio would be progressive and continuously increasing
  for increasing radii.

- The problems arising from the lack of correlations in binary
  galaxies would naturally disappear. The velocity observed at the
  higher radius where the signal is detected would be the result of
  internal magnetic fields, clearly unrelated to any orbital
  velocity, whether or not the pair lies in a common halo.

- The agreement with theoretical hierarchical CDM models is
  better, as these models have ``no responsibility'' to directly explain rotation
  curves, unless they include magnetic fields. For instance, 
  the theoretical prediction that irregulars
  orbiting a bright galaxy (like ours) would be embedded in its halo
  and would not have their own halo, would not contradict the
  standard interpretation of observations that irregulars are
  particularly dark matter dominated.

c) Other alternatives.

Theories assuming galactic dark matter to be undetected gas must
eventually answer two basic questions. How have these galaxies formed?
and, if $\Omega_M \sim 0.3$ and $\Omega_B \sim 0.03$, where is the
non-baryonic dark matter? (could the answer to the second
question be the existence of common cluster halos?).

Theories proposing a modification of Newton's Second Law should clarify
whether we should also reject General Relativity. Modifying Newtonian
Mechanics is a tolerable sacrifice, but physics would probably
require more solid proof before abandoning General Relativity.

Summarizing this summary, we are beginning to understand galaxy
formation, the nature and
distribution of dark matter in galaxies and rotation in what could
be called the standard scenario. But there are other interesting
alternatives that should not be disregarded without an intense
debate. MOND is one of them. Gaseous dark matter is another. 
The magnetic alternative is
not frontally opposed to CDM hierarchical scenarios, but is, rather, 
complementary. Only secondary phenomena are in clear
contradiction. It is unrealistic to attempt to deal with rotation  curves
while ignoring magnetic fields. This could constitute a particular flaw in
the standard model for rotation curves.

If after this review, the topic of
galactic dark matter is less clear, we will have accomplished our mission. 

\vskip 2cm

{\bf{Acknowledgements}}

We are very grateful to Kor Begeman, Adrick Broeils, Jordi Cepa,
Carlos Frenk, Mareki Honma, Julio Navarro, Michael Merrifield,
Yoshiaki Sofue, Rob Swaters and Juan Carlos Vega Beltr\'an, for their
permission to include some of their figures and for their valuable
help. We are specially grateful to Constantino Ferro-Font\'an, for his
encouragement and stimulus to undertake this review. Our thanks also
to Jorge Jim\'enez-Vicente and Ana Guijarro. Support for this
work was provided by the Ministerio de Educaci\'on y Cultura (PB96-1428)
and by Junta de Andaluc\'{\i}a (FQM-108).

\newpage
\noindent {\bf Bibliography}

\vskip 1.5cm

The literature about dark matter is really copious. Some reviews have
been particularly useful in preparing this one, such as those  by
Trimble (1982) ``Existence of dark matter in the Universe'',
Ashman (1992) ``Dark matter in galaxies'',
Bosma (1998) ``Dark matter in disk galaxies'',
Fich and Tremaine (1991) ``The mass of the Galaxy''
and those appeared in the book ``Dark and visible matter in
galaxies'' (1997), as those by Salucci and Persic (1997) ``Dark matter
halos around galaxies'', Pfenniger (1997) ``Galactic Dynamics and the
nature of dark matter'', Frenk et al. (1997) ``Numerical and
analytical modelling of galaxy formation and evolution'', Navarro
(1997) ``Cosmological constraints from rotation curves of disk
galaxies'' and others. The whole book was decisive to prepare this
review, as it is a recent complete and updated compilation by many workers on the
field. The PhD theses of Begeman (1987) ``HI rotation curves of spiral
galaxies'', Broeils (1992) ``Dark and visible matter in spiral
galaxies'' and Swaters (1999) ``Dark matter  in late-type dwarf
irregulars'' were also special sources of information. For the section
devoted to magnetic fields, some reviews were specially detachable:
those by Kronberg (1994) ``Extragalactic magnetic fields'', Lesch and
Chiba (1997) ``On the origin and evolution of galactic magnetic
fields'', and Vall\'ee (1997) ``Observations of the magnetic fields
inside and outside the Milky Way'', starting with globules ($\sim$ 1
parsec), filaments, clouds, superbubbles, spiral arms, galaxies,
superclusters, and ending with the cosmological Universe's Background
Surface (at $\sim$ 8 Teraparsec).

\vskip 1.5cm

\noindent References
\vskip 0.5cm 

\noindent Aaronson, M. (1983), {\it Astrophys. J.} {\bf 266}, L11\\
Adams, J., Danielson, U.H., Grasso, D. and Rubinstein, H.R. (1996),
  {\it Phys. Lett. B} {\bf 388}, 253\\
Alcock, C. et al. (1993), {\it Nature} {\bf 365}, 621\\
Alderberger, K., Steidel, C.C., Giavalisco, M., Dickinson, M.,
  Pettini, M. and Kellog, M. (1998), {\it Astrophys. J.} {\bf 505}, 18\\
Andredakis, Y.C., Peletier, R. and Balcells, M. (1995),
  {\it Mon. Not. Roy. Ast. Soc.} {\bf 275}, 874\\
Aparicio, A., Dalcanton, J., Gallart, C. and Mart\'{\i}nez-Delgado,
  D. (1997), {\it Astron. J.} {\bf 114}, 1447\\
Ashman, K.M. (1992), {\it PASP} {\bf 104}, 409\\
Athanassoula, E., Makino, J. and Bosma, A. (1997),
  {\it Mon. Not. Roy. Astr. Soc.} {\bf 286}, 825\\
Atrio-Barandela, F., Einasto, J., Gottl\"{o}ber, S., M\"{u}ller,
  V. and Starobinsky, A. (1997), {\it J. Exper. Theor. Phys.} {\bf 66}, 397\\
Aubourg, E. et al. (1993), {\it Nature} {\bf 365}, 623\\
Avila-Reese, V., Firmani, C. and Hernandez, X. (1998),
  {\it Astrophys. J.} {\bf 505}, 37\\
Avila-Reese, V., Firmani, C., Klypin, A. and Kravtsov, A.V. (1999),
  {\it Mon. Not. Royal Astr. Soc.} {\bf 310}, 527\\
Babcock, H.W. (1939), {\it Lick Obs. Bull.} {\bf 19}, 41\\
Bahcall, J.N. and Casertano, S. (1985), {\it Astrophys. J.} {\bf 293}, L7\\
Bahcall, N.A., Lubin, L.M. and Dorman, V. (1995), {\it Astrophys. J.} {\bf 447}, L81\\
Barnes, J.E. (1987), in {\it Nearly normal galaxies. From the Planck
  time to the present}, Proc. 8th Santa Cruz Summer Workshop in
  A\&A. Springer Verlag, p. 154\\
Barnes, J.E. (1989), {\it Nature} {\bf 338}, 123\\
Barnes, J. and White, S.D.M. (1984), {\it Mon. Not. Roy. Astr. Soc.} {\bf 211}, 753\\
Barrow, J.D. (1976), {\it Mon. Not. Roy. Astr. Soc.} {\bf 175}, 339\\
Barrow, J.D., Ferreira, P.G. and Silk, J. (1977), astro-ph/9701063\\
Barteldrees, A. and Dettmar, R.J. (1994), {\it Astrophys. J.} {\bf 427}, 155\\
Batchelor, G.K. (1950), in {\it Proc. R. Soc. London} {\bf 201}, 405\\
Battaner, E. (1995), in  E. Alfaro and G. Tenorio-Tagle (Eds.), {\it The formation of the Milky Way},
  Cambridge University Press\\
Battaner, E. (1996), {\it Astrophysical Fluid Dynamics}, Cambridge
  Univ. Press\\
Battaner, E. (1998), {\it Astron. Astrophys.} {\bf 334}, 770\\
Battaner, E. and Florido, E. (1995), {\it Mon. Not. Roy. Ast. Soc.} {\bf
  277}, 1129\\
Battaner, E. and Florido, E. (1997), in T.A. Agekian, A.A. Mullari and
  V. Orlov. (Eds.), {\it Structure and evolution of
  stellar systems},  S. Petersburgh Univ. Press\\
Battaner, E. and Florido, E. (1998), {\it Astron. Astrophys.} {\bf 338}, 383\\
Battaner, E., Florido, E. and Garc\'{\i}a-Ruiz, J.M. (1997),
  {\it Astron. Astrophys.} {\bf 327}, 8\\
Battaner, E., Florido, E. and Jim\'enez-Vicente, J. (1997),
  {\it Astron. Astrophys.} {\bf 326}, 13\\
Battaner, E., Florido, E., S\'anchez-Saavedra, M.L. (1990),
  {\it Astron. Astrophys.} {\bf 236}, 1\\
Battaner, E., Florido, E. and S\'anchez-Saavedra, M.L. (1991), in
   S. Casertano, P. Sackett and F. Briggs (Eds.), {\it Warped disks
   and inclined rings around galaxies},
   Cambridge University Press\\
Battaner, E., Garrido, J.L., Membrado, M. and Florido, E. (1992),
  {\it Nature} {\bf 360}, 652\\
Battaner, E., Garrido, J.L., S\'anchez-Saavedra, M.L. and Florido,
  E. (1991), {\it Astron. Astrophys.} {\bf 251}, 402\\
Battaner, E., Jim\'enez-Vicente, J. (1998), {\it Astron. Astrophys.} {\bf 332},
  809\\
Battaner, E. and Lesch, H. (2000), in preparation\\
Battaner, E., Lesch, H. and Florido, E. (1999), {\it An. F\'{\i}sica} {\bf
  94}, 98\\
Baugh, C.M., Benson, A.J., Cole, S., Frenk, C.S. and Lacey,
  C.G. (1999), {\it Mon. Not. Roy. Astr. Soc.} {\bf 305}, L21\\
Baugh, C.M., Cole, S., Frenk, C.S. and Lacey, C.G. (1998),
  {\it Astrophys. J.} {\bf 498}, 504\\
Baym, G., Boedecker, D. and McLerran, L. (1996), {\it Phys. Rev. D} {\bf
  53}, 662\\
Beck, R. (1982), {\it Astron. Astrophys.} {\bf 106}, 121\\
Beck, R., Carilli, C.L., Holdaway, M.A. and Klein, U. (1994),
  {\it Astron. Astrophys.} {\bf 292}, 409\\
Begeman, K. (1987), Ph. Thesis, univ. Groningen\\
Begeman, K.G., Broeils, A.H. and Sanders, R.H. (1991), {\it Mon. Not. Roy.
  Ast. Soc.} {\bf 249}, 523\\
Becquaert, J.F. and Combes, F. (1997), {\it Astron. Astrophys.} {\bf 325}, 41 \\
Benson, A.J., Cole, S., Frenk, C.S., Baugh, C.M. and Lacey,
  C.G. (1999), astro-ph/9903343\\
Bertschinger, E. (1998), {\it Ann. Rev. Astron. Astrophys.} {\bf 36}, 599\\
Bertin, G., Saglia, R.P. and Stiavelli, M. (1992), {\it Astrophys. J.} {\bf 384},
  427\\
Bertin, G. and Stavielli, M. (1993), {\it Rep. Prog. Phys.} {\bf 56}, 493\\
Bertin, G. et al. (1994), {\it Astron. Astrophys.} {\bf 292}, 381\\
Bertola, F., Pizella, A., Persic, M. and Salucci, P. (1993),
  {\it Astrophys. J.} {\bf 416}, 248\\
Biermann, L. (1950), {\it Zeit. Naturforschung} {\bf 5a}, 65\\
Biermann, L. and Schluter, A. (1951), {\it Phys. Rev.} {\bf 82}, 863\\
de Block, W.J.G. and McGaugh, S.S. (1998), {\it Astroph. J.} {\bf 508}, 132D\\
Binney, J. (1991), in B. Sundelius (Ed.), {\it Dynamics of disk
  galaxies}, p. 297,  G\"{o}teborg\\
Binney, J. (1992), {\it Nature} {\bf 360}, 624\\
Binney, J. (1992), {\it Ann. Rev. Astron. Astrophys.} {\bf 30}, 51\\
Binney, J.J., Davies, R.L. and Illingworth, G.D. (1990),
  {\it Astrophys. J.} {\bf 361}, 78\\
Binney, J.J. and Cowie, L.L. (1981), {\it Astrophys. J.} {\bf 247}, 464\\
Binney, J. and Tremaine, S. (1987), {\it Galactic Dynamics}, Princeton
  Univ. Press \\
Blumenthal, G.R., Faber, S.M., Flores, R. and Primack, J.R. (1986),
  {\it Astrophys. J.} {\bf 301}, 27\\
Bosma, A. (1978), Phy. D. Thesis, Univ. of Groningen\\
Bosma, A. (1981a), {\it Astron. J.} {\bf 86}, 1791\\
Bosma, A. (1981b), {\it Astron. J.} {\bf 86}, 1825\\
Bosma, A. (1993), in G. Meylan
  \& P. Prugniel (Eds.), {\it ESO/ESA Workshop on Dwarf Galaxies} p. 187, Garching\\
Bosma, A. (1998), in {\it Galaxy Dynamics}, Rutgers University, ASP
  Conf. Series {\bf 182}, p. 339\\
Bradamonte, F., Matteucci, F. and d'Ercole, A. (1998),
  {\it Astron. Astrophys.} {\bf 337}, 338\\
Brandenberger, R.H., Davis, A.C., Matheson, A.M. and Thodden,
  M. (1992), {\it Phys. Lett. B} {\bf 293}, 287\\
Brandenburg, A., Enqvist, K. and Olesen, P. (1996), {\it Phys. Rev. D} {\bf
  54}, 1291\\
Breimer, T.G. and Sanders, R.H. (1993), {\it Astron. Astrophys.} {\bf 274}, 96\\
Breitschwerdt, D. and Komossa, S. (1999), astro-ph/9908003\\
Breitschwerdt, D., McKenzie, J.F. and V\"{o}lk, H.J. (1991),
  {\it Astron. Astrophys.} {\bf 245}, 79\\
Broadhurst, T.J., Ellis, R.S., Koo, D.C., Szalay, A.S. (1990), {\it Nature}
  {\bf 343}, 726\\
Broeils, A. (1992), Ph. Thesis, univ. Groningen\\
Broeils, A.H. (1992), {\it Astron. Astrophys.} {\bf 256}, 19\\
Broeils, A.H. and Courteau, S. (1997), in M. Persic and P. Salucci
  (Eds.),  {\it Dark and visible matter in Galaxies},
  ASP Conference Series {\bf 117}\\
Buote, D.A. and Canizares, C.R. (1998), {\it Mon. Not. Roy.
  Ast. Soc.} {\bf 298}, 811\\
Burkert, A. (1997), {\it Astrophys. J.} {\bf 474}, L99\\
Burton, W.B. (1976), {\it Ann. Rev. Astron. Astrophys.} {\bf 14}, 275\\
Burton, W.B. (1992), SAAS-FEE Proc. Springer-Varlag\\
Carignan, C. (1985), {\it Astrophys. J.} {\bf 299}, 59\\
Carignan, C. and Beaulieu, S. (1989), {\it Astrophys. J.} {\bf 347}, 760\\
Carignan, C., Charbonneau, P., Boulanger, F. and Viallefond,
  F. (1990), {\it Astron. Astrophys.} {\bf 234}, 43\\
Carignan, C. and Freeman, K.C. (1988), {\it Astrophys. J.} {\bf 332}, L33\\
Carignan, C., Sancisi, R. and van Albada, T.S. (1988), {\it Astron. J.} {\bf
  95}, 37\\
Carollo, C.M., de Zeeuw, P.T., van der Marel, R.P., Danzinger,
  I.J. and Qian, E.E. (1995), {\it Astrophys. J.} {\bf 441}, L25\\
Casertano, S. (1983), {\it Mon. Not. Roy. Ast. Soc.} {\bf 203}, 735\\
Casertano, S. and van Gorkom, J.H. (1991), {\it Astron. J.} {\bf 101}, 1231\\
Cavaliere, A. and Busco-Femiano, R. (1976), {\it Astron. Astrophys.} {\bf 49}, 137\\
Charlton, J.C. and Salpeter, E.E. (1991), {\it Astrophys. J.} {\bf 375}, 517\\
Cheng, B. and Olinto, A. (1994), {\it Phys. Rev. D} {\bf 50}, 2421\\
Cheng, B., Olinto, A.V., Schramm, D.M. and Truran, J.W. (1997),
  preprint Los Alamos Nat. Observatory\\
Chiba, M. and Lesch, H. (1994), {\it Astron. Astrophys.} {\bf 284}, 731\\
Ciardullo, R. and Jacobi, G.H. (1993), {\it Amer. Astron. Soc. Meet.} {\bf 182}\\
Ciotti, L., Pellegrini, S., Renzini, A. and D'Ercole, A. (1991),
  {\it Astrophys. J.} {\bf 376}, 380\\
Clemens, D.P. (1985), {\it Astrophys. J.} {\bf 295}, 422\\
Cole, S. (1991), {\it Astrophys. J.} {\bf 367}, 45\\
Cole, S., Arag\'on-Salamanca, A., Frenk, C.S., Navarro, J.F. and Zepf,
  S. (1994), {\it Mon. Not. Roy. Ast. Soc.} {\bf 271}, 744\\
Coles, P. (1992), {\it Comments Astrophys.} {\bf 16}, 45\\
Colin, P., Klypin, A., Kravtsov, A. and Khokhlov, A. (1999),
  {\it Astrophys. J.} {\bf 523}, 32\\
Combes, F. and Arnaboldi, M. (1996), {\it Astron. Astrophys.} {\bf 305}, 763\\
Corbelli, E. and Salucci, P. (1999), astro-ph/9909252\\
C\^{o}t\'e, S. (1995), Ph. D. Thesis, Australian Nat. Univ.\\
Cowie, L.L., Songaila, A., Hu, E.M. and Cohen, J.G. (1996),
  {\it Astron. J.} {\bf 112}, 839\\
Cr\'ez\'e, M., Chereul, E., Bienaym\'e, O. and Pichon, C. (1998),
  {\it Astron. Astrophys.} {\bf 329}, 920\\
Cuddeford, P. and Binney, J.J. (1993), {\it Nature} {\bf 365}, 20\\
Dahlem, M., Dettmar, R.J. and Hummel, E. (1994),
  {\it Astron. Astrophys.} {\bf 290}, 384\\
Danziger, I.J. (1997), in M. Persic and P. Salucci (Eds.), {\it Dark
  and visible matter in galaxies}, ASP Conf. Series {\bf 117},, p. 28\\
Davidson, S. (1996), {\it Phys. Lett. B} {\bf 380}, 253\\
Davies, A.C. and Dimopoulos, K. (1995), cern-th/95-175\\
Davis, D.S. and White, R.E. III (1996), {\it Astrophys. J.} {\bf 470}, L35\\
Diaferio, A., Geller, M.J. and Ramella, M. (1994), {\it Astron. J.} {\bf 107}, 868\\
Diplas, A. and Savage, B.D. (1991), {\it Astrophys. J.} {\bf 377}, 126\\
Dolag, K., Bartelmann, M. and Lesch, H. (1999), astro-ph/9906329\\
Dolgov, A.D. (1993), {\it Phys. Rev. D} {\bf 48}, 2499\\
Dolgov, A.D. and Silk, J. (1993), {\it Phys. Rev. D} {\bf 47}, 3144\\
Dubinski, J. and Kuijken, K. (1995), {\it Astrophys. J.} {\bf 442}, 492\\
Eilek, J. (1999), astro-ph/9906485\\
Einasto, J. et al. (1997a), {\it Mon. Not. Roy. Ast. Soc.} {\bf 289},
  801\\
Einasto, J. et al. (1997b), {\it Mon. Not. Roy. Ast. Soc.} {\bf 289},
  813\\
Einasto, J. et al. (1997c), {\it Nature} {\bf 385}, 139\\
Einasto, M., Einasto, J., Tago, E., Dalton, G.B. and Andernach,
  H. (1994), {\it Mon. Not. Roy. Ast. Soc.} {\bf 269}, 301\\
Eisenstein, D.J., Loeb, A. and Turner, E.L. (1997), {\it Astrophys. J.} {\bf
  475}, 421\\
Enqvist, K. (1998), in L. Roszkowski (Ed.), COSMO-97, World
  Scientific, p. 446\\
Enqvist, K. and Olesen, P. (1993), {\it Phys. Lett. B} {\bf 319}, 178\\
Enqvist, K., Semikoz, V., Shukurov, A. and Sokoloff, D. (1993),
  {\it Phys. Rev. D} {\bf 48}, 4557\\
Fabbiano, G. (1989), {\it Ann. Rev. Astron. Astrophys.} {\bf 27}, 87\\
Faber, S.M. and Jackson, R.E. (1976), {\it Astrophys. J.} {\bf 204}, 668\\
Faber, S.M. and Lin, D.N.C. (1983), {\it Astrophys. J.} {\bf 266}, L17\\
Fabian, A.C., Thomas, P.A., Fall, S.M. and White, R.E. (1986),
  {\it Mon. Not. Roy. Ast. Soc.} {\bf 221}, 1049\\
Fabricant, D. and Gorenstein, P. (1983), {\it Astrophys. J.} {\bf 267}, 535\\
Fabricant, D., Lecar, M. and Gorenstein, P. (1980), {\it Astrophys. J.} {\bf
  241}, 552\\
Felten, J.E. (1984), {\it Astrophys. J.} {\bf 286}, 3 \\
Feretti, L., Dallacasa, D., Govoni, F., Giovannini, G., Taylor,
  G.B. and Klein, U. (1999), astro-ph/9902019\\
Fich, M. and Tremaine, S. (1991), {\it Ann. Rev. Astron. Astrophys.} {\bf 29}, 409\\
Field, G.B. (1974), in Stars and Stellar Systems, {\bf 9}, Chicago
  Univ. Press\\
Flores, R., Primack, J.R., Blumenthal, G.R. and Faber, S.M. (1993),
  {\it Astrophys. J.} {\bf 412}, 443\\
Florido, E. and Battaner, E. (1997), {\it Astron. Astrophys.} {\bf 327}, 1\\
Ford, H.C., Ciardullo, R., Jacoby, G.H. and Hui, X. (1989), in
  S. Torres-Peimbert (Ed.), Planetary Nebulae, IAU Symp. {\bf 131}, Kluwer, 335\\
Forman, W., Jones, C. and Tucker, W. (1985), {\it Astrophys. J.} {\bf 293}, 102\\
Freeman, K.C. (1970), {\it Astrophys. J.} {\bf 160}, 881\\
Freeman, K.C. (1997), in M. Persic and P. Salucci (Eds.), {\it Dark
  and visible matter in Galaxies},
  ASP Conference Series {\bf 117}, 242\\
Frenk, C.S., Baugh, C.M., Cole, S. and Lacey, C. (1997), in M. Persic and P. Salucci (Eds.), {\it Dark
  and visible matter in Galaxies},
  ASP Conference Series {\bf 117}, 335\\
Garretson. W.D., Field, G.B. and Carroll, S.M. (1992), {\it Phys. Rev. D}
  {\bf 46}, 5346\\
Gasperini, M., Giovannini, M. and Veneziano, G. (1995a),
  cern-th/95-85\\
Gasperini, M., Giovannini, M. and Veneziano, G. (1995b),
  cern-th/95-102\\
Gasperini, M. and Veneziano, G. (1993a), {\it Mod. Phys. Lett. A} {\bf 8},
  3701\\
Gasperini, M. and Veneziano, G. (1993b), {\it Astropart. Phys.} {\bf 1}, 317\\
Gerhard, O.E. (1994), in G. Meyan and P. Prugnid (Eds.), {\it Dwarf Galaxies},
  Proc.  of the ESO/OHP Workshop, Garching, 335\\
Gerhard, O.E. and Silk, J. (1996), {\it Astrophys. J.} {\bf 472}, 34\\
Gerhard, O.E. and Spergel, D.N. (1992), {\it Astrophys. J.} {\bf 397}, 38\\
Glazebrook, K., Ellis, R.S., Colles, M., Broadhurst, T.,
  Allington-Smith, J. and Tanvir, N. (1995),
  {\it Mon. Not. Roy. Ast. Soc.} {\bf 273}, 157\\
Gobernato, F., Baugh, C.M., Frenk, C.S., Cole, S., Lacey, C.G., Quinn,
  T. and Stadel, J. (1998), {\it Nature} {\bf 392}, 359\\
G\'omez-Flechoso, M.A. and Dominguez-Tenreiro, R. (1997), in M. Persic and P. Salucci (Eds.), {\it Dark
  and visible matter in Galaxies},
  ASP Conference Series {\bf 117}\\
Gonz\'alez-Serrano, I. and Valentijn, E.A. (1991),
  {\it Astron. Astrophys.} {\bf 242}, 334\\
Grasso, D. and Rubinstein, H.R. (1995), {\it Astropart. Phys.} {\bf 3}, 95\\
Grasso, D. and Rubinstein, H.R. (1996), {\it Phys. Lett. B} {\bf 379}, 73\\
Greenstein, G. (1980), {\it Nature} {\bf 233}, 938\\
Gunn, J.E. (1977), {\it Astrophys. J.} {\bf 218}, 592\\
Gunn, J. and Gott, J.R. (1972), {\it Astrophys. J.} {\bf 176}, 1\\
Han, J.L., Manchester, R.N., Berkhuijsen, E.M. and Beck, R. (1997),
  {\it Astron. Astrophys.} {\bf 322}, 98\\
Han, J.L. and Qiao, G.J. (1994), {\it Astron. Astrophys.} {\bf 288}, 759\\
Harrison, E.H. (1973), {\it Mon. Not. Roy. Ast. Soc.} {\bf 165}, 185\\
Haynes, M.P. and Broeils, A.H. (1997), in J.M. van der Hulst (Ed.),
  {\it The interstellar medium in
  galaxies}, Kluwer, p. 75\\
Hickson, P. (1982), {\it Astrophys. J.} {\bf 225}, 382\\
Hirashita, H., Kamaya, H. and Takeuchi, T.T. (1998), in P. Whitelock
  and R. Cannon (Eds.), {\it The stellar content of Local Group
  Galaxies}, IAU Symp. {\bf 192}, 28, Cape Town\\
Hodge, P.W. and Michie, R.W. (1969), {\it Astron. J.} {\bf 74}, 587\\
Hofner, P. and Sparke, L.S. (1994), {\it Astrophys. J.} {\bf 428}, 466\\
Hogan, C.J. (1983), {\it Phys. Rev. Lett.} {\bf 51}, 1488\\
Honma, M. (1999), astro-ph/9904079\\
Honma, M. and Kan-Ya, Y. (1998), {\it Astrophys. J.} {\bf 503}, L139\\
Honma, M. and Sofue, Y. (1996), {\it PASJ} {\bf 48L}, 103\\
Honma, M. and Sofue, Y. (1997), {\it PASJ} {\bf 49}, 539\\
Howard, A.M. and Kulsrud, R.M. (1996), {\it Astrophys. J.} {\bf 483}, 648\\
Huchra, J. and Brodie, J. (1987), {\it Astron. J.} {\bf 93}, 779\\
van der Hulst, J.M., Skillman, E.D., Smith, T.R., Bothun, G.D.,
  Humason, M.L., Mayall, N.U. and Sandage, A.R. (1956), {\it Astron. J.} {\bf 61},
  97\\
van der Hulst, J.M., Skillman, E.D., Smith, T.R., Bothun, G.D.,
  McGaugh, S.S. and de Block, W.J.G. (1993), {\it Astron. J.} {\bf 106}, 548\\
Hummel, E., Beck, R. and Dahlem, M. (1991), {\it Astron. Astrophys.} {\bf
  248}, 23\\
Hummel, E., Dahlem, M., van der Hulst, J.M. and Sukumar, S. (1991),
  {\it Astron. Astrophys.} {\bf 246}, 10\\
Illingworth, G. (1981), in S.M. Fall and D. Lynden-Bell (Eds.), {\it The
  structure of Normal Galaxies}, Cambridge Univ. Press, 27\\
Innanen, K.A., Harris, W.E. and Webbink, R.F. (1983), {\it Astron. J.} {\bf 88}, 338\\
Jedamzik, K., Katalinic, V. and Olinto, A. (1996), astro-ph/9606080\\
Jenkins, A. et al. (1997), in M. Persic and P. Salucci (Eds.), {\it Dark
  and visible matter in Galaxies},
  ASP Conference Series {\bf 117}\\
Jim\'enez-Vicente, J., Battaner, E., Rozas, M., Casta\~neda, H. and
  Porcel, C. (1999), {\it Astron. Astrophys.} {\bf 342}, 417\\
Jones, C. and Forman, W. (1994), {\it AIP Conference Proc.} {\bf 331}, 129\\
Kahn, F.D. (1994), {\it Astrophy. Space Sci.} {\bf 216}, 325\\
Kahn, F.D. and Woltjer, L. (1959), {\it Astrophys. J.} {\bf 130}, 705\\
Kalnajs, A. (1983), in E. Athanassoula (Ed.), {\it Internal Kinematics and
  Dynamics of Galaxies}, IAU Symp. {\bf 100}, 87, Dordrecht\\
Karachentsev, I.D. (1983), {\it Soviet Astron. Letters} {\bf 9}, 36\\
Karachentsev, I.D. (1985), {\it Soviet Astron. Letters} {\bf 29}, 243\\
Karachentsev, I.D. (1996), {\it Astron. Astrophys.} {\bf 305}, 33\\
Katz, N. and White, S.D.M. (1993), {\it Astrophys. J.} {\bf 412}, 455\\
Kauffmann, G., Guiderdoni, B. and White, S.D.M. (1994),
  {\it Mon. Not. Roy. Ast. Soc.} {\bf 267}, 981\\
Kay, S.T. et al. (1999), astro-ph/9908107\\
Kent, S. (1986), {\it Astron. J.} {\bf 91}, 1301\\
Kent, S.M. (1990), in R.G. Kron (Ed.), {\it Evolution of the Universe of
  Galaxies}, ASP Conference Series, {\bf 10}, 109,  Brigham Young Univ.\\ 
Kerr, F.J., Hindman, J.V. and Robinson, B.J. (1954),
  {\it Ann. J. Phys.} {\bf 7}, 297\\
Kerr, F.J. and de Vaucouleurs, G. (1955), {\it Ann. J. Phys.} {\bf 8}, 508\\
Kibble, T.W.D. and Vilenkin, A. (1995), {\it Phys. Rev. D} {\bf 52}, 1995\\
Killeen, N.E.B. (1995), {\it Pub. Astron. Soc. Australia} {\bf 12}, 124\\
Kim, E., Olinto, A. and Rosner, R. (1996), {\it Astrophys. J.} {\bf 468}, 28\\
Klessen, R.S. and Kroupa, P. (1998), {\it Astrophys. J.} {\bf 498}, 143\\
Kochanek, C. (1995), {\it Astrophys. J.} {\bf 445}, 559\\
Kolb, E.W. and Turner, M.S. (1990), {\it The Early Universe}, Addison-Wesley
  Pub.\\
Kormendy, J. and Freeman, K.C. (1998), {\it American
  Astron. Soc. Meeting} {\bf 193}\\
Kosowsky, A. and Loeb, A. (1997), {\it Astrophys. J.} {\bf 469}, 1\\
Kronberg, P.P. (1994), {\it Reports on Progress in Physics} {\bf 57}, 325\\
Kronberg, P.P. (1995), {\it Nature} {\bf 374}, 404\\
Kronberg, P.P., Lesch, H. and Hopp, U. (1998), {\it American
  Astron. Soc. Meeting} {\bf 193}, 85.05\\
Kronberg, P.P. and Perry, J.J. (1982), {\it Astrophys. J.} {\bf 263}, 518\\
Kronberg, P.P., Perry, J.J. and Zukowski, E.L.H. (1992),
  {\it Astrophys. J.} {\bf 387}, 528\\
Kronberg, P.P. and Simard-Normandin, M. (1976), {\it Nature} {\bf 263}, 653\\
Kroupa, P. (1997), {\it New Astronomy} {\bf 2}, 139\\
Kuhn, J.R. and Miller, R.H. (1989), {\it Astrophys. J.} {\bf 341}, L41\\
Kuijken, K. (1997), in M. Persic and P. Salucci (Eds.), {\it Dark
  and visible matter in Galaxies},
  ASP Conference Series {\bf 117}\\
Kulessa, A.S. and Lynden-Bell, D. (1992), {\it Mon. Not. Roy.
  Ast. Soc.} {\bf 225}, 105\\
Kulsrud, R. (1986), Proc. Joint Varenna-Abastumani Int. School \&
  Workshop on Plasma Astrophys.\\
Kulsrud, R. and Anderson, S.W. (1992), {\it Astrophys. J.} {\bf 396}, 606\\
Kulsrud, R.M., Cowley, S.C., Gruzinov, A.V. and Sudan, R.N. (1997),
  {\it Phys. Rep.} {\bf 283}, 213\\
Lacey, C.G. and Cole, S. (1993), {\it Mon. Not. Roy. Ast. Soc.} {\bf 262}, 627\\
Lake, G. and Schommer, R.A. (1984), {\it Astrophys. J.} {\bf 279}, L19\\
Landy, S.D. et al (1996), {\it Astrophys. J.} {\bf 456}, L1\\
Larson, R.B. (1974), {\it Mon. Not. Roy. Ast. Soc.} {\bf 169}, 229\\
Larson, R.B. (1976), {\it Mon. Not. Roy. Ast. Soc.} {\bf 176}, 31\\
Lee, S., Olinto, A. and Siegl, G. (1995), {\it Astrophys. J. Lett.} {\bf
  455}, L21\\
Lemoine, D. and Lemoine, M. (1995), {\it Phys. Rev. D} {\bf 52}, 1995\\
Lequeux, J., Allen, R.J. and Guilloteau, S. (1993),
  {\it Astron. Astrophys.} {\bf 280}, L23\\
Lesch, H. and Birk, G.T. (1998), {\it Phys. Plasmas} {\bf 5}, 2773\\
Lesch, H. and Chiba, M. (1995), {\it Astron. Astrophys.} {\bf 297}, 305\\
Lesch, H. and Chiba, M. (1997), {\it Fundamentals of Cosmic Physics} {\bf 18}, 273\\
Lifshitz, E.M. (1946), {\it J. Phys. USSR} {\bf 10}, 116\\
Lindner, U. et al. (1996), {\it Astron. Astrophys.} {\bf 314}, 1\\
Lisenfeld, U. (2000), Personal communication\\
Lisenfeld, U., Alexander, P., Pooley, G.G. and Wilding, T. (1996),
  {\it Mon. Not. Roy. Ast. Soc.} {\bf 281}, 301\\
Little, B. and Tremaine, S. (1991), {\it Astrophys. J.} {\bf 320}, 493\\
Loewenstein, M. and White, R. (1999), astro-ph/9901242\\
L\'opez-Corredoira, M., Beckman, J.E., Casuso, E. (1999), to be
  published in Astron. Astrophys., astro-ph/9909389\\
Loveday, J., Peterson, B.A., Efstathiou, G. and Maddox, S.J. (1992),
  {\it Astrophys. J.} {\bf 390}, 338\\
Lynden-Bell, D. (1983), {\it Kinematics, Dynamics and Structure of the
  Milky Way}, Reidel Pub. Co. Dordrecht\\
Lynden-Bell, D., Cannon, R.D. and Godwin, P.J. (1983),
  {\it Mon. Not. Roy. Ast. Soc.} {\bf 204}, 87\\
Magueijo, J.C.R. (1994), {\it Phys. Rev. D} {\bf 49}, 671\\
Maller, A.H., Simard, L., Guhathakurta, P., Hjorth, J., Jaunsen, A.O.,
  Flores, R.A. and Primack, J.R. (1999), astro-ph/9910207\\
Mamon, G.A. (1995), in O.G. Richter and K. Borne (Eds.), {\it Groups
  of galaxies}, ASP Conf. Ser. {\bf 70}, p. 83\\ 
van der Marel, R.P., Binney, J.J. and Davies, R.L. (1990),
  {\it Mon. Not. Roy. Ast. Soc.} {\bf 245}, 582\\
Maoz, D. and Rix, H.W. (1993), {\it Astrophys. J.} {\bf 416}, 425\\
Mart\'{\i}nez-Delgado, D. (1999), Ph. D. Thesis, Univ. de La Laguna\\
Marzke, R.O., Huchra, J.P. and Geller, M.J. (1994), {\it Astrophys. J.} {\bf
  428}, 43\\
Mateo, M. (1994), in G. Meyan and P. Prugnid (Eds.), {\it Dwarf Galaxies},
  Proc.  of the ESO/OHP Workshop, Garching, 309\\
Mateo, M. (1997), in M. Arnaboldi, G.S. da Costa and P. Saha (Eds.), {\it The 
  nature of elliptical galaxies}, 259. ASP Conf. Series
  {\bf 116}\\
Mateo, M. et al. (1992), in B. Barbury and A. Renzini (Eds.), {\it The 
  stellar population of galaxies},
  IAU {\bf 149}, Kluwer\\
Matese, J.J. and O'Connell, R.F. (1970), {\it Astrophys. J.} {\bf 160}, 451\\
Mathewson, D.S., Ford, V.L. and Buckhorn, M. 1992,
  {\it Astrophys. J. Suppl. Ser.} {\bf 81}, 413\\
Matsuda, T., Sato, H. and Takeda, H. (1971), {\it Pub. Astr. Soc. Japan}
  {\bf 43}, 1\\
Matsushita, K. (1997), Ph. D. Thesis. University of Kyoto\\
Matteucci, F. (1992), {\it Astrophys. J.} {\bf 397}, 32\\
McGaugh, S.S. and de Block, W.J.G. (1993), {\it Astron. J.} {\bf 103}, 548\\
Merrifield, M.R. (1992), {\it Astron. J.} {\bf 103}, 1552\\
Meurer, G.R., Staveley-Smith, L. and Killeen, N.E.B. (1998),
  {\it Mon. Not. Roy. Ast. Soc.} {\bf 300}, 705\\
Milgrom, M. (1983a), {\it Astrophys. J.} {\bf 270}, 365\\
Milgrom, M. (1983b), {\it Astrophys. J.} {\bf 270}, 371\\
Milgrom, M. (1983c), {\it Astrophys. J.} {\bf 270}, 384\\
Milgrom, M. (1995), {\it Astrophys. J.} {\bf 455}, 439\\
Milgrom, M. (1997), {\it Astrophys. J.} {\bf 478}, 7\\
Milgron, M. (1999), astro-ph/9805346\\
Mishustin, I.N. and Ruzmaikin, A.A. (1973), {\it Sov. Phys. JETP} {\bf 34}, 233\\
Moore, B., Ghigna, S., Governato, F., Lake, G., Quinn, T., Stadel,
  J. and Tozzi, P. (1999), {\it to be published in Astrophys. J.}, astro-ph/9907411\\
More, B. (1996), {\it Astrophys. J.} {\bf 461}, L13\\
Mould, J.R., Oke, J.B., de Zeeuw, P.T. and Nemec, J.M. (1990),
  {\it Astron. J.} {\bf 99}, 1823\\
Mushotzky, R.F., Loewenstein, M., Awaki, H., Makishima, K.,
  Matsushita, K. and Matsumoto, H. (1994), {\it Astrophys. J.} {\bf 436}, L79\\
Navarro, J.F. (1998), astro-ph/9807084\\
Navarro, J.F., Frenk, C.S. and White, S.D.M. (1996),
  {\it Mon. Not. Roy. Ast. Soc.} {\bf 275}, 56\\
Navarro, J.F., Frenk, C.S. and White, S.D.M. (1997),
  {\it Astrophys. J.} {\bf 462}, 563\\
Navarro, J.F. and Steinmetz, M. (1999), astro-ph/9908114\\
Navarro, J.F. and White, S.D.M. (1993),
  {\it Mon. Not. Roy. Ast. Soc.} {\bf 262}, 271\\
Nelson, A.H. (1988), {\it Mon. Not. Roy. Ast. Soc.} {\bf 233}, 115\\
Nelson, R.W. and Tremaine, S. (1995), {\it Mon. Not. Roy. Ast. Soc.} {\bf
  275}, 897\\
New, K.C.B., Tohline, J.E., Frank, J. and V\"{a}th, H.M. (1998),
  {\it Astrophys. J.} {\bf 503}, 632\\
Olesen, P. (1997), in {\it NATO advanced research workshop on Theoretical
  Physics}, Zakopane, Poland\\
Olling, R.R. (1996), {\it Astron. J.} {\bf 112}, 480\\
Olling, R.P. and Merrifield, M.R. (1997), in D. Zaritzky (Ed.), {\it
  Galaxy Halos}, ASP {\bf 136}\\
Oort, J.H. (1960), {\it Bull. Astron. Inst. Netherland} {\bf 15}, 45\\
Oort, J.H. (1965), in A. Blaauw and M. Schmidt (Eds.), {\it Galactic structure},
  Univ. Chicago Press, p. 455\\
Ostriker, J.P. (1993), {\it Ann. Rev. Astron. Astrophys.} {\bf 31}, 689\\
Ostriker, J.P. and Peebles, P.J.E. (1973), {\it Astrophys. J.} {\bf 186}, 467\\
Pacholczyk, A.G. (1970), {\it Radio Astrophysics}, W.H. Freeman Co., San
  Francisco\\
de Paolis, F., Ingrosso, G., Jetzer, Ph. and Roucadelli, M. (1995),
  {\it Phys. Rev. Lett.} {\bf 74}, 14\\
de Paolis, F., Ingrosso, G., Jetzer, Ph. and Roucadelli, M. (1997), in
  M. Persic and P. Salucci (Eds.), {\it Dark
  and visible matter in Galaxies},
  ASP Conference Series {\bf 117}, 240\\
de Paolis, F., Ingrosso, G. and Strafella, F. (1995), {\it Astrophys. J.} {\bf
  438}, 83\\
Peebles, P.J.E. (1965), {\it Astrophys. J.} {\bf 142}, 1317\\
Peebles. P.J.E. (1974), {\it Astrophys. J.} {\bf 189}, L54\\
Peebles, P.J.E. (1982), {\it Astrophys. J.} {\bf 263}, L1\\
Perley, R.A. and Taylor, G.B. (1991), {\it Astron. J.} {\bf 101}, 1623\\
Perlmutter, S., Turner, M.S. and White, M. (1999), astro-ph/9901052\\
Perlmutter, S. et al. (1999), {\it Astrophys. J.} {\bf 517}, 565\\
Perea, J. et al. (1999), in M. Valtonen and C. Flynn (Eds.), {\it
  Small galaxy Groups}, ASP Conf. Ser.\\
Persic, M. and Salucci, P. (1988), {\it Mon. Not. Roy. Ast. Soc.} {\bf 234},
  131\\
Persic, M. and Salucci, P. (1990), {\it Mon. Not. Roy. Ast. Soc.} {\bf 245},
  577\\
Persic, M. and Salucci, P. (1993), {\it Mon. Not. Roy. Ast. Soc.} {\bf 261}, L21\\
Persic, M. and Salucci, P. (1995), {\it Astrophys. J. Supp. Ser.} {\bf 99}, 501\\
Persic, M., Salucci, P. and Stel, F. (1996),
  {\it Mon. Not. Roy. Ast. Soc.} {\bf 281}, 27\\
Peterson, R.C. and Latham, D.W. (1989), {\it Astrophys. J.} {\bf 336}, 178\\
Petrovskaya, I.V. (1987), {\it Pub. Astron. Inst. Czechoslovak,
  Acad. Sciences} {\bf 69}, 307\\
Pfenniger, D. (1997), in M. Persic and P. Salucci (Eds.), {\it Dark
  and visible matter in Galaxies},
  ASP Conference Series {\bf 117}\\
Pfenniger, D. and Combes, F. (1994), {\it Astron. Astrophys.} {\bf 285}, 94\\
Pfenniger, D., Combes, F. and Martinet, L. (1994),
  {\it Astron. Astrophys.} {\bf 285}, 79\\
Piddington, J.H. (1969), {\it Cosmic Electrodynamics}, Wiley Interscience,
  New York\\
Pignatelli, E. and Galletta, G. (1997), in M. Persic and P. Salucci (Eds.), {\it Dark
  and visible matter in Galaxies}, ASP Conference Series {\bf 117}\\
Plaga, R. (1995), {\it Nature} {\bf 374}, 430\\
Ponman, T.J. and Bertram, D. (1993), {\it Nature} {\bf 363}, 51\\
Porcel, C., Battaner, E. and Jim\'enez-Vicente, J. (1997),
  {\it Astron. Astrophys.} {\bf 322}, 103\\
Postman, M., Lauer, T.R., Szapudi, I. and Oegerle, W. (1998),
  {\it Astrophys. J.} {\bf 506}, 33\\
Press, W.H. and Schechter, P. (1974), {\it Astrophys. J.} {\bf 187}, 425\\
Pryor, C.T. (1998), in D.R. Merritt, M. Valluri and J.A. Sellwood
  (Eds.), {\it Galaxy Dynamics}, Rutgers Univ. Pub.\\Qin, B., Wu, X.P. and Zou, Z.L. (1995), {\it Astron. Astrophys.} {\bf 296}, 264\\
Quashnock, J.M., Loeb, A. and Spergel, D.N. (1989), {\it Astrophys. J.} {\bf
  344}, L49\\
Quashnock, J.M., van den Berk, D.E. and York, D.G. (1996),
  {\it Astrophys. J.} {\bf 472}, L69\\
Ratra, B. (1992), {\it Astrophys. J.} {\bf 391}, L1\\
Rausher et al. (1997), in M. Persic and P. Salucci (Eds.), {\it Dark
  and visible matter in Galaxies},
  ASP Conference Series {\bf 117}\\ 
Rees, M.J. (1987), {\it Q. Jl. R. astr. Soc.} {\bf 28}, 197\\
Rees, M.J. and Reinhardt, M. (1972), {\it Astron. Astrophys.} {\bf 19}, 104\\
Retzlaff, J., Borgani, S., Gottloeber, S. and Mueller, V. (1998),
  astro-ph/9709044\\
Reuter, H.P., Klein, U., Lesch, H., Wielebinski, R. and Kronberg,
  P.P. (1992), {\it Astron. Astrophys.} {\bf 256}, 10\\
Richstone, D.O. and Tremaine, S. (1986), {\it Astron. J.} {\bf 92}, 72\\
del Rio, M.S., Brinks, E., Carral, P. and Cepa, J. (1999), in J. Cepa
  and P. Carral (Eds.), {\it Star formation in early-type galaxies}, ASP
  {\bf 163}, 95\\
Rix, H.W. and Zaritsky, D. (1995), {\it Astrophys. J.} {\bf 447}, 82\\
Rodrigo-Blanco, C. and P\'erez-Mercader, J. (1998),
  {\it Astron. Astrophys.} {\bf 330}, 474\\
Rubin, V.C., Burstein, D., Ford, W.K. and Thonnard, N. (1985),
  {\it Astrophys. J.} {\bf 289}, 81\\
Rubin, V.C., Ford, W.K. and Thonnard, N. (1980), {\it Astrophys. J.} {\bf 238}, 471\\
Rubin, V.C., Waterman, A.H. and Kenney, J.D.P. (1999), astro-ph/9904050\\
Ruzmaikin, A.A., Shukurov, A.M. and Sokoloff, D.D. (1988), {\it Magnetic
  Fields of Galaxies}, Kluwer Ac. Press. Dordrecht\\
Ruzmaikin, A.A., Sokoloff, D. and Shukurov, A.M. (1989),
  {\it Mon. Not. Roy. Ast. Soc.} {\bf 241}, 1\\
Sackett, P.D. (1999) in D. Merrit, J.A. Sellwood and M. Valluri
  (Eds.), {\it Galaxy Dynamics}, ASP\\
Sackett, P.D., Morrison, H.L., Harding, P. and Boronson, T.A. (1994),
  {\it Nature} {\bf 370}, 441\\
Saglia, R.P., Bertin, G. and Stiavelli, M. (1992), {\it Astrophys. J.} {\bf 384},
  433\\
Saglia, R.P. et al (1993), {\it Astrophys. J.} {\bf 403}, 567\\
Salucci, P. and Frenk, C.S. (1989), {\it Mon. Not. Roy. Ast. Soc.} {\bf 237}, 247\\
Salucci, P. and Persic, M. (1997), in M. Persic and P. Salucci (Eds.), {\it Dark
  and visible matter in Galaxies},
  ASP Conference Series {\bf 117}, 1\\
Salvador-Sol\'e, E., Solanes, J.M. and Manrique, A. (1998),
  {\it Astrophys. J.} {\bf 499}, 542\\
Sanchez-Salcedo, F.J. (1996), {\it Astrophys. J.} L21\\
Sandage, A. (1961), {\it The Hubble Atlas of Galaxies}, Carnegie Institution
  of Washington\\
Sanders, R.H. (1990), {\it Astron. Astrophys. Rev.} {\bf 2}, 1\\
Sanders, R.H. (1996), {\it Astrophys. J.} {\bf 473}, 117\\
Sanders, R.H. (1998), {\it Mon. Not. Roy. Ast. Soc.} {\bf 296}, 1009\\
Sanders, R.H. and Verheijen, M.A.W. (1998), {\it Astrophys. J.} {\bf 503}, 97\\
Sanders, R.H. (1999), {\it Astrophys. J.} {\bf 512}, L23\\
Schmidt, M. (1965), in A. Blaauw and M. Schmidt (Eds.), {\it Galactic Structure},
  p. 513, Chicago Univ. Press\\
Schramm, D.N. (1992), {\it Nuclear Phys. B}, {\bf 28A}, 243\\
Schweizer, F., van Gorkom, J.H. and Seitzer, P. (1989),
  {\it Astrophys. J.} {\bf 338}, 770\\
Schweizer, F., Whitmore, B.C. and Rubin, V.C. (1983), {\it Astron. J.} {\bf 88}, 909\\
Schweizer, L.Y. (1987), {\it Astrophys. J. Suppl.} {\bf 64}, 427\\
Sellwood, J.A. (1985), {\it Mon. Not. Roy. Ast. Soc.} {\bf 217}, 127\\
Sensui, T., Funato, Y. and Makino, J. (1999), astro-ph/9906263\\
Shapiro, P.R. and Field, G.B. (1976), {\it Astrophys. J.} {\bf 205}, 762\\
Shapiro, S.L. and Wasserman, I. (1981), {\it Nature} {\bf 289}, 657\\
Sharp, N.A. (1990), {\it Astrophys. J.} {\bf 354}, 418\\
Shu, F.H. (1982), {\it The Physical Universe: An Introduction to
  Astronomy}, Univ. Science Books, Mill Valley, California\\
Silk, J. (1968), {\it Astrophys. J.} {\bf 151}, 459\\
Sofue, Y. (1999) in F. Combes, G.A. Mamon and V. Charmandaris (Eds.),
  {\it Galaxy Dynamics: from the early Universe to the
  Present}, ASP Conference Series\\
Sofue, Y., Tomita, A., Honma, M. and Tutui, Y. (1999), astro-ph/9910105\\
Sofue, Y., Tutui, Y., Honma, M., Tomita, A., Tokamiya, T., Koda,
  J. and Takeda, Y. (1999), {\it Astrophys. J.} {\bf 523}, 136\\
Sofue, Y., Wakamatsu, K. and Malin, D.F. (1994), {\it Astron. J.} {\bf 108}, 2102\\
Sokoloff, D. and Shukurov, A.M. (1990), {\it Nature} {\bf 347}, 51\\
Soneira, R.M. and Peebles, P.J.E. (1977), {\it Astrophys. J.} {\bf 211}, 1\\
Sparke, L. and Casertano, S. (1998), {\it Mon. Not. Roy. Ast. Soc.} {\bf 234}, 873\\
Stanev, T. (1997), {\it Astrophys. J.} {\bf 479}, 290\\
Stanev, T., Biermann, P.L., Lloyd-Evans, J., Rachen, J.P. and Watson,
  A. (1995), {\it Phys. Rev. Lett.} {\bf 75}, 3056\\
Steidel, C., Giavalisco, M., Pettini, M., Dickinson, M. and
  Adelberger, K. (1996), {\it Astrophys. J.} {\bf 462}, L17\\
Steinmetz, M. (1999), astro-ph/9910002\\
Stewart, G.C., Ca\~nizares, C.R., Fabian, A.C. and Nulsen,
  P.E.J. (1984), {\it Astrophys. J.} {\bf 278}, 536\\
Stiavelli, M., Sparke, L.S. (1991), {\it Astrophys. J.} {\bf 382}, 466\\
Subramanian, K., Cen, R. and Ostriker, P. (1999), astro-ph/9909279\\
Summers, F.J., Davis, M. and Evrard, A.E. (1995), {\it Astrophys. J.} {\bf
  454}, 1\\
Sunyaev, R.A. and Zel'dovich, Ya.B. (1972), {\it Astron. Astrophys.} {\bf 20}, 189\\
Sutherland, W. et al. (1999), {\it Mon. Not. Roy. Ast. Soc.} {\bf 308}, 289\\
Swaters, R. (1999), Ph. D. Thesis, Univ. of Groningen\\
Swaters, R.A., Madore, B.F. and Trewhella, M. (2000), to be published
  in Astrophys. J. Lett.\\
Swaters, R.A., Sancisi, R., van Albada, T.S. and van der Hulst,
  J.M. (1998), {\it American Astron. Soc. Meeting} {\bf 193}\\
Sylos-Labini, F., Montuori, M. and Pietronero, L. (1998),
  {\it Phys. Rep.} {\bf 293}, 66\\
Tadros, H., Efstathiou, G. and Dalton, G. (1998),
  {\it Mon. Not. Roy. Ast. Soc.} {\bf 296}, 995\\
The, L.S. and White, S.D.M. (1988), {\it Astron. J.} {\bf 95}, 1642\\
Toomet, O., Andernach, H., Einasto, J., Einasto, M., Kasak, E.,
  Starobinsky, A.A. and Tago, E. (1999), astro-ph/9907238\\
Trimble, V. (1987), {\it Ann. Rev. Astron. Astrophys.} {\bf 25}, 425\\
Trinchieri, G., Fabbiano, G. and Canizares, C.R. (1986),
  {\it Astrophys. J.} {\bf 310}, 637\\
Tucker, D.L. et al (1997), {\it Mon. Not. Roy. Ast. Soc.} {\bf 285}, L5\\
Tully, R.B. and Fisher, J.R. (1978), in M.S. Longair and J. Einasto
  (Eds.), {\it The Large Scale Structure of the Universe}, IAU Symp. {\bf 79}, 31\\
Tully, R.B., Scaramella, R., Vettolani, G. and Zamorani, G. (1992),
  {\it Astrophys. J.} {\bf 388}, 9\\
Turner, M.S. (1999a), astro-ph/9901168\\
Turner, M.S. (1999b), astro-ph/9904051\\
Turner, M.S. and Widrow, L.M. (1988), {\it Phys. Rev. Lett.} {\bf 67}, 1057\\
Vachaspati, T. (1989), {\it Phys. Lett. B} {\bf 265}, 258\\
Valentijn, E.A. (1991), in H. Bloemen (Ed.), {\it The interstellar disk-halo connection in
  galaxies}, IAU, Dordrecht\\ 
Valentijn, E.A. and van der Werf, P.P. (1999), {\it Astrophys. J.} {\bf
  522}, L29\\
Vall\'ee, J.P. (1990), {\it Astrophys. J.} {\bf 360}, 1\\
Vall\'ee, J.P. (1990), {\it Astron. J.} {\bf 99}, 459\\
Vall\'ee, J.P. (1991), {\it Astron. Astrophys.} {\bf 251}, 411\\
Vall\'ee, J.P. (1991), {\it Astrophys. Space Sci.} {\bf 178}, 41\\
Vall\'ee, J.P. (1993), {\it Astrophys. J. Lett.} {\bf 23}, 87\\
Vall\'ee, J.P. (1994), {\it Astrophys. J.} {\bf 437}, 179\\
Vall\'ee, J.P. (1997), {\it Fundamentals of Cosmic Physics} {\bf 19}, 1-89\\
Vall\'ee, J.P., Mac Leod, J.M. and Broten, N.W. (1987),
  {\it Astrophys. Lett.} {\bf 25}, 181\\
van Albada, T.S., Bahcall, J.N., Begeman, K. and Sancisi, R. (1985),
  {\it Astrophys. J.} {\bf 295}, 305\\
van Albada, T.S. and Sancisi, R. (1986), {\it Phil. Trans. Roy. Soc. London}
  {\bf A320}, 447\\
van den Bergh, S. (1999), astro-ph/9904251, to be published in PASP\\
van den Bergh, S. (2000), to be published in ARAA \\
van den Bosch, F.C. (1999), astro-ph/9909298. To be published  in Astrophys. J.\\
van Driel, W. and Combes, F. (1997), in M. Persic and P. Salucci (Eds.), {\it Dark
  and visible matter in Galaxies}, ASP Conference Series {\bf 117}, 133\\
van Driel, W. and van Woerden, H. (1991), {\it Astron. Astrophys.} {\bf 243}, 71\\
van Driel, W. and van Woerden, H. (1997), in M. Persic and P. Salucci (Eds.), {\it Dark
  and visible matter in Galaxies}, ASP Conference Series {\bf 117}\\
van Driel et al. (1995), {\it Astron. J.} {\bf 109}, 942\\
van Gorkom, J.H. (1992), in J.M. Shull and H.A. Thronson (Eds.), {\it The environment and evolution of
  galaxies}, Kluwer, p. 345\\
van der Kruit, P.C. (1979), {\it Astron. Astrophys. Suppl.} {\bf 38}, 15\\
van der Kruit, P.C. and Searle, L. (1981a), {\it Astron. Astrophys.} {\bf
  95}, 105\\
van der Kruit, P.C. and Searle, L. (1981b), {\it Astron. Astrophys.} {\bf
  95}, 116\\
van der Kruit, P.C. and Searle, L. (1982a), {\it Astron. Astrophys.} {\bf
  110}, 61\\
van der Kruit, P.C. and Searle, L. (1982b), {\it Astron. Astrophys.} {\bf
  110}, 79\\
van Moorsel, G.A. (1987), {\it Astron. Astrophys.} {\bf 176}, 13\\
de Vaucouleurs, G. (1975), in A.R. Sandage, M. Sandage and J. Kristian
  (Eds.), {\it Stars and stellar systems}, Chicago Univ. Press, p. 14\\
de Vaucouleurs, G. and Stiavelli, M. (1993), {\it Rep. Prog. Phys.} {\bf 56}, 493\\
Vega-Beltr\'an, J.C. (1997), Ph. D. Thesis, Univ. Of La Laguna\\
Veneziano, G. (1991), {\it Phys. Lett. B} {\bf 265}, 287\\
Verdes-Montenegro, L., Yun, M. and del Olmo, A. (1997), in M. Persic and P. Salucci (Eds.), {\it Dark
  and visible matter in Galaxies}, ASP Conference Series {\bf 117}\\
Verdes-Montenegro et al. (1999), in M. Valtonen and C. Flynn (Eds.),
  {\it Small galaxy Groups}, ASP Conf. Ser.\\
Verheijen, M. (1997), Ph. D. Thesis, Univ. Groningen\\
Vogt, S.S., Mateo, M., Olszewski, E.W., Keane, M.J. (1995),
  {\it Astron. J.} {\bf 109}, 151\\
Wasserman, I. (1978), {\it Astrophys. J.} {\bf 224}, 337\\
Watson, A.M. and Perry, J.J. (1991), {\it Mon. Not. Roy. Ast. Soc.} {\bf
  248}, 58\\
Waxman, E. and Miralda-Escud\'e, J. (1995), {\it Astrophys. J.} {\bf 451}, 451\\
Wevers, B.M.H.R., van der Kruit, P.C. and Allen, R.J. 1986,
  {\it Astron. Astrophys. Suppl. Ser.} {\bf 66}, 505\\
Weinberg, S. (1972), {\it Gravitation and Cosmology}, John Wiley \& Sons,
  Inc. New York\\
White, S.D.M. and Frenk, C.S. (1991), {\it Astrophys. J.} {\bf 379}, 52\\
White, S.D.M., Huchra, J., Latham, D. and Davis, M. (1983),
  {\it Mon. Not. Roy. Ast. Soc.} {\bf 203}, 701\\
White, S.D.M. and Rees, M.J. (1978), {\it Mon. Not. Roy. Ast. Soc.} {\bf 183}, 341\\
Whitemore, B.C., Forbes, D.A. and Rubin, V.C. (1988),
  {\it Astrophys. J.} {\bf 333}, 542\\
Wielebinski, R. (1993), in F. Krause et al. (Eds.), {\it The Cosmic
  Dynamo}, p. 271\\
Wielebinski, R. and Krause, F. (1993), {\it Astron. Astrophys. Rev.} {\bf
  4}, 449\\
Wilkinson, M.I. and Evans, N.W. (1999), astro-ph/9906197\\
Wolfe, A.M. (1988), in C.V. Blader et al. (Eds.), {\it QSO Absorption 
  Lines: Probing the Universe}, Cambridge Univ. Press, p. 297\\
Wolfe, A.M., Lanzetta, K.M. and Oren, A.L. (1992), {\it Astrophys. J.} {\bf
  387}, 17\\
Zaritsky, D. (1997), in M. Persic and P. Salucci (Eds.), {\it Dark
  and visible matter in Galaxies}, ASP Conference Series {\bf 117}\\
Zaritsky, D., Olszewski, E.W., Schommer, R.A., Peterson, R.C. and
  Aaronson, M. (1989), {\it Astrophys. J.} {\bf 345}, 759\\
Zaritsky, D., Smith, R., Frenk, C. and White, S.D.M. (1993),
  {\it Astrophys. J.} {\bf 405}, 464\\
Zaritsky, D. and White, S.D.M. (1994), {\it Astrophys. J.} {\bf 435}, 599\\
de Zeeuw, P.T. (1992), in B. Barbuy and A. Renzini
  (Eds.), {\it The stellar population of galaxies}, IAU {\bf 149}\\
Zel'dovich, Ya.B. (1967), {\it Soviet Phys. USPEKHI} {\bf 9}, 602\\
Zel'dovich, Ya.B. (1970), {\it Astron. Astrophys.} {\bf 5}, 84\\
Zeldovich, Ya B. and Novikov, I.D. (1975), {\it The structure and evolution
  of the Universe}, Univ. Press, Chicago\\
Zucca, E. et al. (1997), {\it Astron. Astrophys.} {\bf 326}, 477\\
Zurita, A. and Battaner, E. (1997), {\it Astron. Astrophys.} {\bf 322}, 86\\
Zweibel, B.G. and Heiles, C. (1997), {\it Nature} {\bf 385}, 131\\
Zwicky, F. (1937), {\it Phys. Rev. Lett.} {\bf 51}, 290\\

\newpage

\noindent {\bf Figure captions}

{\bf Figure 1.-} Upper panel: The luminosity profile of NGC 2403
observed by Wevers, van der Kruit and Allen (1986). Lower panel: The
observed rotation curve of NGC 2403 (dots) and the rotation curves of
the individual mass components (lines). From Begeman (1987) PhD thesis.

{\bf Figure 2.-} A position-velocity map along the major axis at a
position angle of 21$^o$.  The
filled circles show the adopted rotation curve of NGC 1560. The cross
in the lower right corner indicates the angular and velocity
resolutions. From Broeils (1992). Courtesy of Astronomy and Astrophysics.

{\bf Figure 3.-} The observed rotation curve and disk-halo models. The
top panel shows a halo added to the maximum disk; the middle panel
shows the ``maximum halo'', with almost no mass in the stellar disk
($M/L_B = 0.1$). The bottom panel shows the ``best fit'' disk-halo
model. The dotted and dashed lines indicate the gas and stellar
disks. The dot-dashed line shows the rotational velocities of the
halo; the full line is the contribution of all the components. From
Broeils (1992). Courtesy of Astronomy and  Astrophysics.

{\bf Figure 4.-} Left: Upper panel: The luminosity profile of NGC 5033
observed by Kent (1986). Lower panel: The observed rotation curve of
NGC 5033 (dots) and the rotation curves of the individual mass
components (lines).
Right: Upper panel: The luminosity profile of NGC 5371 observed by
Wevers, van der Kruit and Allen (1986) and its decomposition into a
bulge and disk (lines). Lower panel: The observed rotation curves of
NGC 5371 (dots) and the rotation curves of the individual mass
components (lines). From Begeman (1987) PhD thesis.

{\bf Figure 5.-} Comparison of stellar and gaseous rotation curves by
Vega-Beltr\'an (1999). Upper window: R-band image. V-window: rotation
velocities. The other windows represent the velocity dispersion and
the third and fourth orders Gauss-Hermite of the line-of-sight
velocity distribution of the stars. From Vega-Beltr\'an (1999) PhD
thesis.

{\bf Figure 6.-} Same as for figure 9. From Vega-Beltr\'an (1999) PhD
thesis.

{\bf Figure 7.-} The Milky Way rotation curve in (a) a logarithmic scale
by Sofue et al. (1999) as compared with the linear scale rotation
curve (b). Courtesy of American Astronomical Society. 

{\bf Figure 8.-} The observed rotation curve of NGC 2460 together
with a maximum-disk+halo model. From Broeils (1992) PhD thesis.

{\bf Figure 9.-} Logarithmic slope between two and three disk scale
lengths $S_{(2.3)}^h$ versus absolute R-band magnitude $M_R$. Filled
circles correspond to late-type galaxies in a high quality rotation
curve sample, open circles represent dwarfs in a lower quality rotation
curve sample, open triangles are galaxies in Verheijen's (1997) Ursa
Major sample, filled triangles represent the galaxies from various
sources presented in Broeils (1992). From Swaters (1999) PhD thesis.

{\bf Figure 10.-} The Tully-Fisher relation for spiral and late-type
dwarf galaxies. Symbol coding as in Fig. 13. From Swaters (1999) PhD 
thesis.

{\bf Figure 11.-} Overall rotation curves of the Galaxy for $\theta_0=
220 km s^{-1}$ (filled circles), $\theta_0 = 200 km s^{-1}$ (open
circles), and $\theta_0 =180 km s^{-1}$ (open triangles). The data for
the inner rotation curve were taken from Fich et al. (1989). The outer
rotation curve are those obtained by Merrifield's method. The error
bars are indicated only for $\theta_0 = 220 km s^{-1}$, and are almost
the same for the three cases. From Honma and Sofue (1997).
Courtesy of the Astronomical Society of Japan.

{\bf Figure 12.-} Observed rotation curve of NGC 404, uncorrected for
inclination; crosses: observations, solid squares: Keplerian
decline. From del R\'{\i}o et al. (1999). Courtesy of the Astronomical
Society of the Pacific.

{\bf Figure 13.-} Different possibilities to understand the negative
radial velocity of M31.

{\bf Figure 14.-} The evolution of Jean's mass as a function of $a$
(the cosmological scale factor taking its present value as unit,
i.e. $a = R/R_0$). Points over the curve correspond to unstable
situations leading to gravitational collapse. An inhomogeneity as the
Milky Way, with a rest mass of about $10^{12} M_\odot$ was unstable
until $a$ slightly greater than $10^{-6}$. Then, it underwent acoustic
oscillations until the epoch of Recombination, when it become unstable
again. Adopted from Battaner (1996).

{\bf Figure 15.-} The evolution of the relative overdensity of an
inhomogeneity cloud as a function of $a= R/R_0$, being $R$ the
cosmological scale factor and $R_0$ its present value. Adopted from
Battaner (1996).

{\bf Figure 16.-} The redshift distribution of galaxies with
magnitudes in the range 22.5 $<$ B $<$ 24.0. The solid histogram shows the
data of Glazebrook et al. (1995), while the dash histogram shows the
(more complete) data of Cowie et al. (1996). The lines show the
predictions of the model of Cole et al. (1994) for a Scalo IMF (solid
line) and a Miller-Scalo IMF (dotted line). From Frenk et
al. (1997). Courtesy of the Astronomical Society of the Pacific.

{\bf Figure 17.-} Present-day B-band luminosity functions in a model
with [$\Omega_0$ =1, $\Lambda_0 =0$, h=0.5, $\sigma_8 = 0.67$] (solid
lines) and another with [$\Omega_0$ =0.3, $\Lambda_0 =0.7$, h=0.6, 
$\sigma_8 = 0.97$] (dotted lines). In each case, the extended curve
shows the luminosity function of the galaxy population as a whole. The
shorter curve shows the luminosity function of galaxies that contained
at least one progenitor satisfying the selection criteria for
Lyman-break galaxies at high redshift in the study of Steidel et
al. (1995). The data points show observational determinations of the
luminosity function from Loveday et al (1992) and Marzke et
al. (1994). From Baugh et al. (1998). Courtesy of the Astronomical
Society of the Pacific.

{\bf Figure 18.-} NFW circular velocity profiles for different values
of the concentration. The circular velocity at the virial radius,
$V_{200}$, is chosen to match the rotation speed in the outskirts of
the galaxy, and varies from 100 to 130 km s$^{-1}$ from top to
bottom. As illustrated in this figure for NGC 3198, most galaxy
rotation curves can be adequately parametrized. The best fit value of
the $c$ parameter ($c_{obs} \approx$26 in this case) provides a
quantitative measure of the shape of the rotation curve. Low values of
$c_{obs}$ denote slowly rising curves, while high values of $c_{obs}$
indicate steeply rising, flat or declining rotation
curves. Furthermore, $c_{obs}$ constitutes a firm upper limit to the
concentration of the halo, since the fit neglects the contribution of
the luminous component. From Navarro (1999).

{\bf Figure 19.-} Top-left: Rotation curve shape characterized by the
parameter $c_{obs}$, the value of the concentration deduced by the
observations of the rotation curve, versus effective surface
brightness ($\Sigma_{eff} = L/\pi R^2_{disk}$). Top-right: $V_{200}$
versus $V_{rot}$, commented in text. In all four plots: dotted lines
assume that $V_{rot} = V_{200}$ and only $(M/L)_{disk}$ is adjusted to
match the Tully-Fisher relation. Short-dashed lines assume that
$(M/L)_{disk} = h M_\odot/L_\odot$ in all galaxies; only $V_{200}$ is
varied to match the Tully-Fisher relation. Solid lines correspond to
varying both $(M/L)_{disk}$ and $V_{200}$ so as to match the
Tully-Fisher relation and the $\Sigma_{eff}-c_{obs}$ correlation. From
Navarro (1999).

{\bf Figure 20.-} Top left: Alfv\'en's speed in km s$^{-1}$ for three
values of the free parameter $k$ ($10^{-9}$, upper curve; $3 \times
10^{-9}$, middle curve; $10^{-8}$cm$^{-1}$s, lower curve). Top right:
Magnetic field strength in gauss. Botton left: The flaring function
Z(r) in kpc, for three different values of the free parameter $k$
defined in the text ($10^{-9}$, upper curve; $3 \times
10^{-9}$, middle curve; $10^{-8}$cm$^{-1}$s, lower curve). Botton
right: Density profiles for the value of the free parameter $k =
3\times 10^{-9}$cm$^{-1}$s, for three different values of z=0, 4 and 8
kpc from the galactic plane. From Battaner and Florido
(1995). Courtesy of the Royal Astronomical Society.

{\bf Figure 21.-} Ideal scheme of the egg-carton universe formed with
octohedra only contacting at their vertexes. Adopted from Battaner and
Florido (1997). Courtesy of Astronomy and Astrophysics.

{\bf Figure 22.-} The two large octahedra closer to the Milky Way.

{\bf Figure 23.-} Ideal scheme of the fractal geometry of the
octahedron network. In this figure we plot the case of a fractal
dimension equal to 1.77. A value of 2 is also an interesting
possibility. Adopted from Battaner (1998). Courtesy of Astronomy 
and Astrophysics.

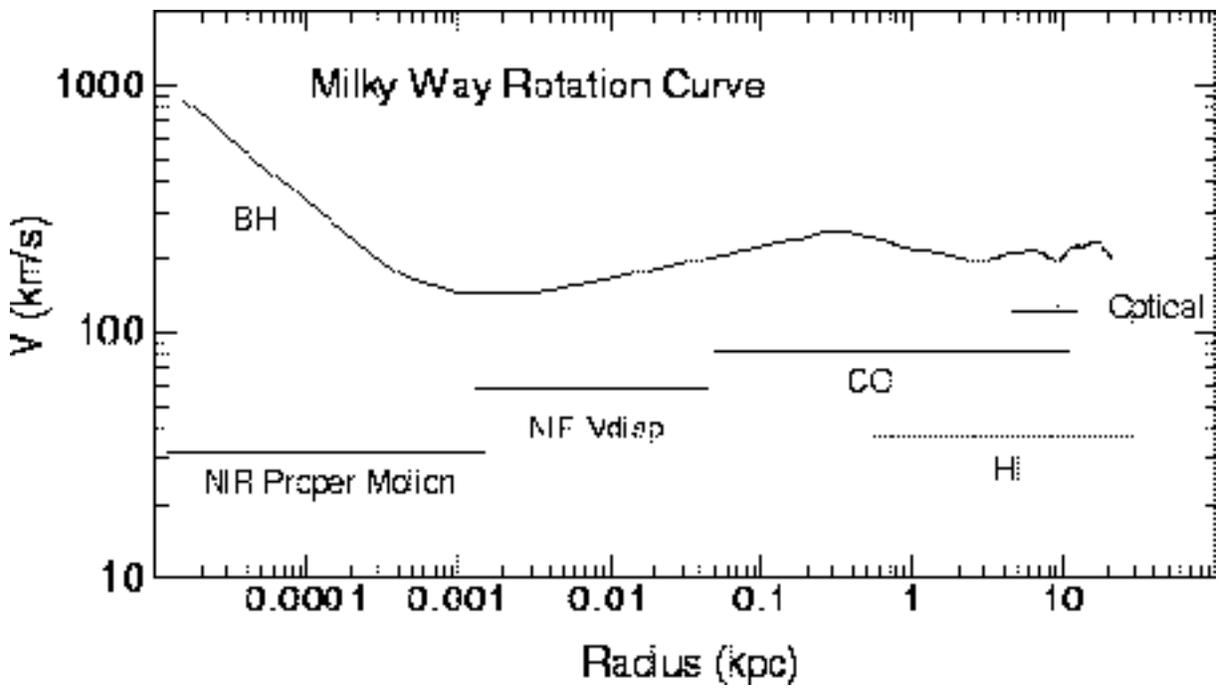
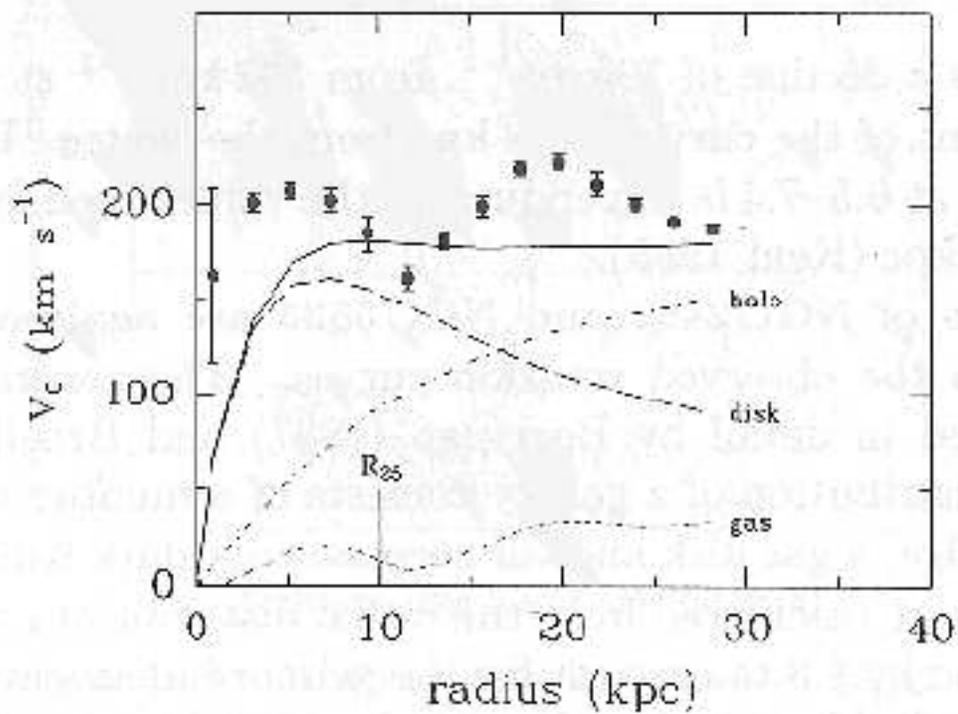
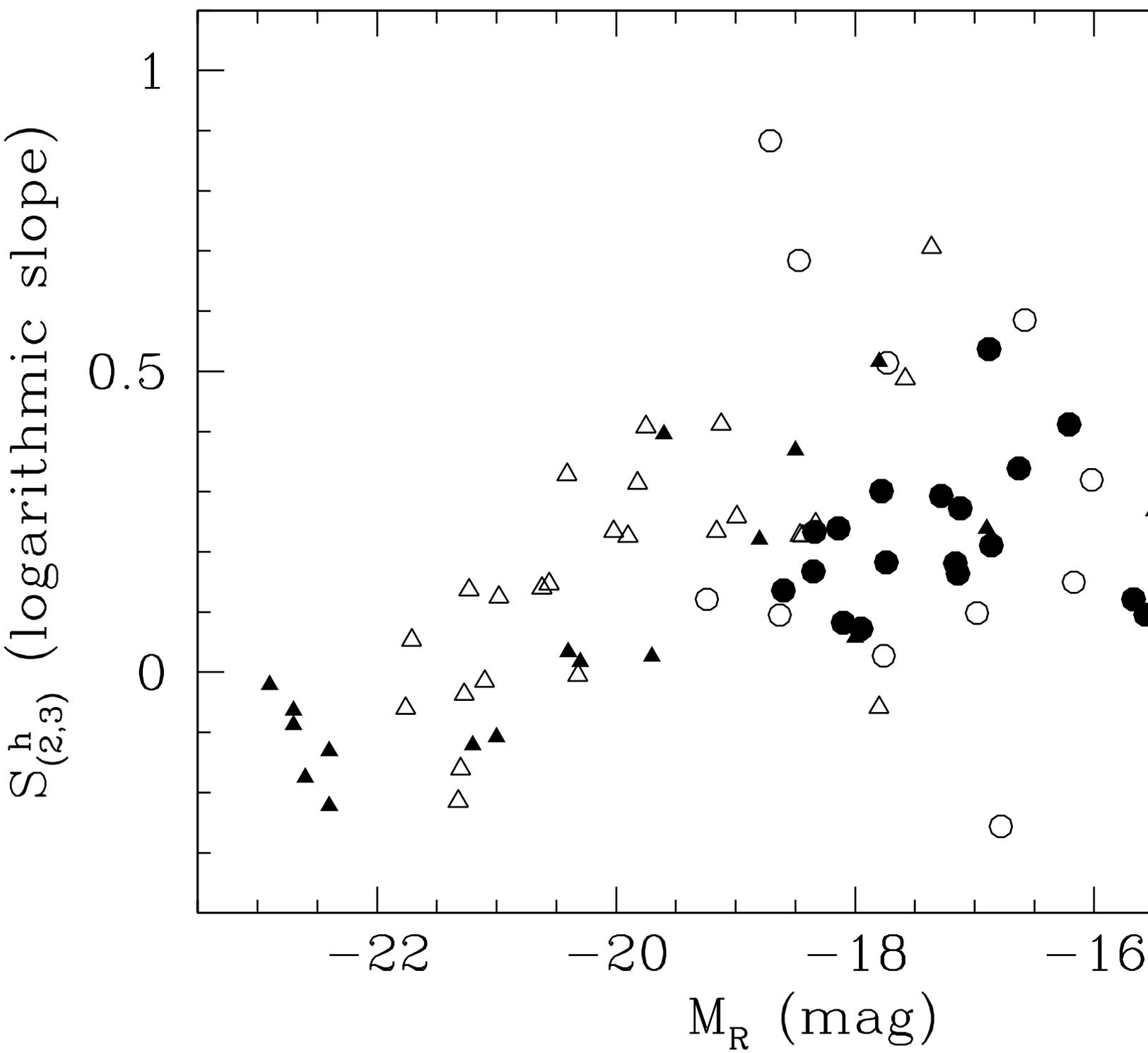
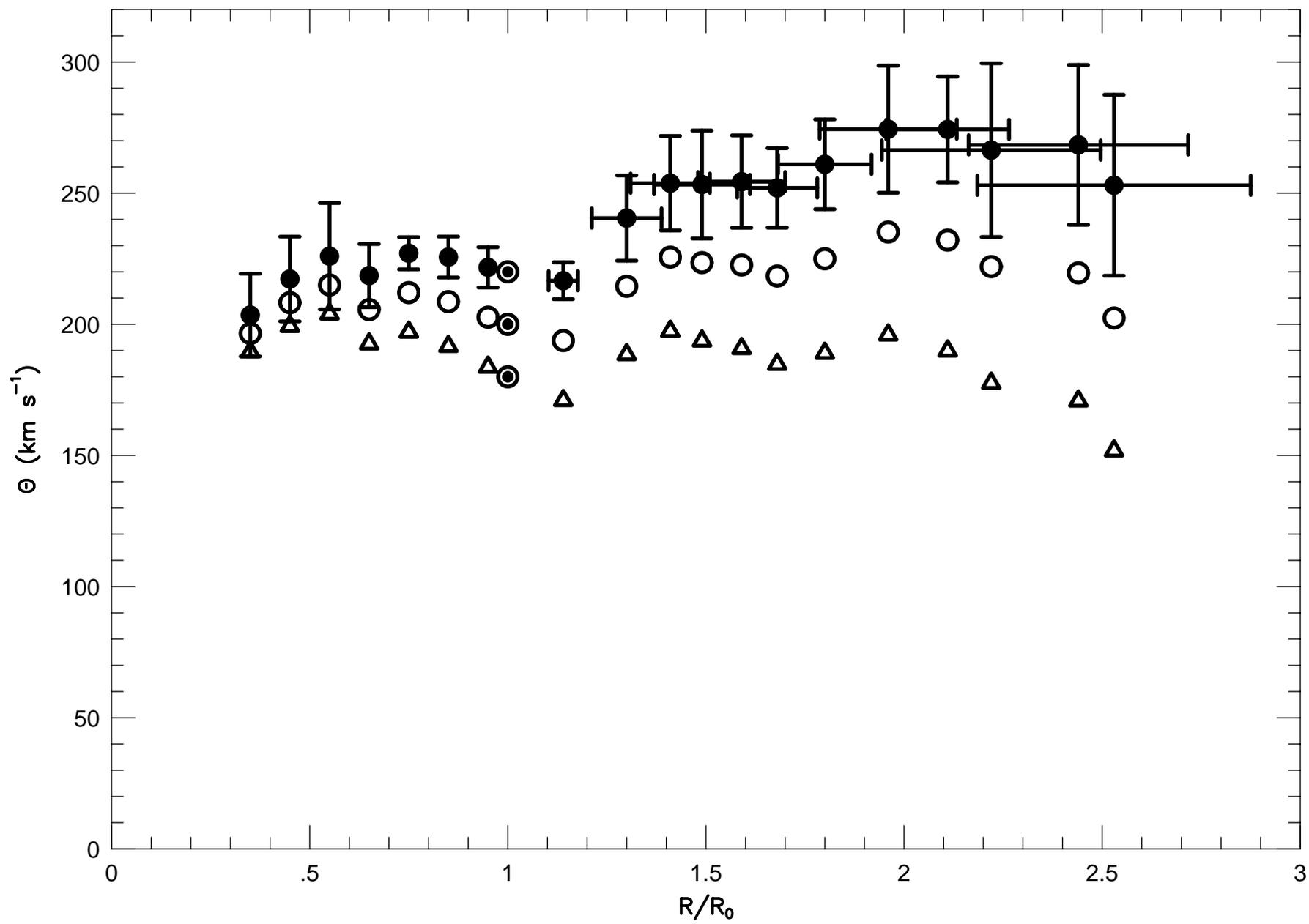
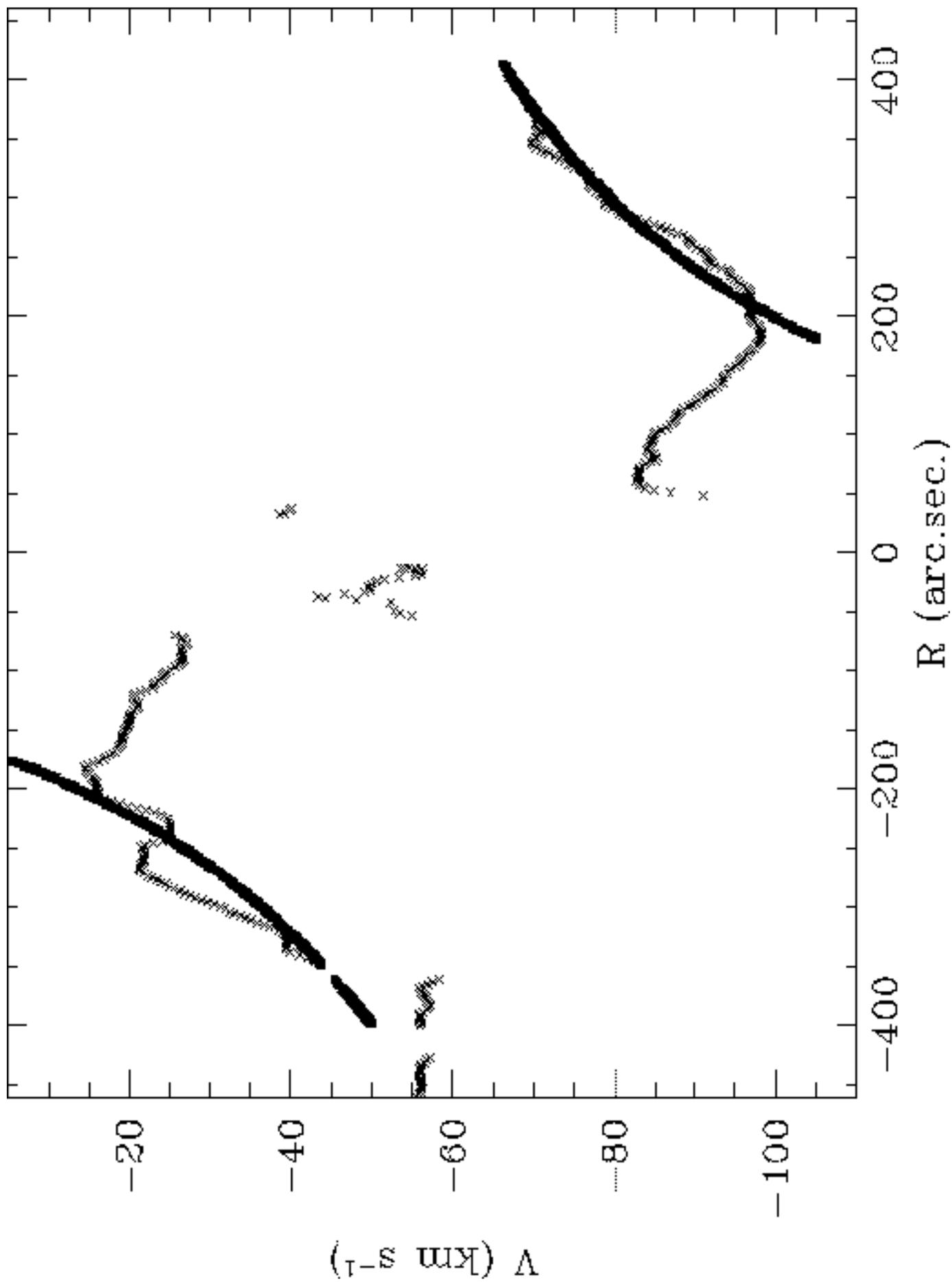
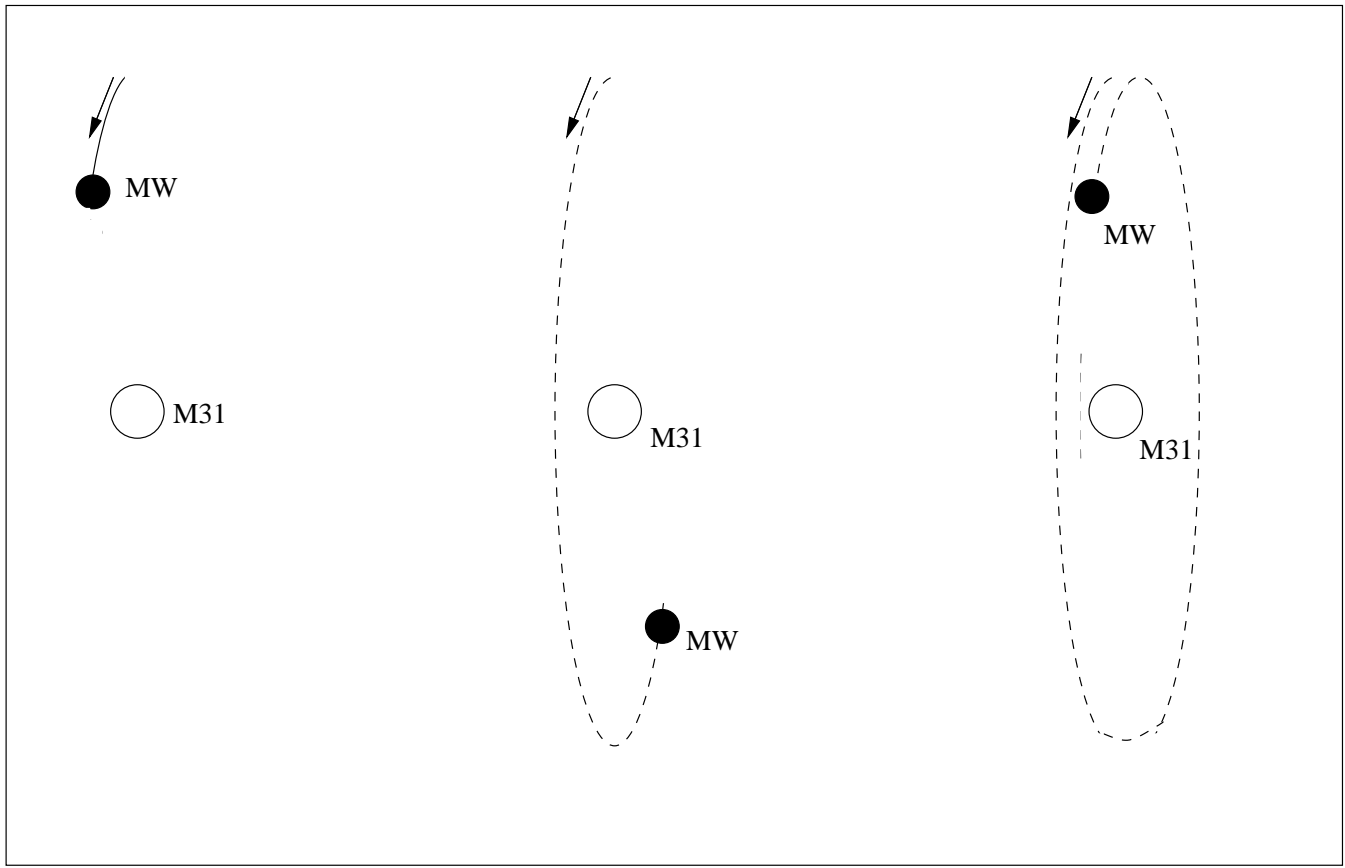
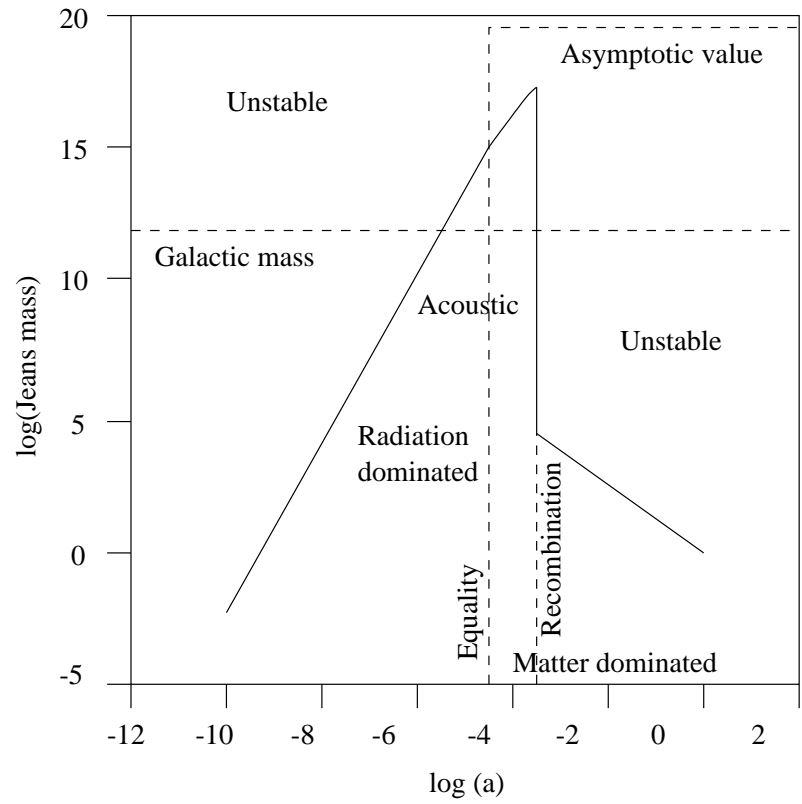
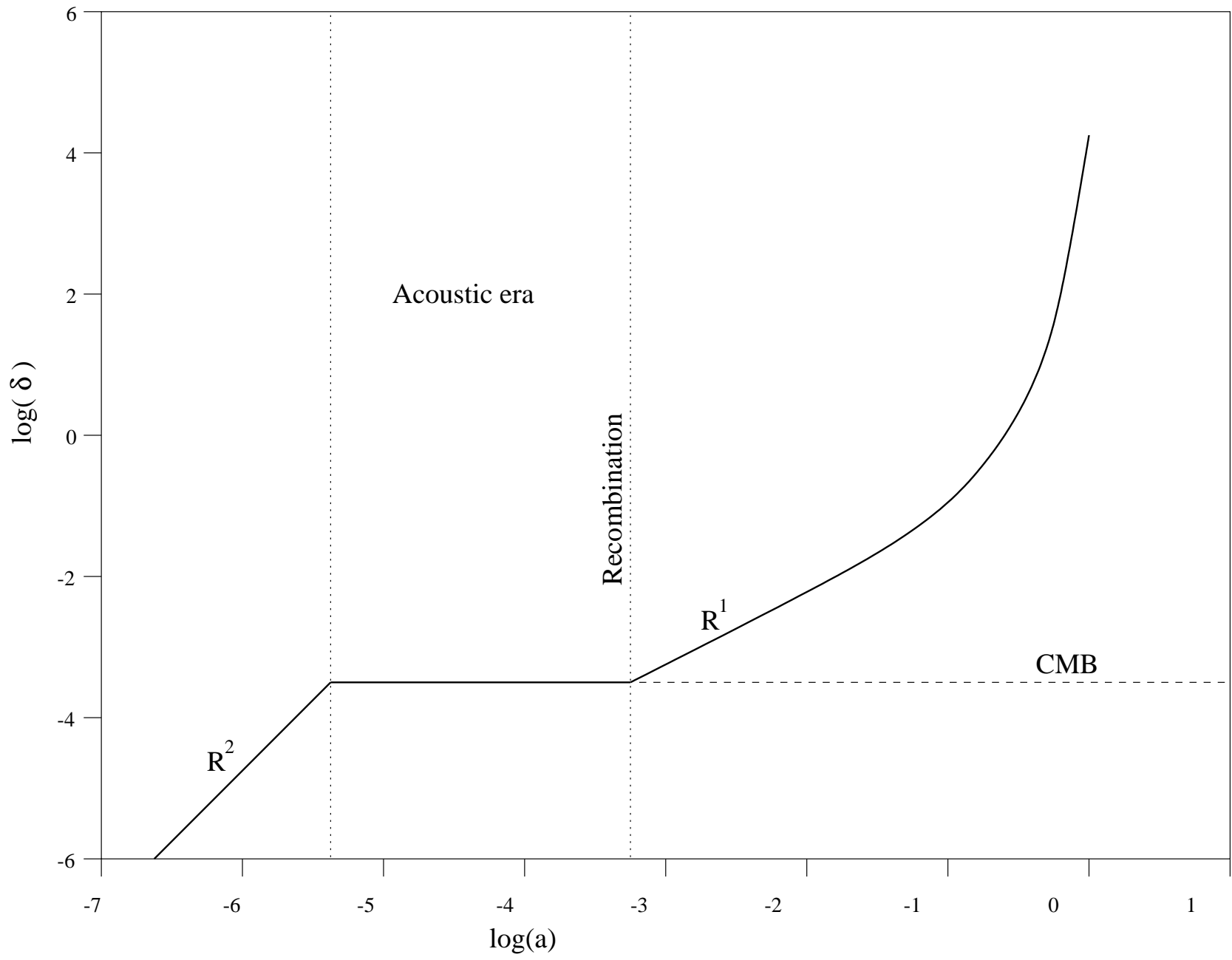
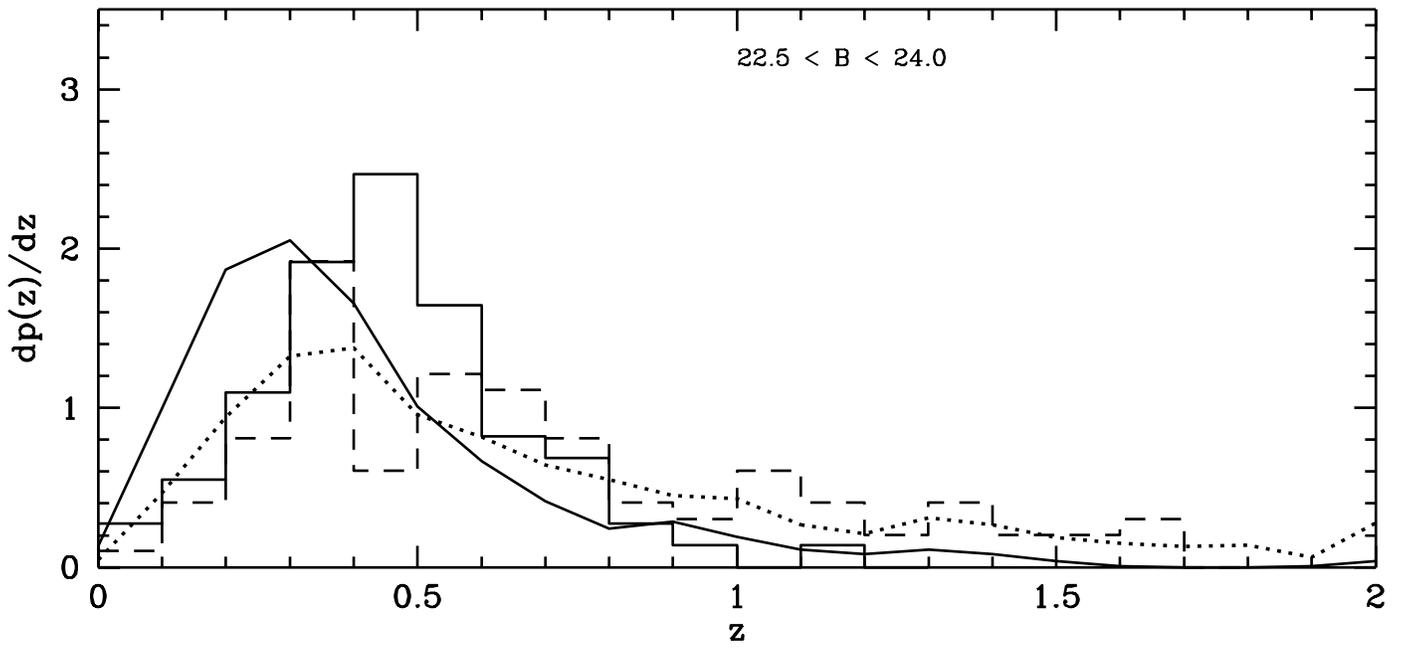
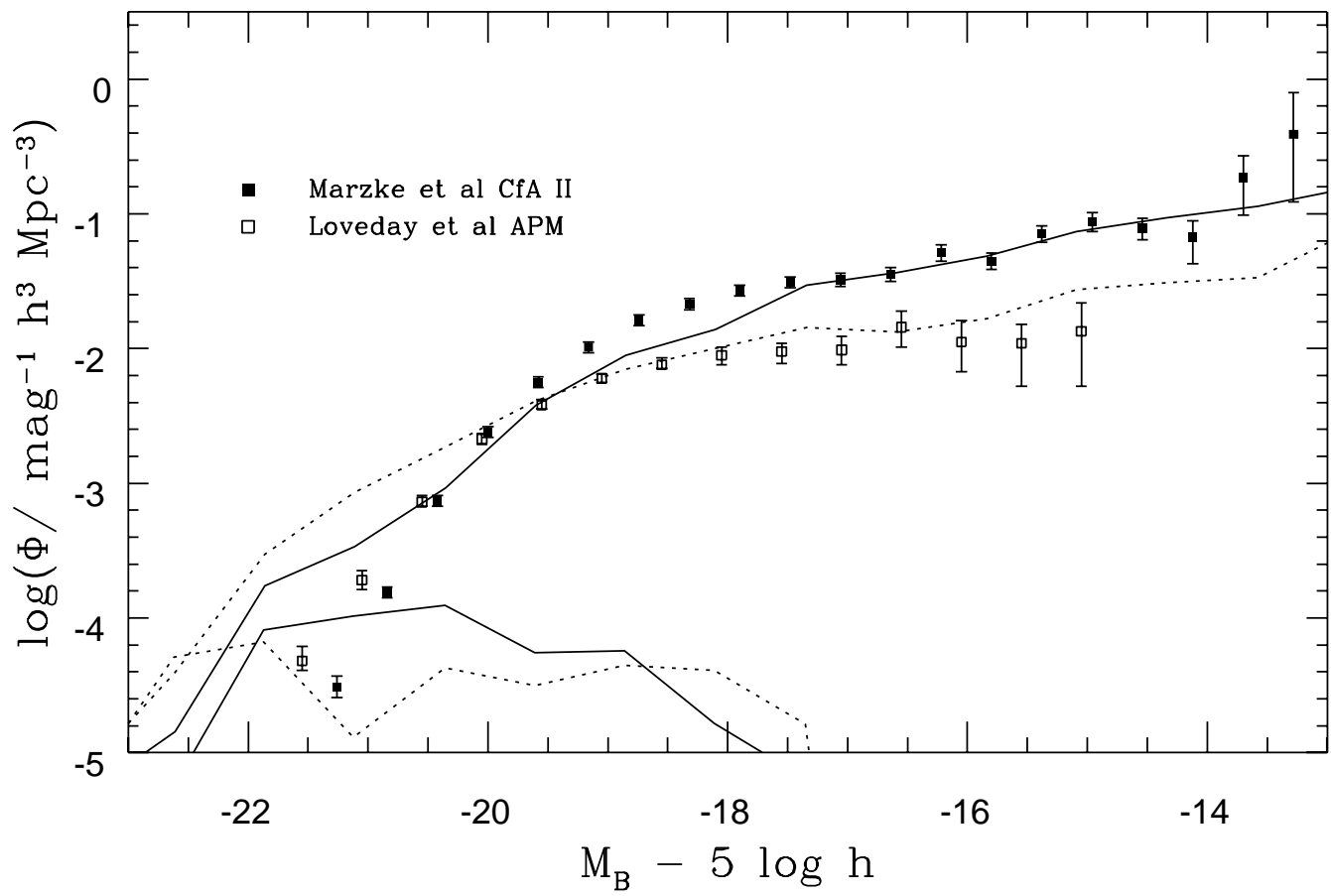
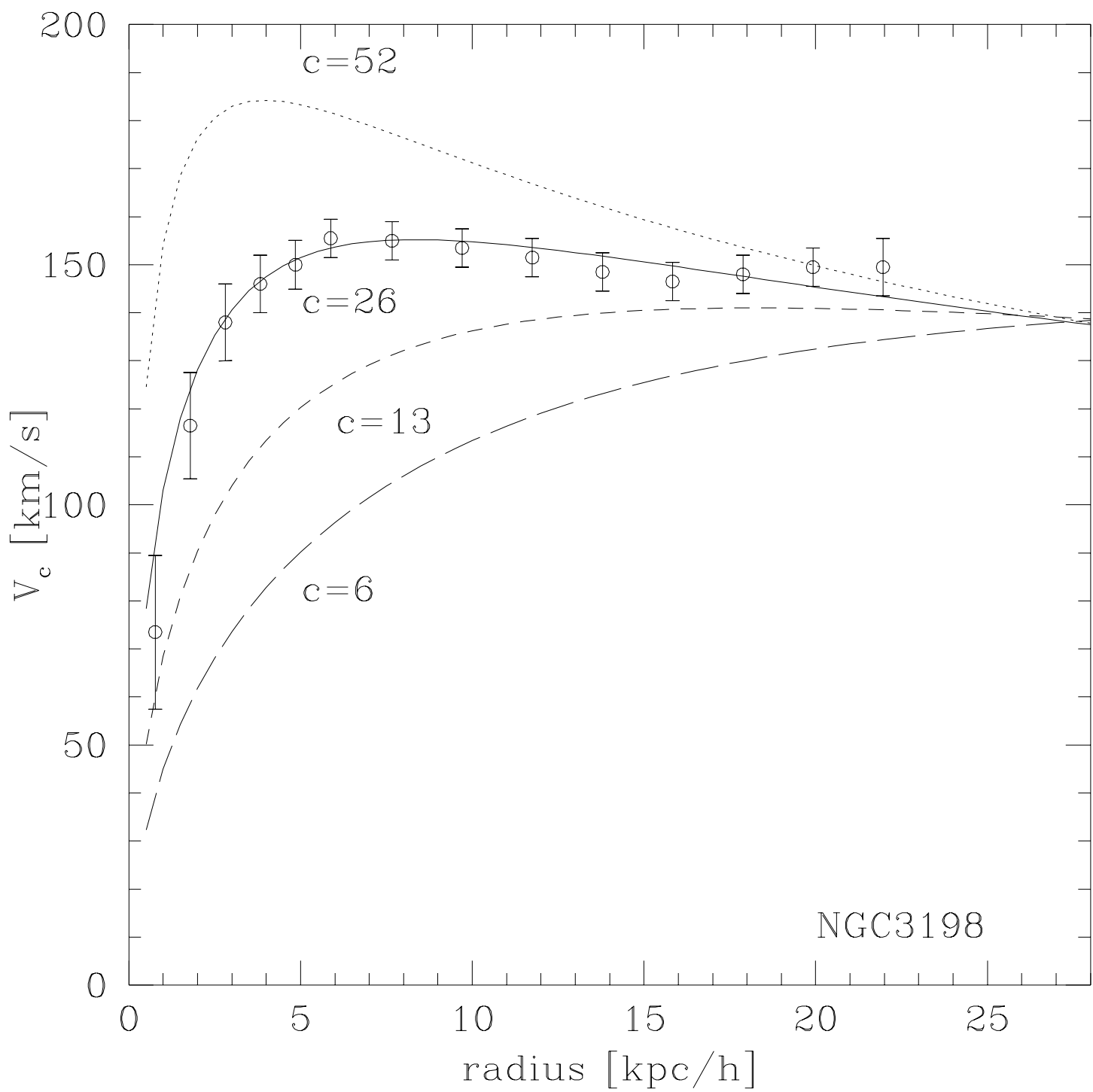
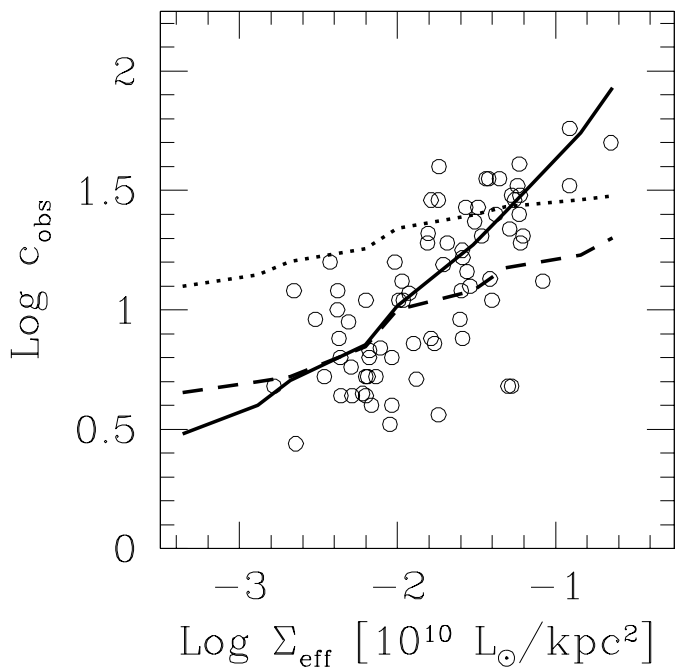
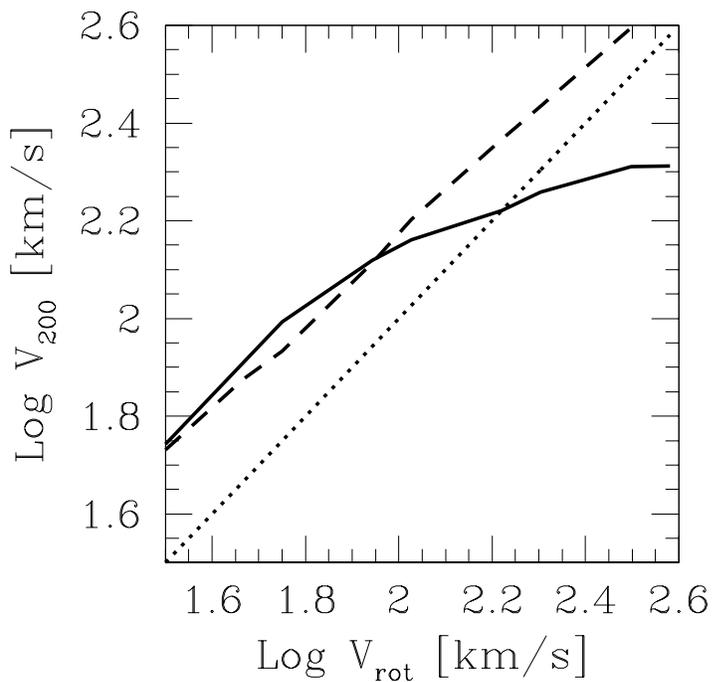
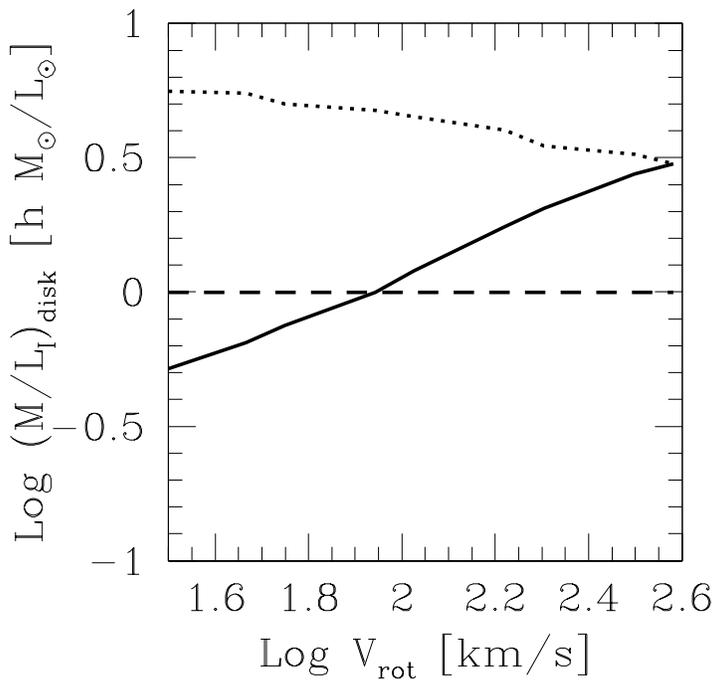
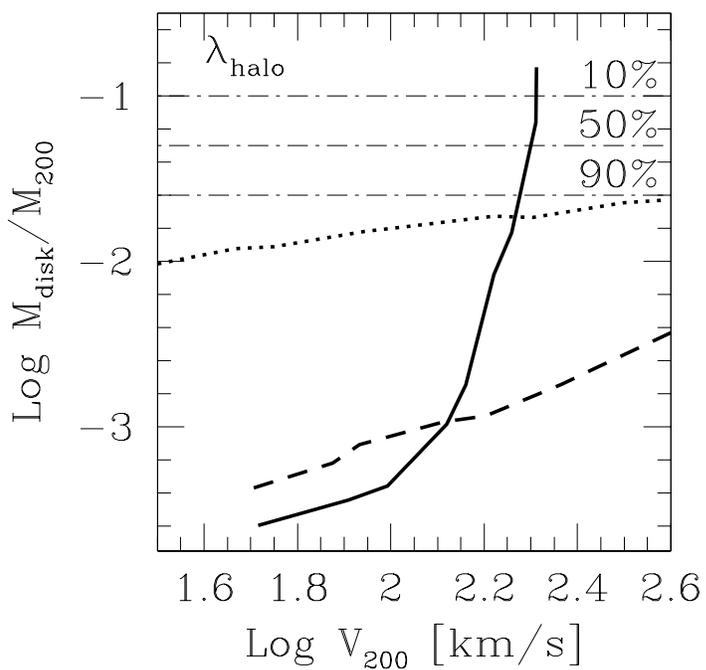
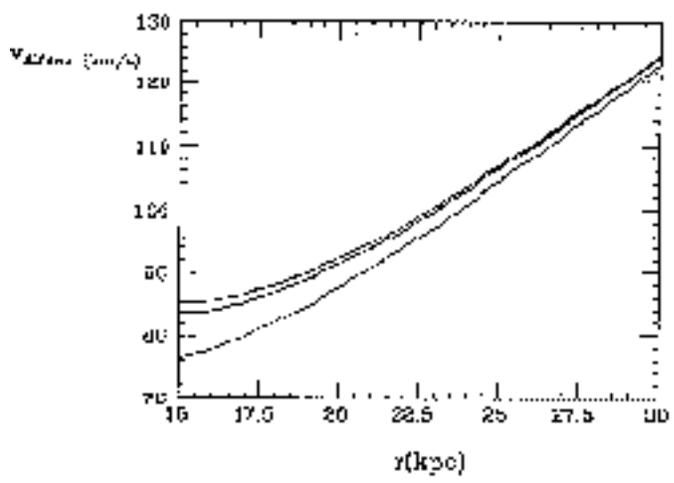
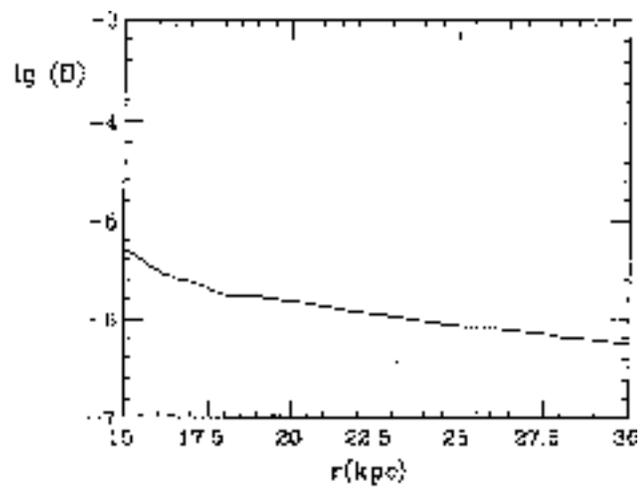
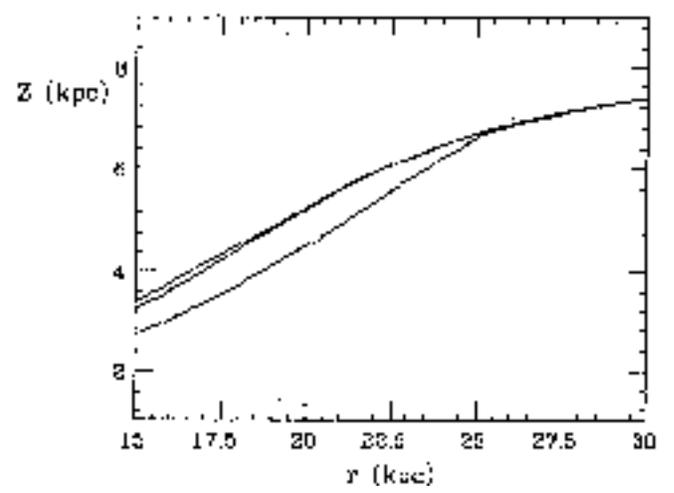
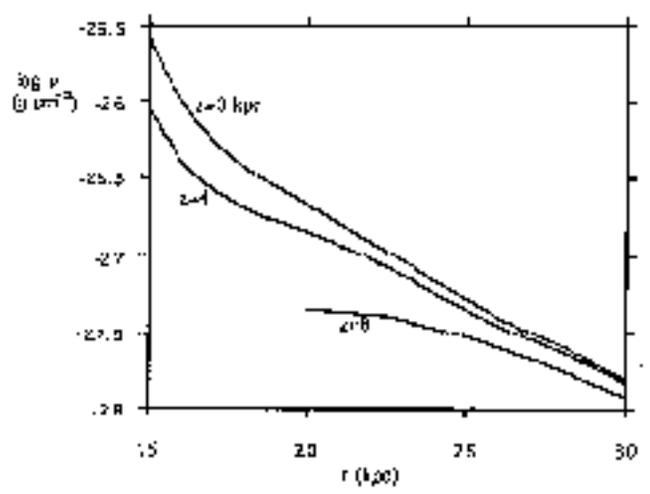
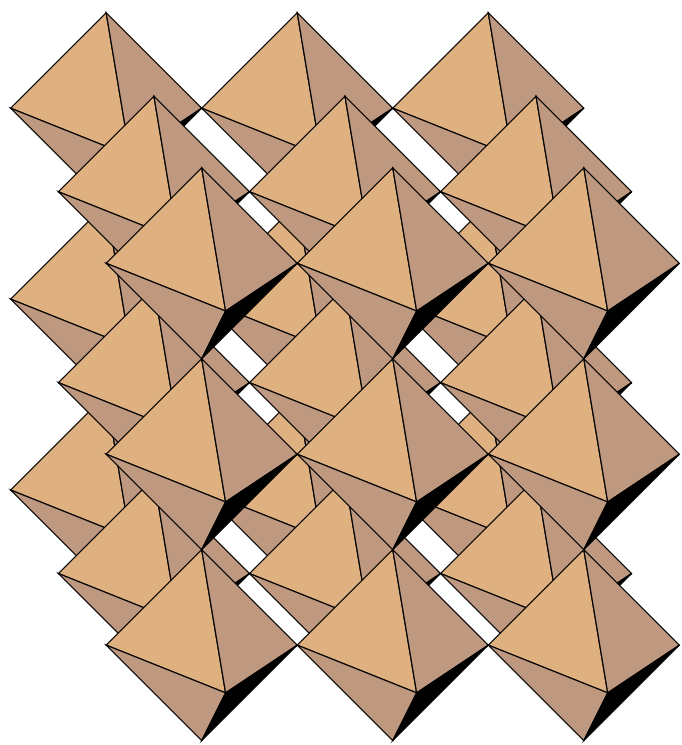
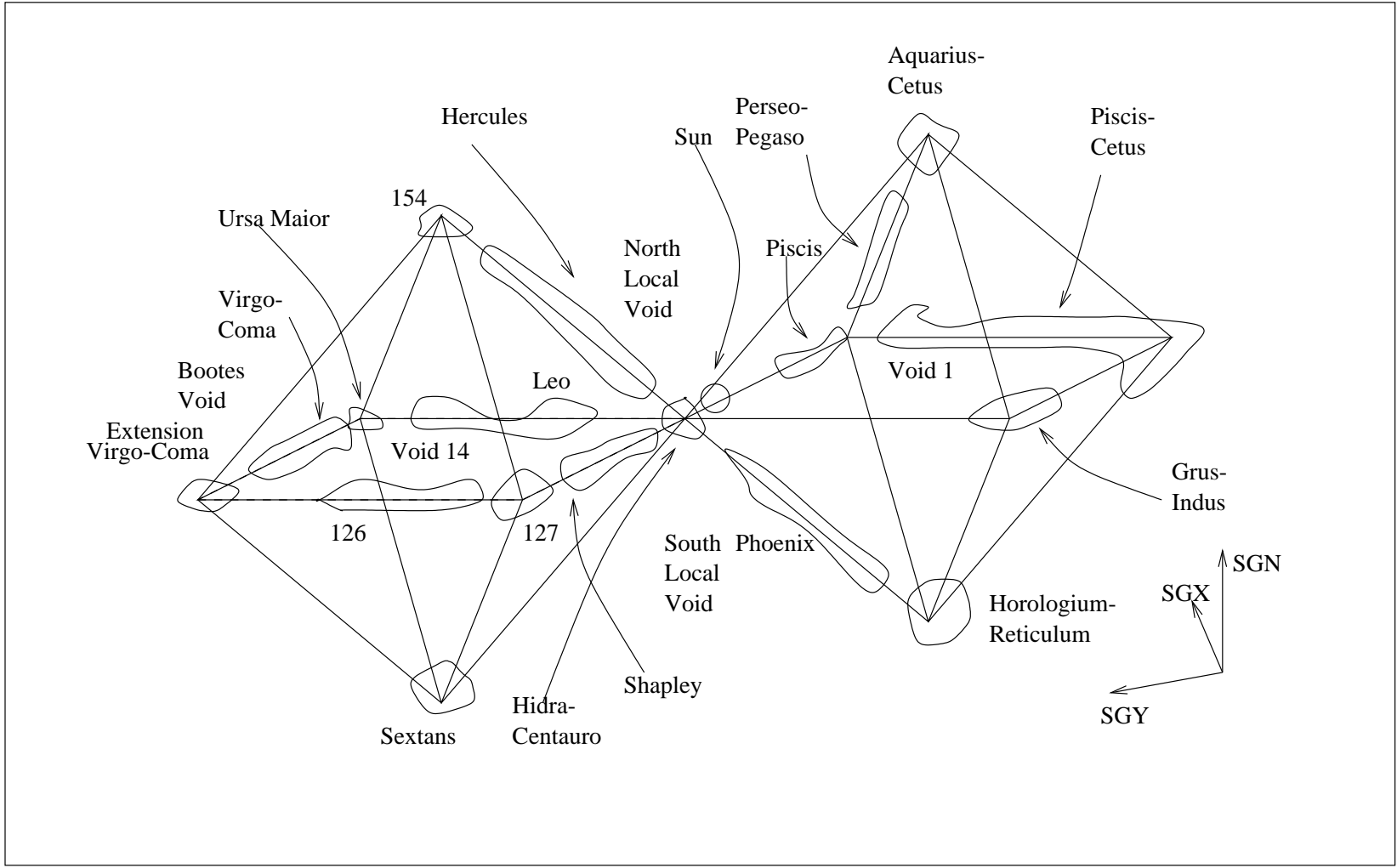
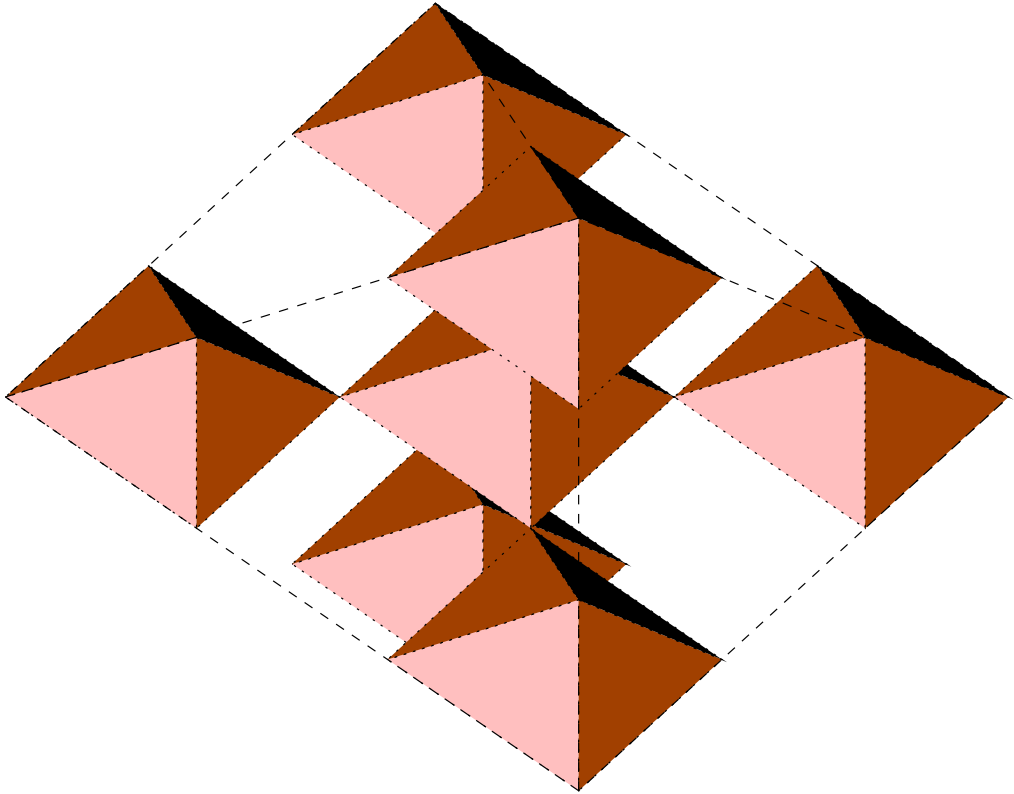
\end{document}